\documentclass[longauth]{aa} 

\usepackage{graphicx}
\usepackage{txfonts}
\usepackage{natbib}
\usepackage[colorlinks,urlcolor=cyan,citecolor=blue,linkcolor=blue]{hyperref}
\usepackage{amssymb,amsmath}
\usepackage{array}
\usepackage{booktabs}

\begin{document}

\title{A super-Earth and a sub-Neptune orbiting the bright, quiet M3 dwarf TOI-1266}

\author{ 
B.-O.~Demory\inst{1}
\and F.J.~Pozuelos\inst{2,3}
\and Y.~G\'omez Maqueo Chew\inst{4}
\and L.~Sabin\inst{5}
\and R.~Petrucci\inst{4,6,7}
\and U.~Schroffenegger\inst{1}
\and S.L.~Grimm\inst{1}
\and M.~Sestovic\inst{1}
\and M.~Gillon\inst{3}
\and J.~McCormac\inst{8,9}
\and K.~Barkaoui\inst{3,10}
\and W.~Benz\inst{1}
\and A. Bieryla\inst{11}
\and F.~Bouchy\inst{12}
\and A.~Burdanov\inst{13,14}
\and K.A.~Collins\inst{11}
\and J.~de Wit\inst{13}
\and C.D.~Dressing\inst{15}
\and L.J.~Garcia\inst{3}
\and S.~Giacalone\inst{15}
\and P.~Guerra\inst{16}
\and J.~Haldemann\inst{1}
\and K.~Heng\inst{1,8}
\and E.~Jehin\inst{2}
\and E.~Jofr\'e\inst{4,6,7}
\and S.R.~Kane\inst{17}
\and J.~Lillo-Box\inst{18}
\and V.~Maign\'e\inst{1}
\and C.~Mordasini\inst{19}
\and B.~M.~Morris\inst{1}
\and P.~Niraula\inst{13}
\and D.~Queloz\inst{20}
\and B.V.~Rackham\inst{13,21}
\and A.B.~Savel\inst{15,22}
\and A.~Soubkiou\inst{10}
\and G. Srdoc\inst{23}
\and K.G.~Stassun\inst{24}
\and A.H.M.J.~Triaud\inst{25}
\and R.~Zambelli\inst{26}
\and G.~Ricker\inst{21}
\and D.W.~Latham\inst{11}
\and S.~Seager\inst{13,21,27}
\and J.N.~Winn\inst{28}
\and J.M.~Jenkins\inst{29}
\and T.~Calvario-Vel\'asquez\inst{5}
\and J.A.~Franco Herrera\inst{5}
\and E.~Colorado\inst{5}
\and E.O.~Cadena Zepeda\inst{5}
\and L.~Figueroa\inst{5}
\and A.M.~Watson\inst{4}
\and E.E.~Lugo-Ibarra\inst{5}
\and L.~Carigi\inst{4}
\and G.~Guisa\inst{5}
\and J.~Herrera\inst{5}
\and G.~Sierra D\'iaz\inst{5}
\and J.C.~Su\'arez\inst{30,31}
\and D.~Barrado\inst{18}
\and N.M.~Batalha\inst{32}
\and Z.~Benkhaldoun\inst{10}
\and A.~Chontos\inst{33}
\and F.~Dai\inst{34}
\and Z.~Essack\inst{13,21}
\and M.~Ghachoui\inst{10}
\and C.X.~Huang\inst{21} 
\and D.~Huber\inst{33}
\and H.~Isaacson\inst{15,36}
\and J.J.~Lissauer\inst{29}
\and M.~Morales-Calder\'on\inst{18}
\and P.~Robertson\inst{37}
\and A.~Roy\inst{34}
\and J.D.~Twicken\inst{29,38}
\and A.~Vanderburg\inst{39}
\and L.M.~Weiss\inst{33}
}

\institute{Center for Space and Habitability, University of Bern, Gesellschaftsstrasse 6, CH-3012, Bern, Switzerland
\and Space Sciences, Technologies and Astrophysics Research (STAR) Institute, Universit\'e de Li\`ege, All\'ee du 6 Ao\^ut 19C, B-4000 Li\`ege, Belgium
\and Astrobiology Research Unit, Universit\'e de Li\`ege, All\'ee du 6 Ao\^ut 19C, B-4000 Li\`ege, Belgium
\and Instituto de Astronom\'ia, Universidad Nacional Aut\'onoma de M\'exico, Ciudad Universitaria, Ciudad de M\'exico, 04510, M\'exico
\and Instituto de Astronom\'ia, Universidad Nacional Aut\'onoma de M\'exico, Apdo. Postal 877, 22800, Ensenada, B.C., M\'exico
\and Universidad Nacional de C\'ordoba - Observatorio Astron\'omico de C\'ordoba, Laprida 854, X5000BGR, C\'ordoba, Argentina
\and Consejo Nacional de Investigaciones Científicas y Técnicas (CONICET), Argentina
\and Department of Physics, University of Warwick, Gibbet Hill Road, Coventry CV4 7AL, UK
\and Centre for Exoplanets and Habitability, University of Warwick, Gibbet Hill Road, Coventry CV4 7AL, UK
\and Oukaimeden Observatory, High Energy Physics and Astrophysics Laboratory, Cadi Ayyad University, Marrakech, Morocco
\and Center for Astrophysics | Harvard \& Smithsonian, 60 Garden Street, Cambridge, MA, 02138, USA
\and Observatoire astronomique de l'Universit\'e de Gen\`eve, 51 chemin des Maillettes, 1290 Versoix, Switzerland
\and Department of Earth, Atmospheric and Planetary Sciences, Massachusetts Institute of Technology, Cambridge, MA 02139, USA
\and Instituto de Astrof\'isica de Canarias, V\'ia L\'actea s/n, 38205 La Laguna, Tenerife, Spain
\and Department of Astronomy, 501 Campbell Hall, University of California at Berkeley, Berkeley, CA 94720, USA
\and Observatori Astron\`omic Albany\`a, Cam\'i de Bassegoda S/N, Albany\`a 17733, Girona, Spain
\and Department of Earth and Planetary Sciences, University of California, Riverside, CA 92521, USA
\and Centro de Astrobiolog\'ia (CAB, CSIC-INTA), Dpto. de Astrof\'isica, ESAC campus 28692 Villanueva de la Ca\~nada (Madrid), Spain
\and Physikalisches Institut, University of Bern, Gesellschaftsstrasse 6, CH 3012 Bern, Switzerland	   \and Cavendish Laboratory, J.J. Thomson Avenue, Cambridge, CB3 0HE, UK
\and Kavli Institute for Astrophysics and Space Research, Massachusetts Institute of Technology, 77 Massachusetts Avenue, Cambridge, MA 02139, USA
\and Department of Physics, University of California, Berkeley, Berkeley, CA
\and Kotizarovci Observatory, Sarsoni 90, 51216 Viskovo, Croatia
\and Department of Physics \& Astronomy, Vanderbilt University, 6301 Stevenson Center Lane, Nashville, TN, 37235, USA
\and School of Physics \& Astronomy, University of Birmingham, Edgbaston, Birmingham B15 2TT, UK
\and Societ\`a Astronomica Lunae, Castelnuovo Magra, via Montefrancio,77, Italy
\and Department of Aeronautics and Astronautics, MIT, 77 Massachusetts Avenue, Cambridge, MA 02139, USA
\and Department of Astrophysical Sciences, Princeton University, NJ 08544, USA
\and NASA Ames Research Center, Moffett Field, CA, 94035, USA
\and Dpt. F\'isica Te\'orica y del Cosmos, Universidad de Granada, Campus de Fuentenueva s/n, 18071, Granada, Spain 
\and Instituto de Astrof\'isica de Andaluc\'ia (CSIC), Glorieta de la Astronom\'ia s/n, 18008, Granada, Spain
\and Department of Astronomy \& Astrophysics, University of California, Santa Cruz, CA 95064, USA
\and Institute for Astronomy, University of Hawaii, 2680 Woodlawn Drive, Honolulu, HI 96822, USA
\and Division of Geological and Planetary Sciences, California Institute of Technology, 1200 East California Blvd, Pasadena, CA,USA 91125
\and Department of Astronomy, California Institute of Technology, Pasadena, CA 91125, USA
\newpage
\and Centre for Astrophysics, University of Southern Queensland, Toowoomba, QLD, Australia
\and Department of Physics \& Astronomy, University of California Irvine, Irvine, CA 92697, USA
\and SETI Institute, Mountain View, CA 94043, USA
\and Department of Astronomy, The University of Texas at Austin, Austin, TX 78712, USA
}

\titlerunning{A super-Earth and a sub-Neptune orbiting the M3V TOI-1266}\authorrunning{Demory et al.}

\date{Accepted for publication on 20 July 2020.}

\abstract{We report the discovery and characterisation of a super-Earth and a sub-Neptune transiting the bright ($K=8.8$), quiet, and nearby (37~pc) M3V dwarf TOI-1266. We validate the planetary nature of TOI-1266\,b and c using four sectors of TESS photometry and data from the newly-commissioned 1-m SAINT-EX telescope located in San Pedro M\'artir (Mexico). We also include additional ground-based follow-up photometry as well as high-resolution spectroscopy and high-angular imaging observations. The inner, larger planet has a radius of $R=2.37_{-0.12}^{+0.16}$ R$_{\oplus}$ and an orbital period of 10.9 days. The outer, smaller planet has a radius of $R=1.56_{-0.13}^{+0.15}$ R$_{\oplus}$ on an 18.8-day orbit. The data are found to be consistent with circular, co-planar and stable orbits that are weakly influenced by the 2:1 mean motion resonance. Our TTV analysis of the combined dataset enables model-independent constraints on the masses and eccentricities of the planets. We find planetary masses of $M_\mathrm{p}$ = $13.5_{-9.0}^{+11.0}$ $\mathrm{M_{\oplus}}$ ($<36.8$ $\mathrm{M_{\oplus}}$ at 2-$\sigma$) for TOI-1266\,b and $2.2_{-1.5}^{+2.0}$ $\mathrm{M_{\oplus}}$ ($<5.7$ $\mathrm{M_{\oplus}}$ at 2-$\sigma$) for TOI-1266\,c. We find small but non-zero orbital eccentricities of $0.09_{-0.05}^{+0.06}$ ($<0.21$ at 2-$\sigma$) for TOI-1266\,b and $0.04\pm0.03$ ($<0.10$ at 2-$\sigma$) for TOI-1266\,c. The equilibrium temperatures of both planets are of $413\pm20$\,K and $344\pm16$\,K, respectively, assuming a null Bond albedo and uniform heat redistribution from the day-side to the night-side hemisphere. The host brightness and negligible activity combined with the planetary system architecture and favourable planet-to-star radii ratios makes TOI-1266 an exquisite system for a detailed characterisation.
}


\maketitle
\section{Introduction}

The science of exoplanets has been historically driven by dedicated astronomical observations. Currently, the Transiting Exoplanet Survey Satellite \citep[TESS;][]{Ricker:2015}
is leading the discovery of multi-planetary transiting systems with relatively small planets (i.e. sub-Neptune or smaller), orbiting around bright M-dwarf stars in the solar neighbourhood \citep[e.g.][]{Gunther2019,Jenkins2019,Kostov2019,Cloutier2020}. Their brightness allows for a detailed characterisation of small planets, and, in the near future, a glimpse into their atmospheric composition with the James Webb Space Telescope (JWST). Furthermore, when multiple transit-like signals are detected from a single star, the signals are likely to be genuine as opposed to false positives such as eclipsing binaries \citep{Latham:2011a,Lissauer2012,Morton:2016}. In some cases, in particular where the planets are near resonant orbits, time-series photometry alone not only allows for the measurement of the planet size, but also places dynamical constraints on the planet mass \citep{Holman:2005,Agol:2005}. Measuring the planet mass and radius allows for the derivation of the bulk density, thus constraining planetary structure models \citep[e.g.][]{Dorn2017}.

There are more than 3000 transiting exoplanets known to date\footnote{\url{http://exoplanet.eu/} retrieved on 10 Aug 2020}, including 499 planetary systems with more than one detected transiting planet. 
This large sample of transiting exoplanets allows for in-depth exploration of the distinct exoplanet populations. One such study by \citet{Fulton:2017} identified a bi-modal distribution for the sizes of super-Earth and sub-Neptune \textit{Kepler} exoplanets, with a peak at $\sim$1.3~R$_\oplus$ and another at $\sim$2.4~R$_\oplus$. The interval between the two peaks is called the radius valley and it is typically attributed to the stellar irradiation received by the planets, with more irradiated planets being smaller due to the loss of their gaseous envelopes. Studying more than one super-Earth or sub-Neptune planet in a single system allows for tighter constraints on formation models \citep[e.g.][]{Owen:2019,Kubyshkina2019} and thus the exploration of the effects of other physical processes such as the core and envelope mass distribution \citep[e.g.][]{Modirrousta2020}.

Here we present the discovery and characterisation of the planetary system TOI-1266, which was first identified from the TESS photometry. We confirmed the planetary nature of the transits through ground-based follow-up observations, including time-series photometry, high-angular resolution images, spectroscopy, and archival imagery.

The paper is structured as follows. Section~\ref{sec:obs} describes all observations. The stellar characterisation of the planet host is described in \S\ref{sec:star}. The validation of the transit signals in the light curves is presented in \S\ref{sec:validation}.  The search for transit signals and the analysis of the light curves to derive physical properties are presented in \S\ref{sec:results}. We also include a stability analysis and mass constraints from a dynamical analysis in \S\ref{subsec:dyn}. Finally, in \S\ref{sec:disc}, we discuss the implications for the formation and evolution of the TOI-1266 planetary system given the measured planet radii, orbital periods, and constraints on the masses, as well as prospects for atmospheric characterisation. A summary of our results and their implications is presented in \S\ref{sec:concl}.

\section{Observations}\label{sec:obs}

In this section, we present all the observations of TOI-1266 obtained with TESS and ground-based facilities. A summary of all ground-based time-series photometric observations of TOI-1266 is shown in Table~\ref{tab:lcs}. 

\begin{table*}
\begin{center}
\begin{tabular}{l c c c c}
\toprule
Date (UT) & Filter &  Facility & Exp. time [s] & Notes\\
\midrule

25 Dec 2019 & $Ic$ & OAA-0.4m & 160 & b full \\
29 Jan 2020 & $z'$ & SAINT-EX-1m & 12 & c partial \\
29 Feb 2020 & $z'$ & SAINT-EX-1m & 12 & b full \\
21 Mar 2020 & $z'$ & TRAPPIST-N-0.6m & 15 & b full \\
21 Mar 2020 & $r'$ & Artemis-1m & 10 & b partial\\
01 Apr 2020 & $V$ &  TRAPPIST-N-0.6m & 60 & b partial \\
23 Apr 2020 & $TESS$ & Kotizarovci-0.3m & 50 & b full \\
23 Apr 2020 & clear & ZRO-0.4m & 200 & b full \\
\hline 
\end{tabular}
\caption{Ground-based time-series photometric observations of TOI-1266.\label{tab:lcs}}
\end{center}
\end{table*}

\subsection{TESS photometry}\label{subsec:tess}

TOI-1266 is a late-type star with a measured parallax that is part of the TESS Candidate Target List \citep{Stassun:2018}. It was observed by TESS with 2-min-cadence in sectors 14\---15 (18 July to 11 Sep 2019) and 21\---22 (21 Jan to 18 Mar 2020). TOI-1266's astrometric and photometric properties from the literature are reported in Table~\ref{tab:starlit}.
The time-series observations of TOI-1266 were processed with the TESS Science Processing Operations Center (SPOC) pipeline \citep{Jenkins2002,Jenkins2016,Jenkins:2017}, which resulted in the detection of two periodic transit signals:
TOI-1266.01 and .02, the latter being at the detection limit using sectors 14-15 data alone, thus requiring additional data to strengthen that signal.

\begin{figure*}
    \centering
    \includegraphics[width=0.24\textwidth]{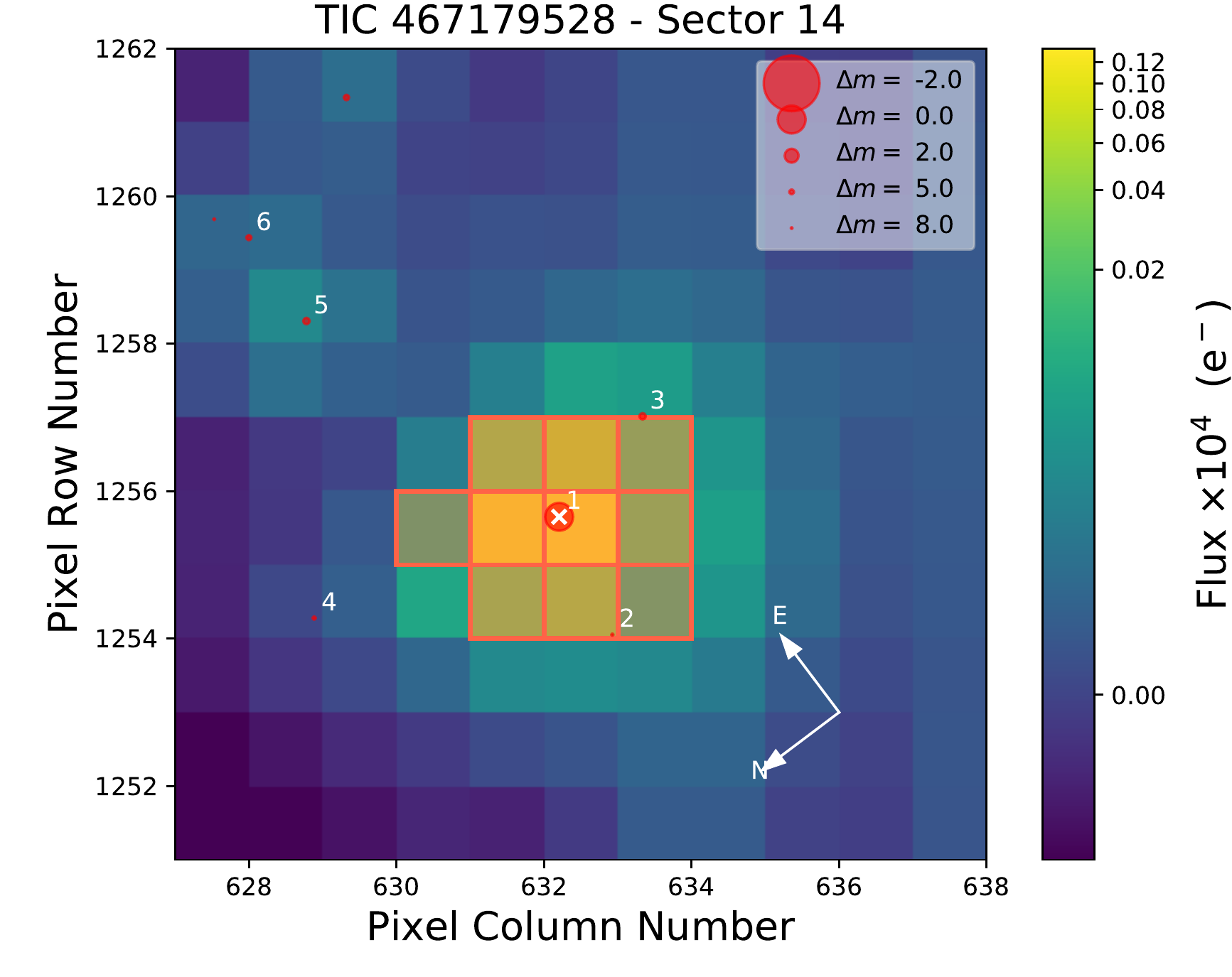}
    \includegraphics[width=0.24\textwidth]{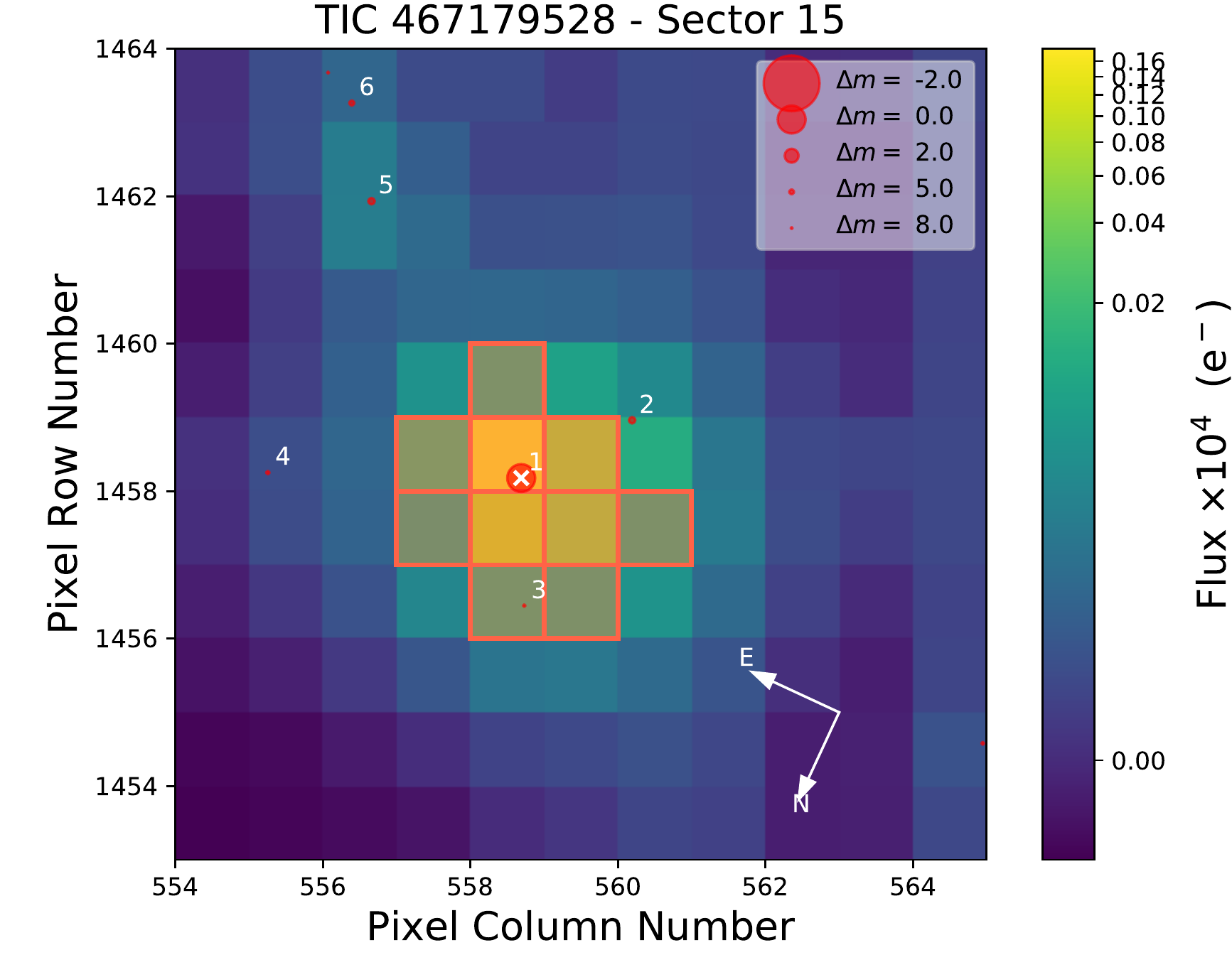}
    \includegraphics[width=0.24\textwidth]{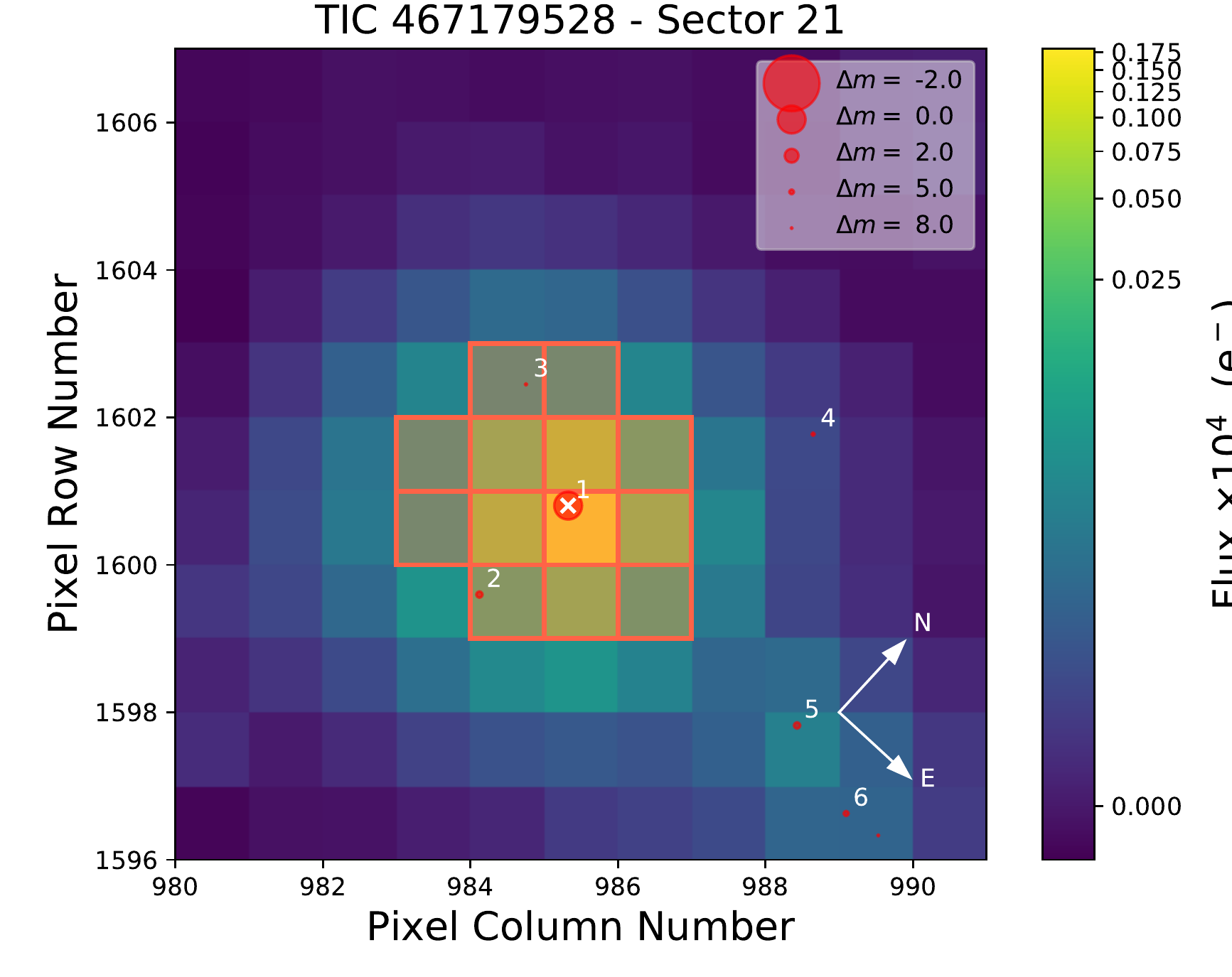}
    \includegraphics[width=0.24\textwidth]{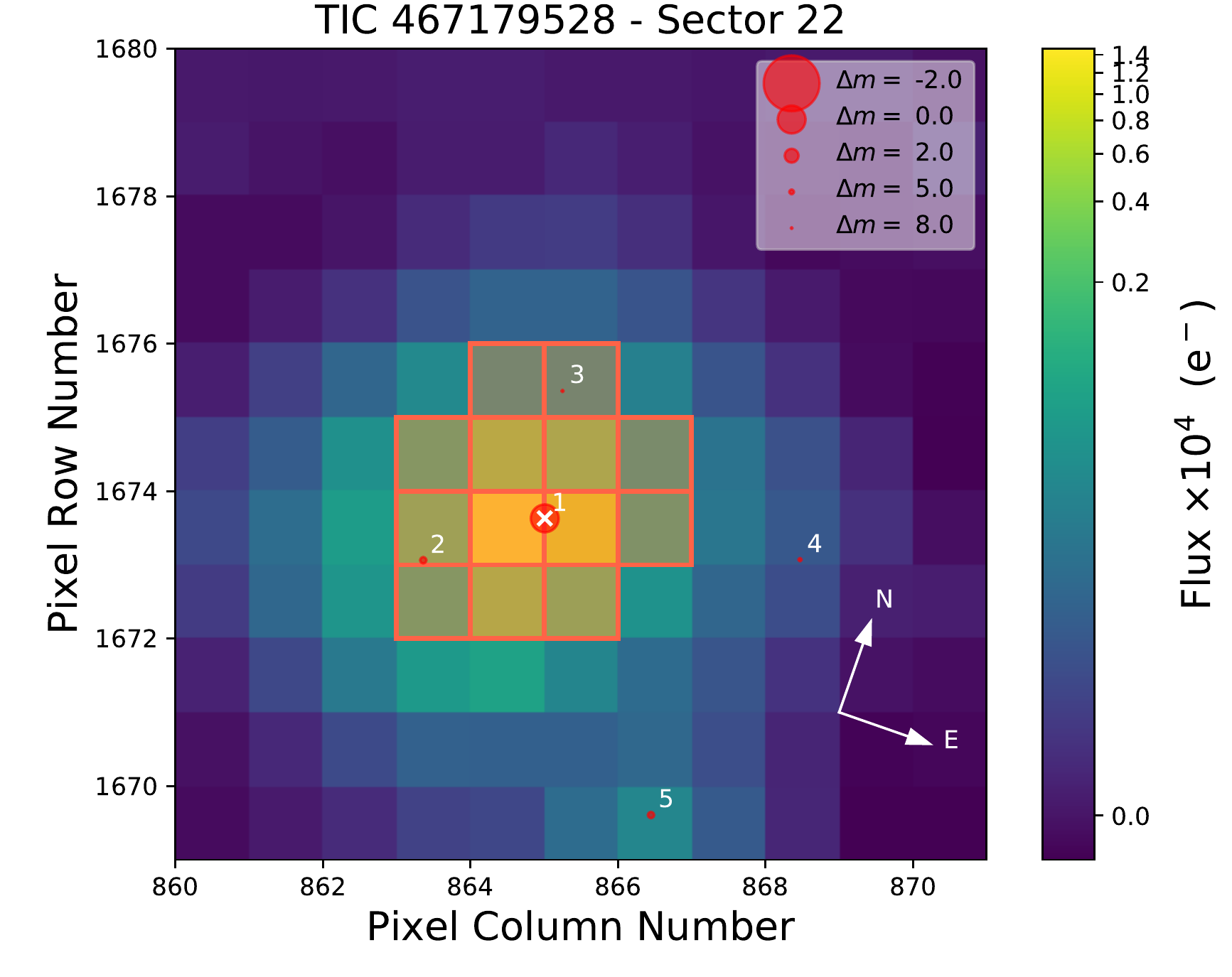}
    \caption{TESS target pixel files (TPFs) of the four sectors that observed TOI-1266. 
    The plots were created with \texttt{tpfplotter}\protect\footnotemark\   \citep{aller20}. 
    The apertures used to extract the photometry \citep{Twicken2010,Morris2017} by the SPOC pipeline are shown as red shaded regions. The {\it Gaia\/} DR2 catalogue \citep{gaia18} is overplotted, with all sources up to 8 magnitudes in contrast with TOI-1266 are shown as red circles. We note that the symbol size scales with the magnitude contrast.}
    \label{fig:tpf}
\end{figure*}

We retrieved the Presearch Data Conditioning Simple  Aperture  Photometry (PDC-SAP) \citep{Stumpe2012,Smith2012,Stumpe2014}
from the Mikulski Archive for Space Telescopes
and removed all datapoints flagged as `bad quality'. We identified 819/19337 such datapoints for sector 14, 912/18757 for sector 15, 1074/19694 for sector 21, and a larger count (5652/19579) for sector 22. Figure~\ref{fig:tpf} shows the TESS fields of view and apertures used for TOI-1266 over each of the four sectors with the location of nearby {\it Gaia\/} DR2 sources superimposed.

\begingroup
\begin{table}
\begin{center}
\renewcommand{\arraystretch}{1.15}
\begin{tabular*}{\linewidth}{@{\extracolsep{\fill}}l c c}
\toprule
Parameter & Value & Source \\
\midrule
\multicolumn{3}{c}{\textit{Target designations}} \\
TIC      & 467179528           & 1 \\
2MASS    & J13115955+6550017   & 2 \\
UCAC 4   & 780-025091          & 3 \\
{\it Gaia\/} DR2 & 1678074272650459008 & 4 \\
\midrule
\multicolumn{3}{c}{\textit{Photometry}} \\
$TESS$	& 11.040 $\pm$ 0.007   & 1 \\
$B$	    & 14.58 $\pm$ 0.05     & 3 \\
$V$	    & 12.94 $\pm$ 0.05     & 3 \\
$Gaia$	& 12.1222 $\pm$ 0.0002 & 4 \\
$u$	    & 16.527 $\pm$ 0.007   & 5 \\
$g$	    & 14.950 $\pm$ 0.005   & 5 \\
$r$	    & 12.584 $\pm$ 0.002   & 5 \\
$i$	    & 11.600 $\pm$ 0.001   & 5 \\
$z$	    & 11.608 $\pm$ 0.005   & 5 \\
$J$	    & 9.71 $\pm$ 0.02      & 2 \\
$H$	    & 9.07 $\pm$ 0.03      & 2 \\
$K$	    & 8.84 $\pm$ 0.02      & 2 \\
WISE 3.4 $\mu$m	& 8.72 $\pm$ 0.02 & 6 \\
WISE 4.6 $\mu$m	& 8.61 $\pm$ 0.02 & 6 \\
WISE 12 $\mu$m	& 8.50 $\pm$ 0.02 & 6 \\
WISE 22 $\mu$m	& 8.23 $\pm$ 0.21 & 6 \\
\midrule
\multicolumn{3}{c}{\textit{Astrometry}} \\
RA  (J2000) & 13 11 59.56   & 4 \\
DEC (J2000) & +65 50 01.70  & 4 \\
RA PM (mas/yr)  & -150.652 $\pm$ 0.041  & 4 \\
DEC PM (mas/yr) & -25.368 $\pm$ 0.039   & 4 \\
Parallax (mas) & 27.7397 $\pm$ 0.0226 & 4 \\
\bottomrule
\end{tabular*}
\end{center}
\caption{
TOI-1266 stellar astrometric and photometric properties. 1. \citet{Stassun:2018}, 2. \citet{Cutri:2003}, 3. \citet{Zacharias:2013}, 4. \citet{Brown:2018}, 5. \citet{Alam:2015}, 6. \citet{Cutri:2013}.
\label{tab:starlit}} 

\end{table}
\endgroup

\subsection{SAINT-EX photometry}
\footnotetext{\url{https://github.com/jlillo/tpfplotter}}

We obtained ground-based photometric time-series observations of TOI-1266 from the SAINT-EX  Observatory\footnote{\url{https://www.saintex.unibe.ch/saint_ex_observatory/}} (Search And characterIsatioN of Transiting EXoplanets), which was commissioned in March 2019.
SAINT-EX  is a 1-m F/8 Ritchey-Chr\'etien telescope built by the ASTELCO company and uses a similar design to the telescopes of the SPECULOOS Southern Observatory \citep{Delrez:2018b,Jehin:2018}. SAINT-EX is located at the Observatorio Astron\'omico Nacional, in the Sierra de San Pedro M\'artir in Baja California, M\'exico (31.04342 N, 115.45476 W) at 2780 m altitude. The telescope is installed on an ASTELCO equatorial NTM-1000 mount equipped with direct-drive motors, which enables operations without meridian flip. The telescope is installed in a 6.25-m wide dome built by the Gambato company. In terms of mount performance, SAINT-EX typically achieves a RMS better than 3\arcsec\ relative to the pointing model and a tracking accuracy \--- without autoguiding \--- better than 2\arcsec\ over 15-min timescales. To improve this figure further, SAINT-EX uses the DONUTS autoguiding software \citep{McCormac:2013}, which increases the guiding precision to 0.2\arcsec\ RMS or better that is less than a pixel. SAINT-EX is equipped with an Andor iKon-L camera that integrates a deep-depletion e2v 2K $\times$ 2K CCD chip with a BEX2-DD coating that is optimised for the near infrared (NIR). The filter wheel includes the Sloan $ugriz'$ broad-band filters, as well as special blue-blocking (transmittance > 90\% beyond 500 nm) and NIR (transmittance > 95\% beyond 705 nm) filters. The detector gives a field of view of 12\arcmin$\times$12\arcmin\ with 0.34\arcsec per pixel. 

SAINT-EX operations are robotic and the data reduction and analysis are automated by a custom pipeline PRINCE (Photometric Reduction and In-depth Nightly Curve Exploration) that ingests the raw science and calibration frames and produces light curves using differential photometry. 
The PRINCE pipeline performs standard image reduction steps, applying bias, dark, and flat-field corrections.
Astrometric calibration is conducted using Astrometry.net \citep{Lang_2010} to derive correct world coordinate system (WCS)  information for each exposure.
\texttt{Photutils} star detection \citep{Bradley_2019_2533376} is run on a median image of the whole exposure stack to create a pool of candidate stars in the field of view. Stars whose peak value in the largest aperture is above the background by a certain threshold, defined by an empirical factor times the median background noise of the night, are kept as reference stars for the differential photometric analysis. 
From the WCS information and the detected stars' coordinates, the pipeline runs centroiding, aperture and annulus photometry on each detected star from the common pool, using LMFit \citep{newville_matthew_2014_11813} and Astropy \citep{astropy:2013, astropy:2018}, and repeats this for each exposure to obtain the measured lightcurves for a list of apertures.
The measured lightcurves for each aperture are corrected for systematics using either a PCA approach \citep{scikit-learn} or a simple differential photometry approach that corrects a star's lightcurve by the median lightcurve of all stars in the pool except for the target star.

SAINT-EX observed one transit of each planet of the TOI-1266 system in early 2020. The observing strategy was to use the $z'$ filter with a slightly-defocused 12-s exposure time to mitigate shutter noise and scintillation. A partial transit of TOI-1266.02 was observed on 29 January 2020 from 7:36 to 12:05 UT. A full transit of TOI-1266.01 was then observed on 29 February 2020 from 6:16 to 11:03 UT. We reduced both datasets with PRINCE using differential aperture photometry. We corrected our differential light curves for variations in both the airmass and the full width at half maximum (FWHM) along both horizontal and vertical axes on the detector. This correction is performed simultaneously to the transit fit in our MCMC framework detailed in Section~\ref{sec:ga}.

\subsection{TRAPPIST-North photometry}

We used the 60-cm TRAPPIST-North telescope located at Oukaimeden Observatory in Morocco \citep{jehin2011,gillon2013,barkaoui2019} to observe one full and one partial transit of TOI-1266.01. TRAPPIST-North is equipped with a thermoelectrically cooled $2K\times2K$ Andor iKon-L BEX2-DD CCD camera with a pixel scale of 0.6\arcsec\ and a field of view of $20\arcmin\times20\arcmin$. We used the TESS Transit Finder tool, which is a  customised version of the Tapir software package \citep{jensen2013}, to schedule the photometric time-series. Data reduction and photometric analysis were performed  with the \textit{TRAPPHOT} pipeline developed in \textit{IRAF} as described in \citet{gillon2013}. The first transit was observed on 21 March 2020 in the Sloan-$z'$ filter with an exposure time of 15 seconds. We took 476 raw images and performed aperture photometry in an uncontaminated aperture of 6.5~pixels (4.0\arcsec) and a FWHM of 2.6\arcsec. The second transit was observed on 01 April 2020 in the Johnson-$V$ filter with an exposure time of 60 seconds. We took 338 raw images and we performed aperture photometry in an uncontaminated aperture of 8.1~pixels (5.0\arcsec) and a FWHM of 3.4\arcsec. During that second observation of TOI-1266, the telescope underwent a meridian flip at HJD 2458941.5431, and we noticed the presence of thin clouds during the ingress.

\subsection{Artemis photometry}

One partial transit of TOI-1266.01 was acquired on 21 March 2020 with the Artemis telescope, which constitutes the SPECULOOS-North facility located at the Teide Observatory (Canary Islands, Spain). The Artemis telescope is a twin of the SPECULOOS-South and SAINT-EX telescopes, which are all operated similarly and utilise Andor iKon-L cameras with e2v 2K $\times$ 2K deep-depletion CCDs. We acquired 668 images in the Sloan-$r'$ filter with an exposure time of 10 seconds. Data reduction consisted of standard calibration steps and subsequent aperture photometry with the \textit{TRAPPHOT} pipeline. Best comparison stars and optimum aperture size were selected on the basis of the  minimisation of the out-of-transit scatter of the light curve. 

\subsection{Observatori Astron\`omic Albany\`a}

A full transit of TOI-1266~b was observed in I$_{\rm C}$ band on 25 Dec 2019 from Observatori Astron\`omic Albany\`a (OAA) in Girona, Spain. The 0.4 m telescope is equipped with a $3056\times3056$ Moravian G4-9000 camera with an image scale of 1$\farcs$44 pixel$^{-1}$, resulting in a $36\arcmin\times36\arcmin$ field of view. The images were calibrated and the photometric data were extracted using the {\tt AstroImageJ} ({\tt AIJ}) software package \citep{Collins:2017}.

\subsection{Kotizarovci Observatory}

A full transit of TOI-1266~b was observed in Baader R longpass 610 nm band on 23 Apr 2020 from Kotizarovci Observatory near Viskovo, Croatia. The 0.3 m telescope is equipped with a $765\times510$ SBIG ST7XME camera with an image scale of 1$\farcs$2 pixel$^{-1}$ resulting in a $15.3\arcmin\times10.2\arcmin$ field of view. The images were calibrated and the photometric data were extracted using {\tt AIJ}.

\subsection{ZRO Observatory}

A full transit of TOI-1266~b was observed without filter on 23 Apr 2020 from the Zambelli Roberto Observatory (ZRO) in Liguria, Italy. The 0.4-m telescope is equipped with a $3072\times2048$ SBIG STXL-6303E camera with an image scale of 1$\farcs$16 pixel$^{-1}$ resulting in a $49\arcmin\times33\arcmin$ field of view. The images were calibrated and the photometric data were extracted using {\tt AIJ}.

\subsection{TRES spectroscopy}
\label{sectionTRES}

We obtained two `reconnaissance' spectra with the Tillinghast Reflector Echelle Spectrograph \citep[TRES;][]{Furesz:2008} instrument mounted on the Fred Lawrence Whipple Observatory's 1.5-m telescope on 25 January 2020 and 30 January 2020. The TRES wavelength coverage spans 385 to 910 nm and the resolving power is 44000. The purpose of these spectra is to discard obvious false positive scenarios and provide basic constraints on the stellar host. The two TRES spectra were extracted using the procedures outlined in \citet{Buchhave:2010}. Radial velocities were determined using special procedures developed by Jonathan Irwin for the analysis of spectra of M dwarfs, rather than the standard pipeline analysis, which was optimised for Solar-type stars, and uses the wavelength region near the Mg b features at 519 nm and a library of calculated templates to yield absolute velocities. Irwin's tools use an observed spectrum of Barnard's star as the template, and the wavelength region near 715 nm that contains TiO features rich in radial-velocity information. For M dwarfs, the S/N per resolution element is significantly stronger at 715 nm, in this case 30, compared to 12 at 519 nm. The two TRES observations were obtained at phases 16.79 and 17.25 of the TESS ephemeris for the inner planet (i.e. at opposite quadratures for a circular orbit), and yielded absolute velocities of -41.511 and -41.662 km/s, thus ruling out a stellar companion or brown dwarf orbiting TOI-1266, as the source of the transits for the inner planet.

\subsection{HIRES spectroscopy}
\label{sectionHIRES}
We obtained a single spectrum with the HIRES \citep{1994SPIE.2198..362V} instrument mounted on the Keck-I 10-m telescope on 15 December 2019. HIRES has a wavelength coverage of 390 to 900 nm and a resolving power of 50\,000. The spectrum with which we conducted these analyses has a S/N per resolution element of 80.

\subsection{High-angular resolution imaging}

We used high-angular resolution imaging to rule out false-positive signals caused by unresolved blended stars in the time-series photometry. This is particularly important for the TESS light curves, given that the TESS pixels are $\sim$21\arcsec. 

\subsubsection{AstraLux at Calar Alto}
\label{subsubsec:astralux}

TOI-1266 was observed on 30 October 2019 with the AstraLux instrument \citep{hormuth08}, a high-spatial resolution camera installed at the 2.2\,m telescope of the Calar Alto Observatory (Almer\'ia, Spain). This fast-readout camera uses the lucky-imaging technique \citep{fried1978} by obtaining thousands of short-exposure frames to subsequently select a small percentage of them showing the best Strehl ratio \citep{strehl1902}, and combining them into a final image. We observed this target using the Sloan Digital Sky Survey $z$ filter (SDSSz), which provides the best resolution and contrast capabilities for the instrument \citep{hormuth08}, and obtained 35\,000 frames with 20\,ms exposure time and a $6\arcsec\times6\arcsec$ field-of-view. The datacube was subsequently reduced using the observatory pipeline \citep{hormuth08}, and we used a 10\% selection rate for the best frames to obtain a final high-resolution imaging with a total exposure time of 70\,s. We then computed the sensitivity curve by using our own \texttt{astrasens} package\footnote{\url{https://github.com/jlillo/astrasens}}, following the procedure described in \cite{lillo-box12,lillo-box14b}. The image allows us to discard stellar companions with magnitude contrasts down to $\Delta m = 4$ at 0.2\arcsec\ (corresponding to 7.2 au at the distance of this system), and hence, establish a maximum contamination in the planet transit better than 3\%. The AstraLux contrast curve is shown in Fig.~\ref{fig:hri}.

\subsubsection{ShARCS}

TOI-1266 was observed on 12 November 2019 with the adaptive-optics-assisted ShARCS camera \citep{mcgurk2014commissioning, gavel2014shaneao} on the Shane 3-m telescope at Lick Observatory. We collected observations in both the $K_s$ and $J$ filters with exposure times of 6 s and 12 s, respectively. Observations were performed with a four-point dither pattern, with the distance between subsequent exposures being 4.00\arcsec\ on a side.

\begin{figure}
    \centering
    \includegraphics[width=0.48\textwidth]{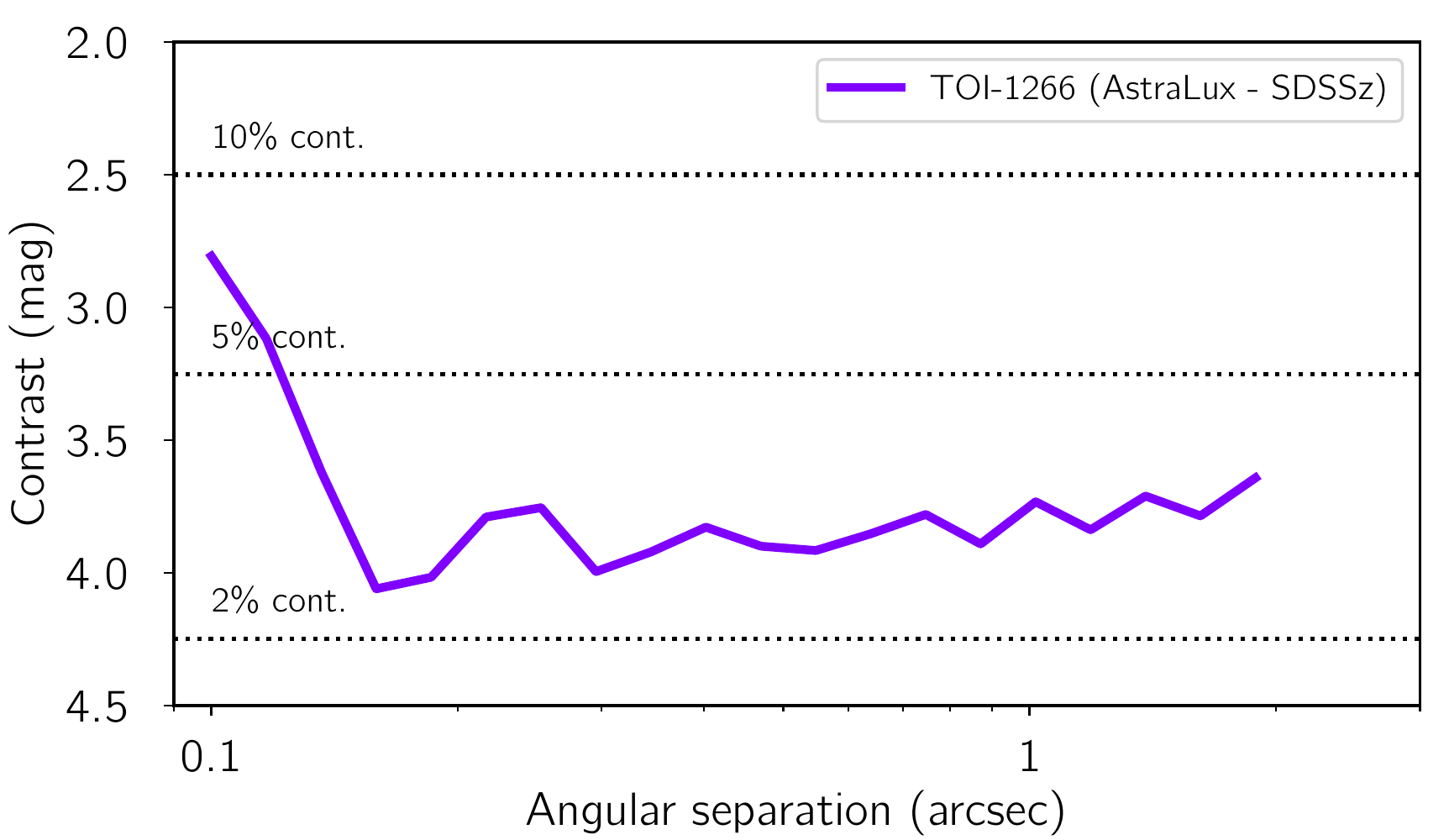}
    \includegraphics[width=0.48\textwidth]{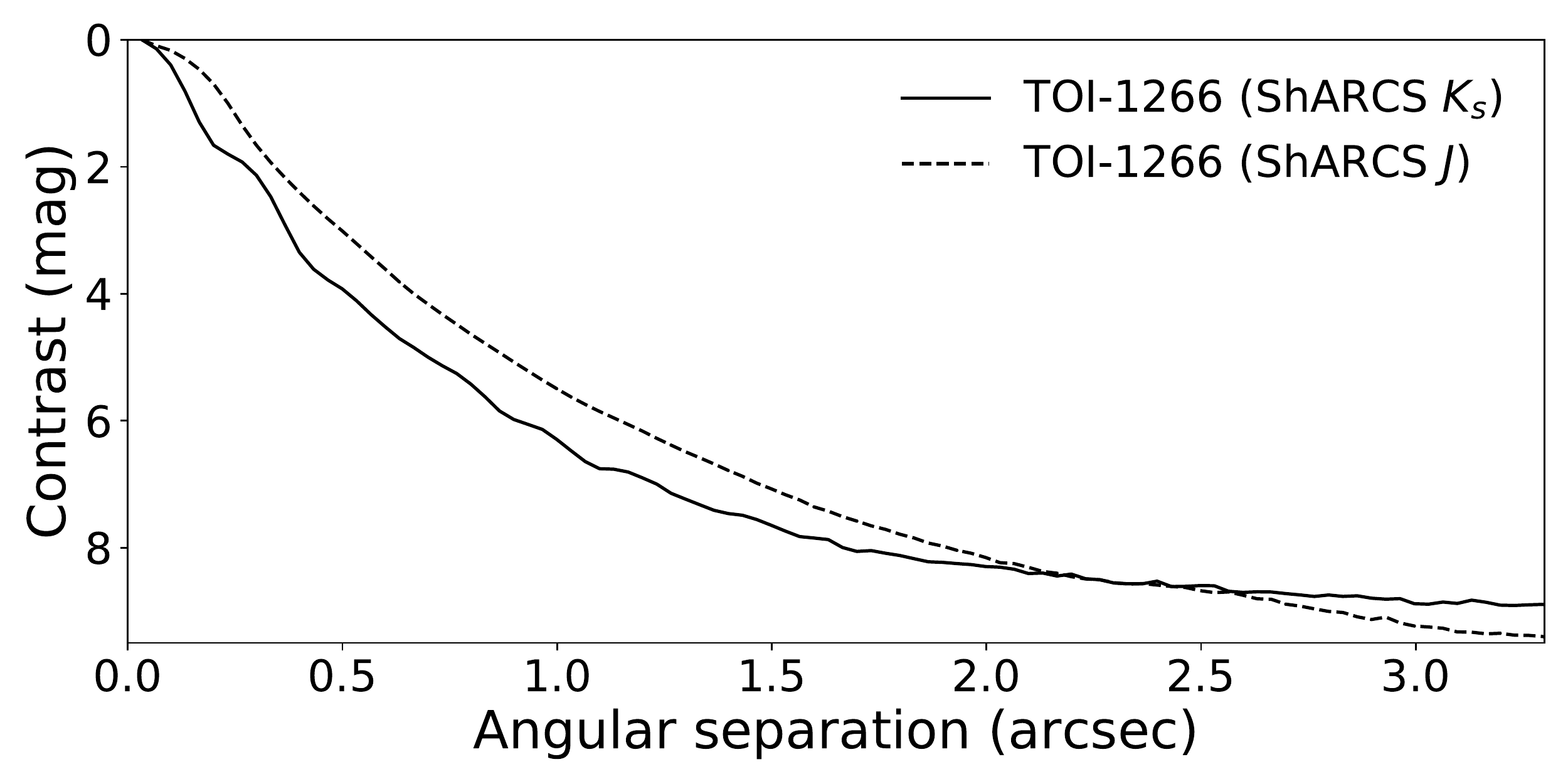}
    \caption{Contrast curves for TOI-1266 resulting from high-angular resolution imaging.
    \textbf{Top:} 5$\sigma$ contrast curve of the AstraLux image and contamination levels (horizontal dotted lines) obtained in SDSS $z$ filter. \textbf{Bottom:}
    5$\sigma$ contrast curves computed from ShARCS images of TOI-1266 taken in the $K_s$ (solid line) and $J$ (dashed line) filters.}
    \label{fig:hri}
\end{figure}

We reduced our data with \texttt{SImMER} (Savel, Hirsch et al., in prep), an open-source, \texttt{Python}-based pipeline\footnote{\url{https://github.com/arjunsavel/SImMER}}. Prior to aligning images, the pipeline implements standard dark-subtraction and flat-fielding. To align our science images for each target, we adapted methods from \citet{morzinski2015magellan}, performing rotations about points within a search radius and minimising the summed residuals from the original image. To determine our sensitivity to undetected stellar companions to TOI-1266, we calculated the minimum detectable companion brightness at increasing angular separations from the target. We performed this step by constructing concentric annuli centred on TOI-1266 and determining the mean and standard deviation of the flux within each annulus (e.g. \citealt{marois2006angular}, \citealt{nielsen2008constraints}, \citealt{janson2011high}, \citealt{wang2015influence}). We subsequently took our contrast curve to be 5$\sigma$ above the mean. Through this method, we find that we are sensitive in the $K_S$ band to companions 4 magnitudes fainter than the host beyond a separation of 0.51\arcsec\ and companions 8 magnitudes fainter than the host beyond a separation of 1.67\arcsec. The full contrast curves are shown in Figure \ref{fig:hri}.

\section{Stellar characterisation}
\label{sec:star}

\subsection{Spectroscopic analysis}
\label{sec:specan}

\subsubsection{TRES spectroscopy}

We used the TRES spectra to derive initial stellar parameters employing the method described in \citet{Maldonado2015}. Briefly, the effective temperature, spectral type and iron abundance were computed from the measurements of pseudo-equivalent widths of several spectral features. This analysis yields $T_{\rm eff}= 3570 \pm 100$ K, spectral type M3, and $[Fe/H]= -0.03 \pm 0.18$ dex. 
We also used the derived temperature and metallicity as inputs in the empirical relations by \citet{Maldonado2015} to derive the stellar mass ($M_{\rm \star}= 0.48 \pm 0.10 M_{\rm \odot}$), the radius ($R_{\rm \star}=0.47 \pm 0.10 R_{\rm \odot}$), and the surface gravity ($\log g = 4.78 \pm 0.10$ dex).

\subsubsection{HIRES spectroscopy}

We analysed the HIRES spectrum with SpecMatch-Empirical \citep{yee2017precision}, which classifies stars by comparing their optical spectra to a library of spectra of well-characterised stars. With this technique, we recovered $T_{\rm eff} = 3548 \pm 70$ K, [Fe/H] $= -0.24 \pm 0.09$ dex, and $R_\star = 0.43 \pm 0.10 \, R_\odot$, confirming the dwarf nature of the host star. Using the spectrum, we also ruled out double-lined spectroscopic binaries with contamination ratios down to $1\%$ \citep{kolbl2014detection}, greatly constraining the parameter space within which false positive scenarios can exist.

\subsection{SED analysis}

\begin{figure}
\centering
\resizebox{\hsize}{!}{\includegraphics[trim=100 70 80 80,clip]{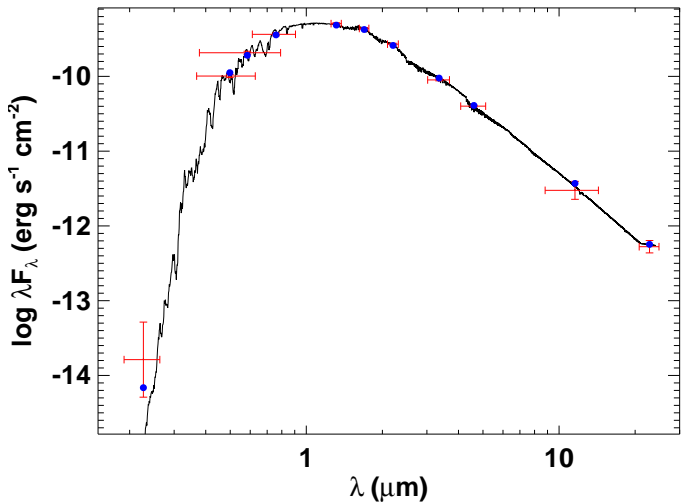}}
\caption{\textbf{Spectral energy distribution (SED) of TOI-1266.} Red symbols represent the observed photometric measurements, where the horizontal bars represent the effective width of the passband. Blue symbols are the model fluxes from the best-fit NextGen atmosphere model (black).}
\label{fig:sed}
\end{figure}

As an independent check on the derived stellar parameters, we performed an analysis of the broadband spectral energy distribution (SED) together with the {\it Gaia\/} DR2 parallax in order to determine an empirical measurement of the stellar radius, following the procedures described in \citet{Stassun:2016,Stassun:2017,Stassun:2018}. We obtained the $BVgri$ magnitudes from APASS, the $JHK_S$ magnitudes from {\it 2MASS}, the W1--W4 magnitudes from {\it WISE}, and the $G G_{BP} G_{RP}$ magnitudes from {\it Gaia} (see Table~\ref{tab:starlit}). In addition, we obtained the NUV flux from {\it GALEX} in order to assess the level of chromospheric activity, if any. Together, the available photometry spans the full stellar SED over the wavelength range 0.2--22~$\mu$m (see Figure~\ref{fig:sed}). 

We performed a fit using NextGen stellar atmosphere models \citep{Hauschildt:1999}, with the fitted parameters being the effective temperature ($T_{\rm eff}$) and metallicity ([Fe/H]). We set the extinction ($A_V$) to zero due to the star being nearby. We used the $T_{\rm eff}$ from the TIC \citep{Stassun:2018} as an initial guess. The broadband SED is largely insensitive to the surface gravity ($\log g$), thus we simply adopted the value from the TIC. The resulting fit is satisfactory (Figure~\ref{fig:sed}) with a reduced $\chi^2$ of 1.9. The best-fit parameters are $T_{\rm eff} = 3600 \pm 150$~K and [Fe/H] = $-0.5 \pm 0.5$ dex. Integrating the (unreddened) model SED gives the bolometric flux at Earth of $F_{\rm bol} =  6.72 \pm 0.16 \times 10^{-10}$ erg~s$^{-1}$~cm$^{-2}$. Taking the $F_{\rm bol}$ and $T_{\rm eff}$ together with the {\it Gaia\/} DR2 parallax, adjusted by $+0.08$~mas to account for the systematic offset reported by \citet{Stassun:2018a}, gives the stellar radius as $R = 0.420 \pm 0.037$~R$_\odot$. Finally, estimating the stellar mass from the empirical relations of \citet{Mann:2019} gives $M = 0.45 \pm 0.03 M_\odot$, which with the empirical radius measurement gives the mean stellar density as $\rho_\star = 8.7 \pm 2.3$~g~cm$^{-3}$. 

In a separate global fit, we used the EXOFASTv2\footnote{\url{https://github.com/jdeast/EXOFASTv2}} analysis package \citep{Eastman:2019} to derive the stellar parameters. EXOFASTv2 simultaneously utilises the SED, MIST stellar models \citep{Dotter:2016,Choi:2016}, {\it Gaia\/} DR2 parallax and enforces an upper limit on the extinction of $A_V= 0.04185$ from the \citet{Schlafly:2011} dust maps. In addition, we set Gaussian priors on $T_{\text{eff}}$ and [Fe/H] from the spectral analysis, presented in Section \ref{sec:specan}.
The resulting EXOFASTv2 fit provides values for $T_{\text{eff}} = 3533 \pm 45 $K, $R= 0.428 \pm 0.012 R_\odot$, and $M = 0.447 \pm 0.023 M_\odot$.

In Table~\ref{tabl:starchar}, we present a summary of the values of the stellar parameters obtained from the different instruments and methods previously described. We also added stellar properties derived in this work. We adopt the mass and metallicity values from the SED+Mann analysis as priors in our global analysis, because it assumes only a simple Gaussian prior on the {\it Gaia} parallax, and thus, is completely empirical.

Additionally, using the radial velocity determined from the HIRES data ($-$41.71 $\pm$ 0.10 kms$^{-1}$), and the proper motion and parallax from {\it Gaia\/} DR2, we derived Galactic space-velocity components (U, V, W) = (3.99 $\pm$ 0.03, $-$43.53 $\pm$ 0.06, $-$32.36 $\pm$ 0.08)  kms$^{-1}$, following the procedure detailed in \citet{Jofre2015}. From the space-velocity components and the membership formulation by \citet{Reddy2006}, we find that TOI-1266 has a probability of $\sim$96\% of belonging to the thin disc population.

\begingroup
\begin{table}
\begin{center}
\renewcommand{\arraystretch}{1.15}
\begin{tabular*}{\linewidth}{@{\extracolsep{\fill}}l c c@{}}
\toprule
Parameter & Value & Source \\
\midrule
$T_{\text{eff}} / K$    & 3548 \text{$\pm$} 70               & HIRES \\
                        & \textbf{3600 \text{$\pm$} 150}     & SED+Mann \\
                        & 3533 \text{$\pm$} 45      & EXOFASTv2 \\
                        & 3570 \text{$\pm$} 100            & TRES \\
\midrule
{[Fe/H]}                & -0.24 \text{$\pm$} 0.09   & HIRES \\
                        & \textbf{-0.5 \text{$\pm$} 0.5}     & SED+Mann \\
                        & -0.03 \text{$\pm$} 0.18   & TRES \\
\midrule
$M_* / M_\odot$         & \textbf{0.45 \text{$\pm$} 0.03}     & Mann \\
                        & 0.447 \text{$\pm$} 0.023   & EXOFASTv2 \\
                        & 0.48 \text{$\pm$} 0.10   & TRES \\
\midrule
$R_* / R_\odot$         & 0.43 \text{$\pm$} 0.02     & HIRES \\       
                        & 0.420 \text{$\pm$} 0.037   & SED+Mann \\
                        & 0.428 \text{$\pm$} 0.012   & EXOFASTv2 \\ 
                        & 0.43 \text{$\pm$} 0.10     & TRES \\
                        & \textbf{0.42 \text{$\pm$} 0.02}     & This work \\
\midrule
$\log g$ / dex          & 4.85 \text{$\pm$} 0.08     & SED+Mann \\
                        & 4.826 \text{$\pm$} 0.024   & EXOFASTv2 \\
                        & 4.78 \text{$\pm$} 0.10     & TRES \\ 
                        & \textbf{4.85 \text{$\pm$} 0.05}     & This work \\ 
\midrule
$\rho_*/\text{g cm}^{-3}$& 8.7 \text{$\pm$} 2.3      & SED+Mann \\
                        & $8.04^{+0.66}_{-0.55}$     & EXOFASTv2 \\
                        & \textbf{8.99\text{$\pm$}1.25}       & This work  \\
\midrule
Spectral Type           & M3     & TRES \\ 
$F_{\text{bol}}/ \text{erg s$^{-1}$cm$^{-2}$}$ & $(6.72 \pm 0.16) \times 10^{-10}$ & SED+Mann \\
\bottomrule
\end{tabular*}
\end{center}
\caption{Stellar Characterisation. SED+Mann use the empirical relations of \citet{Mann:2019}. Parameters in bold are the adopted stellar values. We use the SED+Mann mass and metallicity as priors in our global analysis presented in Sect.~\ref{sec:ga}, where we also derive the other parameters that appear in bold.\label{tabl:starchar}} 
\end{table}
\endgroup

\subsection{Activity}
\label{act}

We first searched for signs of stellar variability in the TESS PDC-SAP data from Sect.~\ref{subsec:tess}, as one might expect given the prevalence of activity in M dwarfs \citep[e.g.][]{Newton:2016, Newton:2018}. We performed a visual inspection of the light curve and found no hints of rotational modulation nor evidence of flaring activity. We then used the tools provided by the \texttt{Lightkurve} Python package \citep{lightkurve:2018} to compute the Lomb-Scargle periodogram  \citep{scargle:1982} for all the photometric data and for the data of each sector, without detecting any significant peak that might indicate periodic variability. To check that the variability was not removed by the PDC pipeline, we also performed an independent data reduction from the TESS Target Pixel Files (TPFs). We used a circular aperture, tracking the star on each image, and we stitched the observations from sectors 14, 15, 21 and 22 together to produce a new lightcurve. We computed a Lomb-Scargle periodogram, and we additionally fit a Gaussian process model with a quasi-periodic kernel \citep[see e.g.][]{aigrain2016}. We did not find any periodic variation consistent with stellar variability, however. This could result from a rotation period much longer than the 120 day observation period, or a small spot coverage. 

Within the total $\sim120$ days of TESS observations across all four sectors, we did not observe any flares from the star, consistent with an inactive early M dwarf \citep{Hawley:2014}. We note that the radius and inferred mass are compatible with an uninflated early M-dwarf, and the NUV photometry is consistent with an unspotted photosphere and negligible chromospheric activity. In summary, TOI-1266 appears to be an old, inactive, slightly metal-poor early M dwarf. 

\section{Target vetting tests}\label{sec:validation}

\subsection{TESS pipeline data validation}

As a first step in the false-positive vetting, we closely examined the data validation (DV) report \citep{Twicken:2018,Li:2019} combining all four sectors (14, 15, 21, and 22) provided by the SPOC pipeline. Both of the TOI-1266 planet candidates successfully passed all the tests, which includes a search for discrepancies between odd and even transit depths, as well as, the search for a shallow secondary eclipse that could both identify a possible eclipsing binary scenario. DV also includes a difference image centroid test to ensure that the source of the transit occurs on the target star, as well as, a `ghost diagnostic test' to discard scattered light as a source for the observed signal. These conclusive tests encouraged us to conduct further analysis to validate both planet candidates.

\subsection{SAINT-EX ground-based photometry}

In the context of the TESS Follow-up Observing Program, we used SAINT-EX to perform observations of the transit of each planet to eliminate possible contamination from nearby stars included in the large TESS aperture. We identified five such sources located between 36 and 132\arcsec\ from TOI-1266 within the {\it Gaia\/} DR2 catalogue \citep{Brown:2018}, with measured delta magnitudes ranging between 4.6 and 7.1 in the $z'$ band. Using AstroImageJ \citep{Collins:2017}, we estimated that eclipse depths between 0.064 and 0.2 occurring on these stars would be necessary to create the 0.0015 deep transit observed on TOI-1266 for the c planet candidate. Visual inspection of the light curves did not show transit signatures with the predicted depths on the nearby stars but revealed a 0.0015 deep transit on the target star as expected. We repeated the same analysis for the deeper transit of the b planet candidate, and confirmed both the lack of deep transits on the other stars and the detection of a $\sim$0.003 transit on TOI-1266. This step of the analysis confirms that both transits occur on TOI-1266 without contamination from known {\it Gaia\/} DR2 sources. We also detected no wavelength dependence of the transit depths between the TESS, $z'$, $r'$, $I_{\rm C}$, and $V$ bandpasses (see Sect.~\ref{sec:ga}).

\subsection{Archival imagery}
\begin{figure*}
    \centering
    \includegraphics[width=\textwidth]{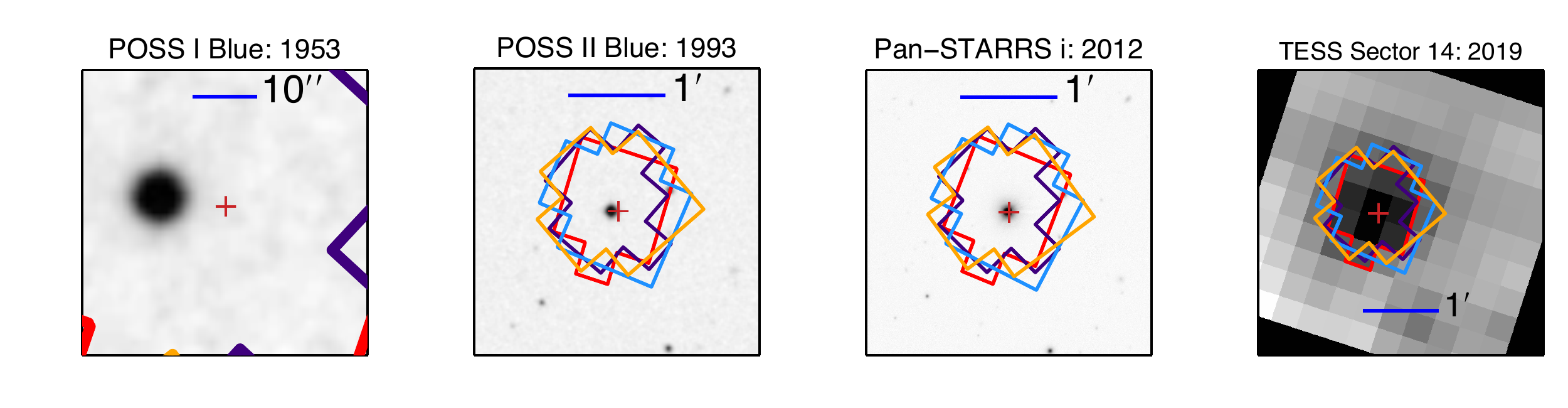}
    \caption{ Archival images of TOI-1266 to assess for current, unresolved blending.
    POSS I (1953), II (1993) and Pan-STARRS (2012) archival images around TOI-1266 with TESS field of view for sectors 14, 15, 21 and 22 superimposed. The red cross marks TOI-1266's location at the 2020 epoch.}
    \label{fig:poss}
\end{figure*}

Archival images are useful for investigating the background contamination of stars with non-negligible proper motion (PM), such as TOI-1266. None of the blue and red POSS1 images from 1953 at the current target location show the presence of a source that would be blended with the target at the epoch of the TESS observations (Fig.~\ref{fig:poss}). TOI-1266's PM, combined with the moderate spatial resolution of ground-based all-sky surveys, does not allow us to constrain background sources from more recent optical imagery such as POSS2, Pan-STARRS and SDSS.

\subsection{Statistical validation using high-resolution imaging with TRICERATOPS \& VESPA}

As an additional vetting step, we calculated the false positive probabilities of these TOIs using \texttt{triceratops} \citep[Tool for Rating Interesting Candidate Exoplanets and Reliability Analysis   of Transits Originating   from Proximate Stars;][]{giacalone2020, 2020ascl.soft02004G} and \texttt{vespa} \citep[Validation of Exoplanet Signals using a Probabilistic Algorithm;][]{morton2012, 2015ascl.soft03011M}. To tighten the constraints we obtain from these calculations, we incorporated  the contrast curves obtained from our high-resolution AO imaging. With \texttt{triceratops}, we compute false positive probabilities of 0.00310 and  0.08287 for TOI-1266.01 and TOI-1266.02, respectively, and with \texttt{vespa}, we compute false positive probabilities of 0.00002 and 0.00993, respectively.\footnote{Discrepancies in these false positive probabilities are due to differences in the calculations performed by the two tools. At the time of this writing, \texttt{triceratops} estimated probability using maximum likelihood estimation, while \texttt{vespa} determines probability by calculating marginal likelihoods. In addition, \texttt{triceratops} considers more false positive scenarios in its procedure. Regardless, both tools require a false positive probability $< 0.05$ to validate a planet candidate.} Because this is a multi-planet system, we were able to apply additional priors to these probabilities using the results of \cite{lissauer2012almost}. This study uses data from the {\it Kepler} mission to estimate the fraction of planet candidates that are false positives when in systems with multiple planet candidates. This is done by assuming that the expected number of targets with $k$ false positives, $E(k)$, is given by the Poisson distribution 
\begin{equation}
    E(k) = \frac{\lambda^k e^{-\lambda}}{k!} N \approx \frac{(R_{\rm FP} \frac{n}{N})^k}{k!} N,
\end{equation}
where $R_{\rm FP}$ is the false positive rate, $n$ is the number of planet candidates, and $N$ is the number of targets from which the sample is drawn. The sample used in \cite{lissauer2012almost} contained $n = 230$ planet candidates in candidate two-planet systems from $N = 160171$ total targets and assumed $R_{\rm FP} = 0.5$. Thus, the expected number of candidate two-planet systems in which both planet candidates are false positives is $E(k=2) \approx 1$ and the expected number of candidate two-planet systems in which only one planet candidate is a false positive is $(1-R_{\rm FP}) \frac{n}{N} \times E(k=1) \approx 2$. Therefore, the prior probability that a planet candidate in a candidate two-planet system is a false positive is $(1+2)/230 \approx 0.01$ (see also Guerrero et al., submitted). Multiplying the probabilities above by this prior, we find a false positive probability $< 0.01$ for both TOIs using both \texttt{triceratops} and \texttt{vespa}.

These probabilities are sufficiently low to rule out astrophysical false positives originating from both resolved nearby stars, and unresolved stars blended within the photometric aperture. 
In conclusion, the combination of TESS photometry, high-resolution spectroscopy, high-precision ground-based photometry, and archival imagery enable us to rule out false-positive scenarios for the observed transit signals of TOI-1266\,b and c.

\subsection{Possible bound stellar companions}

Thanks to our data, we can place some limited constraints on the presence of a binary companion. 
Our high-resolution imaging in Fig.~\ref{fig:hri} limits any object with a $\Delta m > 4$ in $z$-band at a distance $> 0.2\arcsec$ (corresponding to a semi-major axis of 7.2 au, and orbital period of 25 years). Using the \citet{baraffe2015} models, and assuming a system age of 5 Gyr, we can exclude any binary companion with a mass $> 0.10~{\rm M}_\odot$. Using the $K$-band observations, we can exclude the presence of any Hydrogen-burning object beyond $1\arcsec$ ($> 0.07~{\rm M}_\odot$).

Inside of 7.2 au, we have to rely on spectroscopy to place any constraints. A companion could in principle be detected as an extra set of lines, which requires two conditions: it is bright enough, and its orbital velocity is distinct enough from the TOI-1266\,A, the primary star. Our HIRES spectrum reached a S/N of 80. Using models from \citet{baraffe2015} once more, stellar companions with masses $> 0.15~{\rm M}_\odot$ could be detectable with S/N = 5. The velocity resolution of HIRES is $\sim 6~{\rm km~s}^{-1}$. All companions with orbital velocity $< 12~{\rm km~s}^{-1}$ would see their absorption lines blending with the primary's for the majority of their orbital motion, meaning they would likely remain unnoticed. An orbital velocity difference of $>12~{\rm km~s}^{-1}$, corresponds to orbital separations of $<3.8~{\rm au}$ (period $\sim10.7$ years), for a secondary mass of $0.15~{\rm M}_\odot$, and $\sin i = 1$.

Using dynamical arguments, we can also exclude the presence of a secondary star at orbital separations $< 0.3~{\rm au}$ (orbital periods around 0.2 years) or TOI-1266\,c would be unstable for any orbital inclination \citep{Holman1999}.
Similarly, a stable and circular inner planetary system implies that an external companion has not produced Lidov-Kozai cycles that would excite the eccentricities of the planetary orbits \citep{Lidov1962,Kozai1962, Wu2007}. Using the Lidov-Kozai timescale derived in \citet{Kiseleva1998}, we find that companions with a mass of $>0.05~{\rm M}_\odot$, eccentricities of 0.3, and orbital periods of 1000 years ($\sim 82~{\rm au}$), would induce cycles in the eccentricity of TOI-1266\,c, on timescales much shorter than the age of the system (21 Myr for the parameters we quote), but only if that companion has an orbital inclination between $40$ and $140^\circ$. 

In summary, we can exclude detecting the presence {\it within the data} of most stellar companions, except at orbital separations between 3.8 and 7.2 au. However, we cannot exclude any companion should that object be within the line of sight (i.e. high-angular resolution imaging does not detect it, and with a relative radial velocity near 0 km/s). Radial velocities would not detect face-on orbits, however those are ruled out using dynamical arguments. 

\section{Results}\label{sec:results}

\subsection{Photometric analysis}\label{subsec:lcs}

\subsubsection{Planet search and detection limits from the TESS photometry}

\begin{figure*}
\includegraphics[width=0.95\textwidth]{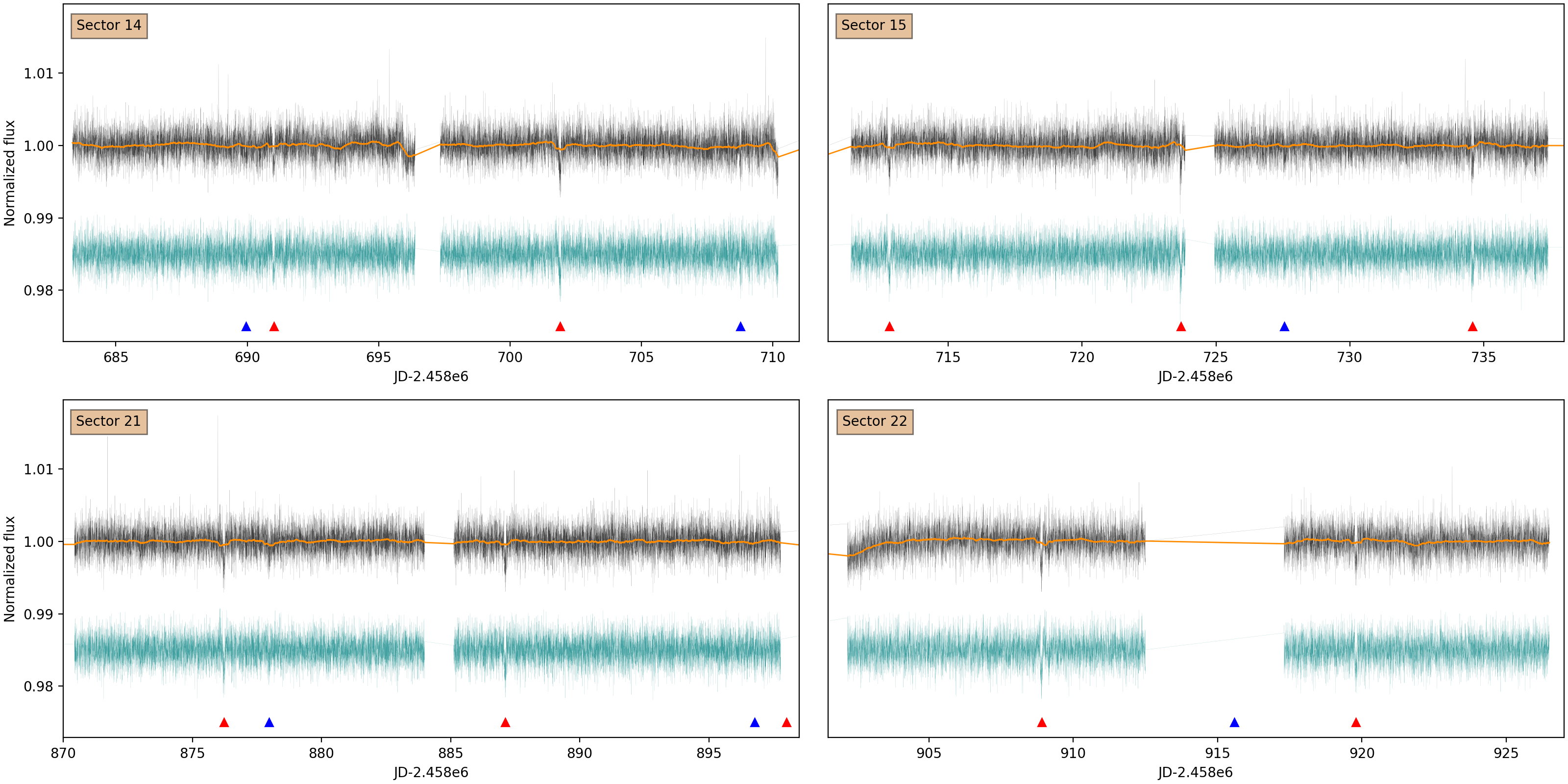}
\caption{TESS data of TOI-1266 from the four sectors in which it was observed. In all cases, the black line corresponds to the PDC-SAP fluxes obtained from SPOC pipeline, the solid-orange 
line corresponds to the best-detrended model, and the teal line is the final detrended lightcurve. The 
red dtriangles mark the 10.8~d period planetary candidate, while the blue triangles mark the 18.8~d candidate.}\label{search}
\end{figure*}

As mentioned previously, TOI-1266 was observed by TESS in sectors 14, 15, 21 and 22. The TESS Science Office issued two alerts for this object based on SPOC DV reports; TOI 1266.01 and TOI 1266.02 correspond to planetary candidates with periods of 10.8 d and 18.8 d respectively.
We performed our own search for candidates using the {\fontfamily{pcr}\selectfont  SHERLOCK}\footnote{{\fontfamily{pcr}\selectfont  SHERLOCK} 
code is available upon request.} (\textbf{S}earching for \textbf{H}ints of \textbf{E}xoplanets f\textbf{R}om \textbf{L}ightcurves \textbf{O}f spa\textbf{C}e-based see\textbf{K}ers) pipeline  presented in \cite{pozuelos2020}. This pipeline makes use of the \texttt{Lightkurve} package \citep{LightkurveCollaboration2018Lightkurve:Python}, which downloads the PDC-SAP flux data from the \emph{NASA} Mikulski Archive for Space Telescope (MAST), and removes outliers defined as data points $>3\sigma$ above the running mean.
In order to remove stellar noise and instrumental drifts, our pipeline uses {\sc{wotan}} \citep{Hippke2019Python} with two different detrending methods: bi-weight and Gaussian process 
with a Mat\'ern 3/2-kernel. In both cases, a number of detrending approaches were applied by
varying the window and kernel sizes to maximise the signal detection efficiency (SDE) of the transit search, which was performed by means of the {\sc{transit least squares}}  
package \citep{Hippke2019TransitPlanets}. The {\sc{transit least squares}} uses an analytical transit model based on the stellar parameters, and is optimised for the 
detection of shallow periodic transits. We properly recovered the two aforementioned candidates, where the best systematics model corresponded to the bi-weight method
with a window-size of 0.4149~d. In addition to these two signals, we found a threshold-crossing event with a period of 12.5~d. However, after an in-depth vetting process \citep{Heller2019TransitK2},
we discarded this signal, which we attributed to systematics in the data set. We show in Fig.~\ref{search} the PDC-SAP flux, the best systematics model, and the final 
detrended lightcurve with the two planet candidates.

\begin{figure}
\includegraphics[width=\columnwidth]{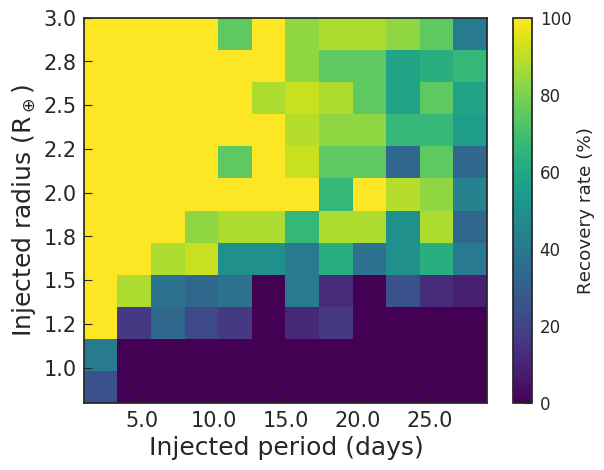}
\caption{Injection-and-recovery test performed to check the detectability of extra planets in the system. We explored a total of 1305 different scenarios. Larger recovery rates are presented in green and yellow colours, while
lower recovery rates are shown in blue and darker hues. Planets smaller than 1.5~R$_{\oplus}$ would remain undetected for almost the full set of periods explored.} \label{recovery}
\end{figure}

In order to assess the detectability of other planets in the data set available from TESS, we performed an injection-recovery test, where we injected synthetic 
planetary signals into the PDC-SAP fluxes corresponding to planets with different radii and periods. We then detrended the lightcurve using the best method 
found previously that is the bi-weight approach with a window size of 0.4149~d. Before searching for planets, we masked the two known candidate planets with periods of 10.8~d and 18.8~d.
We explored the $R_{\mathrm{planet}}$--$P_{\mathrm{planet}}$ parameter space in the ranges of
0.8--3.0~R$_{\oplus}$ with steps of 0.05~R$_{\oplus}$, and 1--30~d with steps of 1~d, for a total of 1305 different scenarios.
In this test, we defined an injected signal as being recovered when a detected epoch matched the injected epoch within one hour, and if a detected period matched any half-multiple 
of the injected period to better than 5$\%$. It is important to note that since we injected the synthetic signals directly in the PDC-SAP lightcurve, they were not affected by the 
PDC-SAP systematic corrections. Therefore, the detection limits found correspond to the most optimistic scenario 
\citep[see e.g.][]{pozuelos2020,Eisner2019PlanetOrbit}. The results are shown in Fig.~\ref{recovery}, and we reached several conclusions from this test: 
(1) We can rule out the presence of planets with sizes larger than $\sim$1.5~R$_{\oplus}$ with orbits shorter than 10~d. 
However, such planets might reside in orbits with longer periods and, thus, remain undetected, as a longer period yields a lower detectability. In fact, 
for periods greater than 10~d, we obtained a recovery rate ranging from 30 to 70~$\%$. (2) For the full set of investigated periods, planets with sizes smaller than 
1.5~R$_{\oplus}$ would remain undetected, with recovery rates lower than 30\%, and close to 0\% for 1.0~R$_{\oplus}$.

\subsubsection{Global analysis}
\label{sec:ga}

We used the full photometric dataset described in Sect.~\ref{sec:obs} along with the transit signals detected in the previous step as input parameters to a global analysis of the TOI-1266 system. We also included the stellar mass and effective temperature derived in Sect.~\ref{sec:star} as Gaussian priors to convert the transit fitted parameters into physical values.

We used the MCMC algorithm implementation already presented in the literature \citep[e.g.][]{Gillon:2012a,Demory:2012,Gillon:2014a}. The inputs to the MCMC are the photometric time-series obtained during the data reduction described above. For each light curve, we fit simultaneously for the instrument baseline model and a transit model of two Keplerian orbits corresponding to TOI-1266\,b and c. This approach ensures that instrumental systematic noise is properly propagated to the system parameters of interest. The photometric baseline model coefficients used for detrending for each instrument are determined at each step of the MCMC procedure using a singular value decomposition method \citep{Press:1992}. The resulting coefficients are then used to correct the raw photometric lightcurves. Such an approach is necessary because the input data originate from multiple instruments with different sources of systematics. We show in Table~\ref{tab:baseline} the baseline model used for each light curve. To derive accurate uncertainties on the system parameters, we computed two scaling factors, $\beta_w$ and $\beta_r$, following \citet{Winn:2008b}, to account for over-  or under-estimated white noise and correlated noise \citep{Pont:2006} in each dataset (these values are computed over 10 to 240-min timescales).

\begin{table*}
\begin{center}
\begin{tabular*}{\textwidth}{l l c l c c c}
\toprule
Planet & Facility & T0 [$BJD_{\rm TDB}$] & Baseline model functional form & Residual RMS (exp. time) & $\beta_w$ & $\beta_r$ \\
\midrule
b & TESS  & 8691.0063 & flux offset & 0.00165 (120 s) & 0.44 & 1.18 \\
  &       & 8701.9011 & flux offset & 0.00186 (120 s) & 0.50 & 1.10 \\
  &       & 8712.7959 & flux offset & 0.00176 (120 s) & 0.81 & 1.16 \\
  &       & 8723.6907 & flux offset & 0.00173 (120 s) & 0.68 & 1.53 \\
  &       & 8734.5855 & flux offset & 0.00190 (120 s) & 0.84 & 1.11 \\
  &       & 8876.2179 & flux offset & 0.00188 (120 s) & 0.91 & 1.00 \\
  &       & 8887.1127 & flux offset & 0.00174 (120 s) & 0.84 & 1.22 \\
  &       & 8908.9023 & flux offset & 0.00178 (120 s) & 0.78 & 1.18 \\
  &       & 8919.7971 & flux offset & 0.00180 (120 s) & 0.79 & 1.08 \\
  & OAA         & 8843.5335 & flux offset + airmass$^2$ & 0.00080 (160 s) & 1.02 & 1.00 \\
  & SAINT-EX    & 8908.9023 & flux offset + airmass$^2$ + FWHM$^2$ & 0.00352 (12 s) & 0.51 & 1.98 \\
  & TRAPPIST-N  & 8930.6920 & flux offset + airmass$^2$ + FWHM$^2$ & 0.00370 (15 s) & 0.71 & 1.35 \\
  & Artemis     & 8930.6920 & flux offset + airmass$^2$ + FWHM$^2$ & 0.00344 (10 s) & 0.79 & 1.32  \\ 
  & TRAPPIST-N  & 8941.5868 & flux offset + airmass$^2$ + FWHM$^2$ & 0.00287 (60 s) & 0.78 & 1.15 \\ 
  & Kotizarovci & 8963.3764 & flux offset + airmass$^2$ & 0.00435 (50 s) & 0.70 & 1.58 \\
  & ZRO         & 8963.3764 & time$^2$    + airmass$^2$ & 0.00101 (200 s) & 0.56 & 1.43 \\
\midrule
c & TESS  & 8689.9612 & flux offset & 0.00189 (120 s) & 0.51 & 1.04 \\
  &       & 8708.7625 &             & 0.00190 (120 s) & 0.49 & 1.00 \\
  &       & 8727.5638 &             & 0.00185 (120 s) & 0.77 & 1.33 \\
  &       & 8877.9741 &             & 0.00174 (120 s) & 0.84 & 1.00 \\
  &       & 8896.7754 &             & 0.00180 (120 s) & 0.87 & 1.00 \\
  & SAINT-EX    & 8877.9741 & flux offset + airmass$^2$ + FWHM$^2$ & 0.00488 (12 s) & 1.04 & 1.10 \\
\hline 
\end{tabular*}
\end{center}
\caption{Baseline model functional forms, residual RMS, and scaling factors of each transit light curve used in the photometric global analysis. For each baseline, a polynomial is used with the indicated variables as parameters.\label{tab:baseline}}
\end{table*}

We computed the quadratic limb-darkening (LD) coefficients $u_1$ and $u_2$ in the {\it TESS}, $z'$, $r'$, $I_{\rm C}$, and Johnson $V$ filters, using the \texttt{PyLDTk} code \citep{Parviainen:2015} and a library of PHOENIX high-resolution synthetic spectra \citep{Husser:2013}. We placed Gaussian priors on each of the quadratic LD parameters, with a 5-fold inflation of the uncertainties computed from model interpolation. All LD parameters used in this analysis are shown in Table~\ref{tab:ld}. For each of the two planets, we fitted for (1) the transit depth (planet-to-star area ratio) $\frac{R_P^2}{R_{\star}^2}$ for each instrument to assess transit depth chromaticity, (2) the transit duration $T_{14}$, (3) the orbital period $P$, (4) the transit centre $T_0$, and (5) the impact parameter $b=\frac{a \cos i}{R_{\star}}$, where $a$ is the orbital semi-major axis and $i$ the orbital inclination. For this MCMC fit, we assumed circular orbits, and fixed $\sqrt{e}\cos \omega$ and $\sqrt{e} \sin \omega$ values to 0. We ran two chains of 100\,000 steps (including 20\% burn-in) each, and checked their efficient mixing and convergence by visually inspecting the autocorrelation functions of each chain, and by using the \citet{Gelman:1992} statistical test, ensuring that the test values for all fitted parameters were $<1.01$.

\begin{table*}
\begin{center}
\begin{tabular}{l c c l}
\toprule
Filter &  $u_1$ & $u_2$ & Notes\\
\midrule
\textit{TESS} & $0.20\pm0.09$ & $0.41\pm0.10$ &  used as well for Kotizarovci\\
$z'$ & $0.17\pm0.08$ & $0.27\pm0.09$ &  \\
$r'$ & $0.39\pm0.10$ & $0.31\pm0.09$ &  \\
$V$  & $0.38\pm0.10$ & $0.32\pm0.09$ &  \\
$Ic$ & $0.20\pm0.09$ & $0.29\pm0.09$ &  \\
Clear & $0.33\pm0.09$ & $0.30\pm0.09$ & ZRO, QE>50\% between 480 and 800 nm\\
\hline 
\end{tabular}
\end{center}
\caption{Quadratic limb-darkening coefficients used in the photometric global analysis for each instrument.\label{tab:ld}}

\end{table*}

\begin{figure*}
    \centering
    \includegraphics[width=0.48\textwidth]{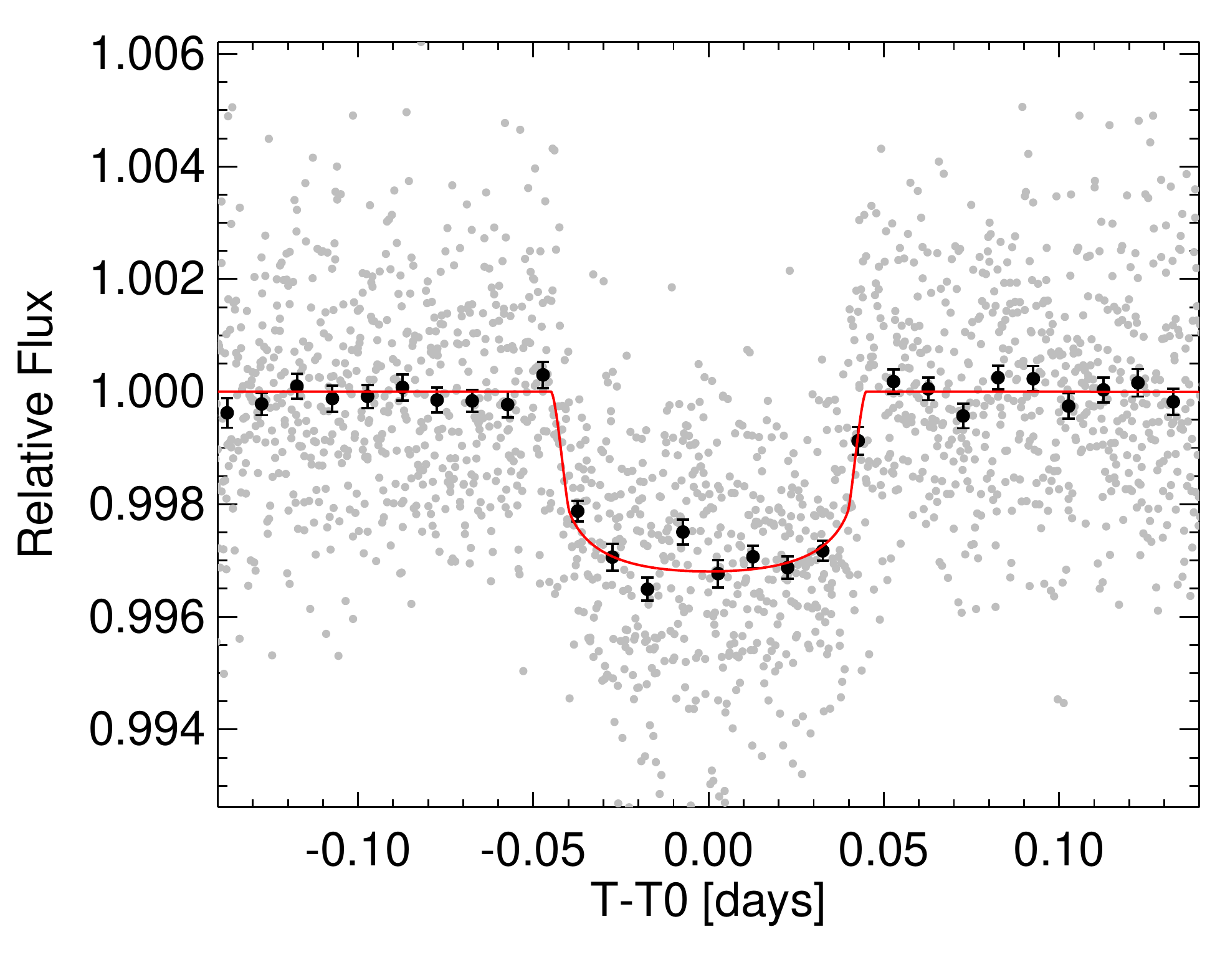}
    \includegraphics[width=0.48\textwidth]{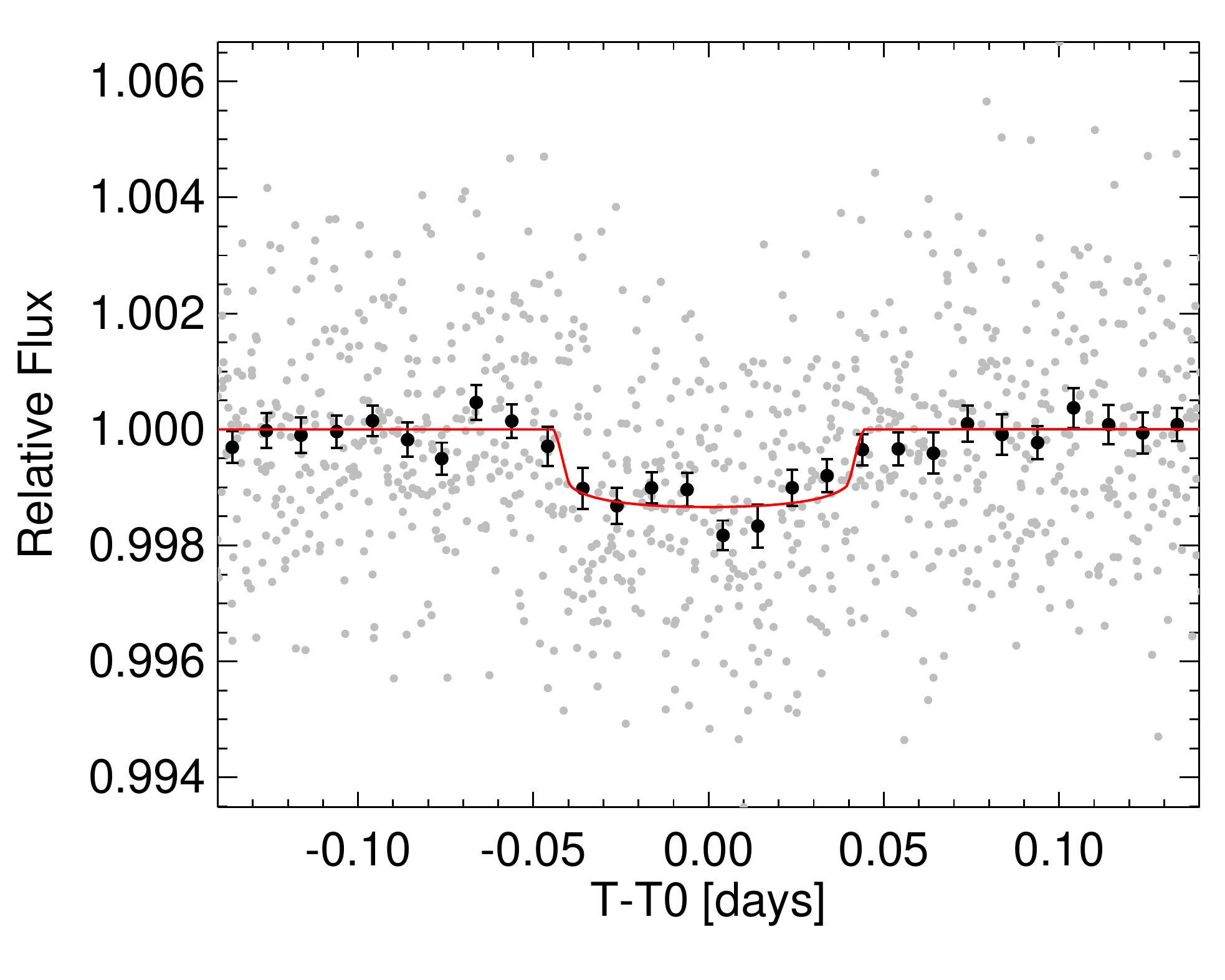}
    \caption{TESS phase-folded transits of TOI-1266\,b (left) and TOI-1266\,c (right) from the global analysis. Grey circles are un-binned data points while 15-min bins are shown as black circles. The best-fit model is shown in red.}
    \label{fig:tess_pf}
\end{figure*}

\begin{figure*}
    \centering
    \includegraphics[width=0.48\textwidth]{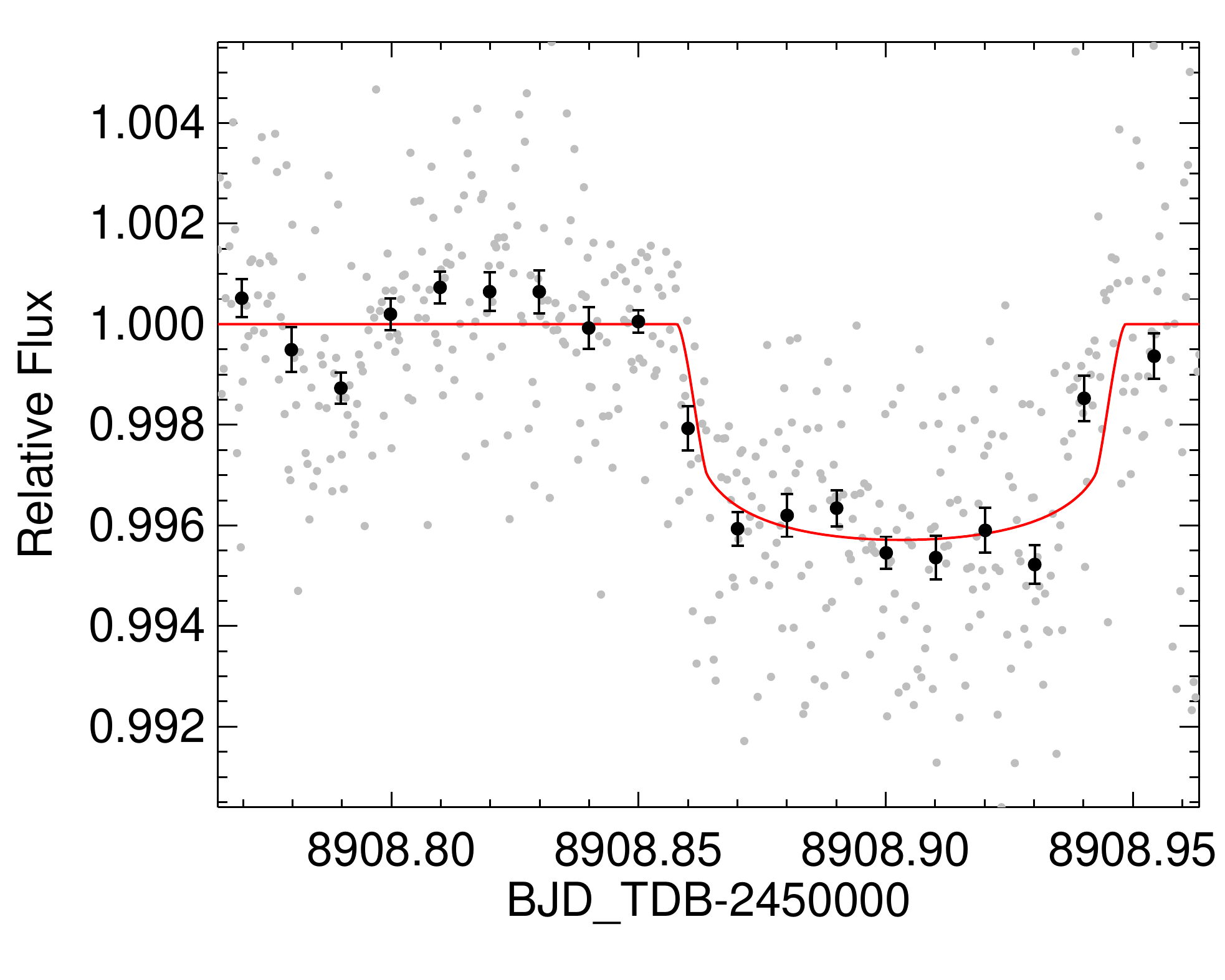}
    \includegraphics[width=0.48\textwidth]{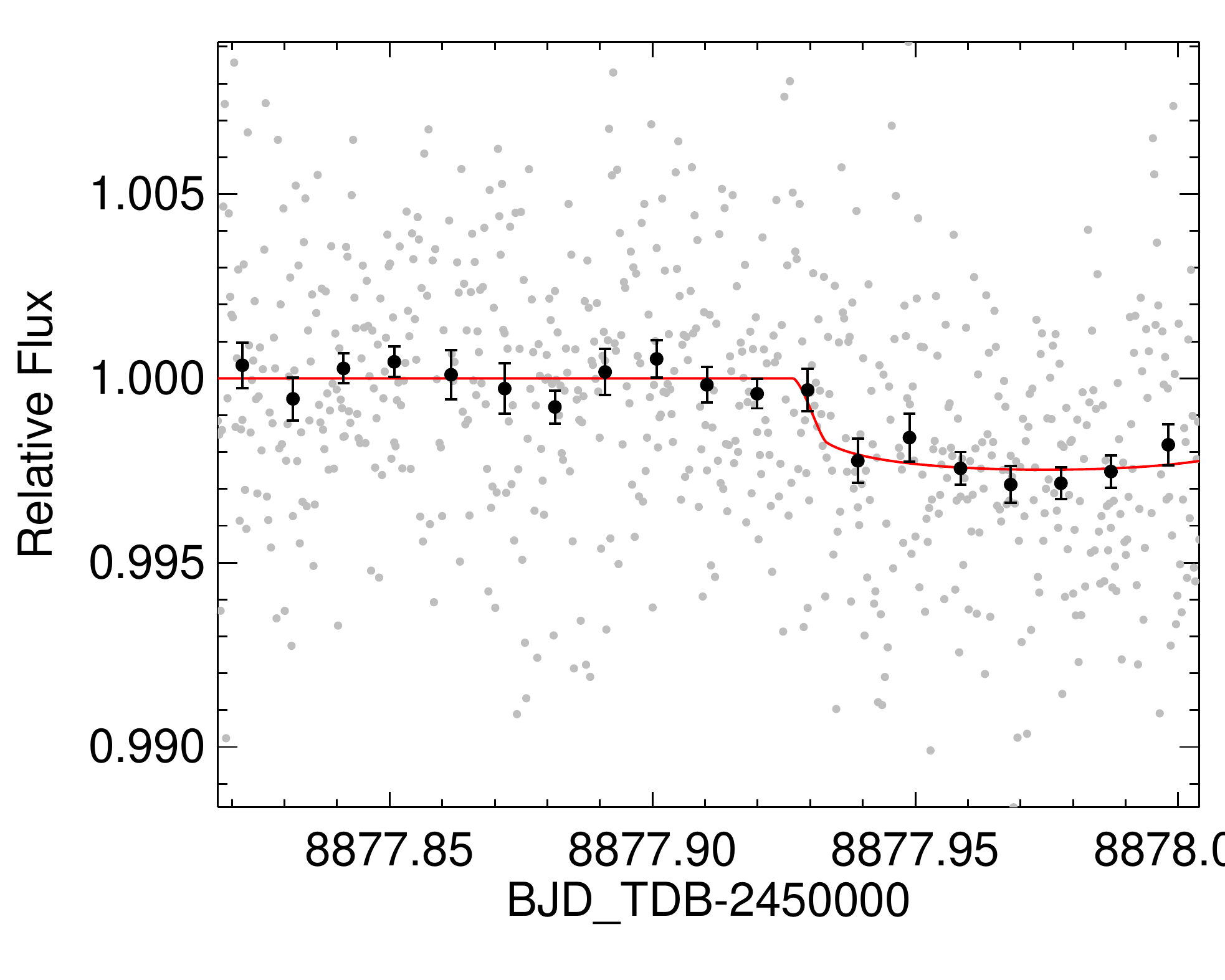}
    \caption{SAINT-EX detrended transits of TOI-1266\,b (left) and TOI-1266\,c (right) from the global analysis, both observed in $z'$. Grey circles are un-binned data points while 15-min bins are shown as black circles. The best-fit model is shown in red.}
    \label{fig:saintex_pf}
\end{figure*}

\begin{figure*}
    \centering
    \includegraphics[width=0.32\textwidth]{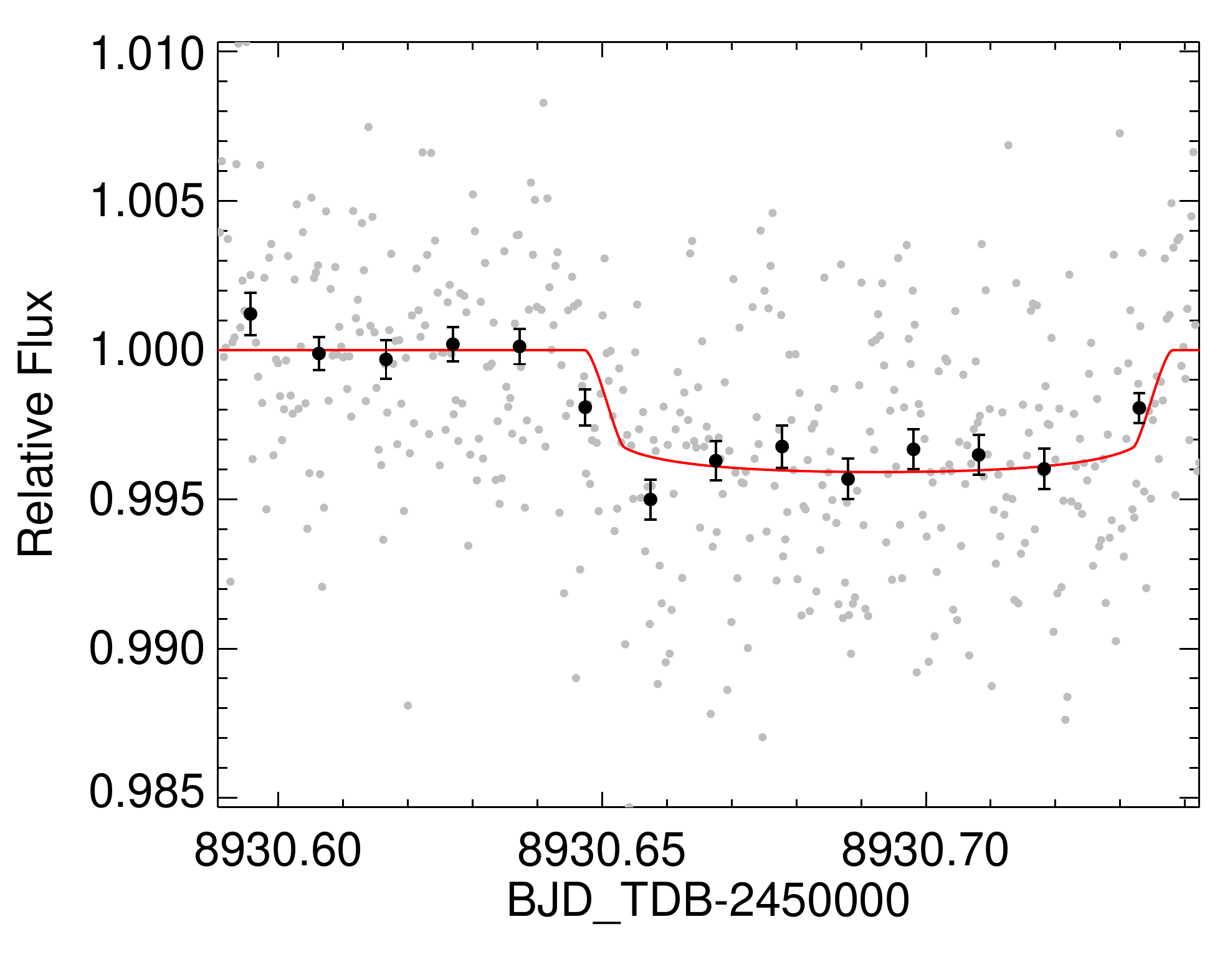}
    \includegraphics[width=0.32\textwidth]{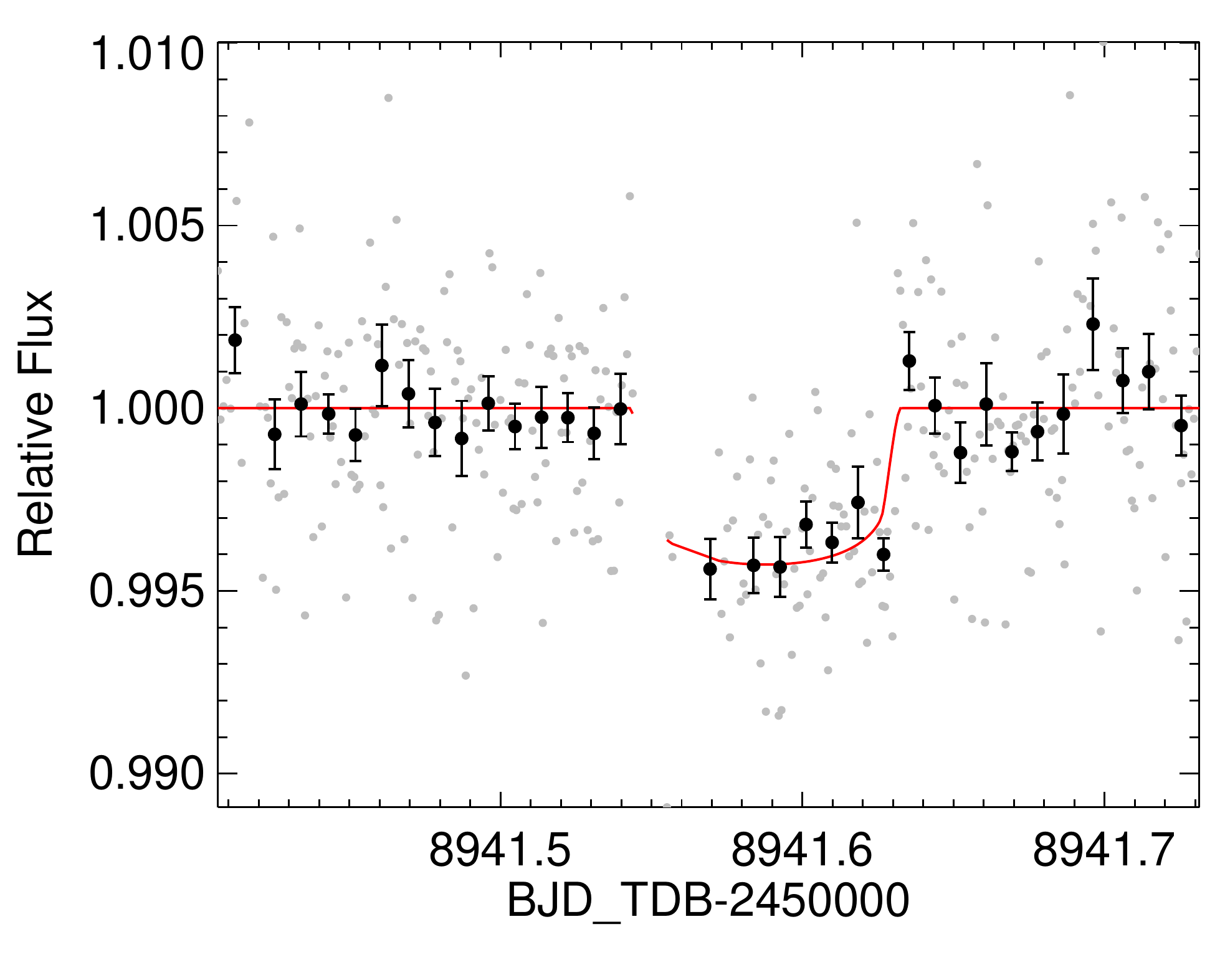}
    \includegraphics[width=0.32\textwidth]{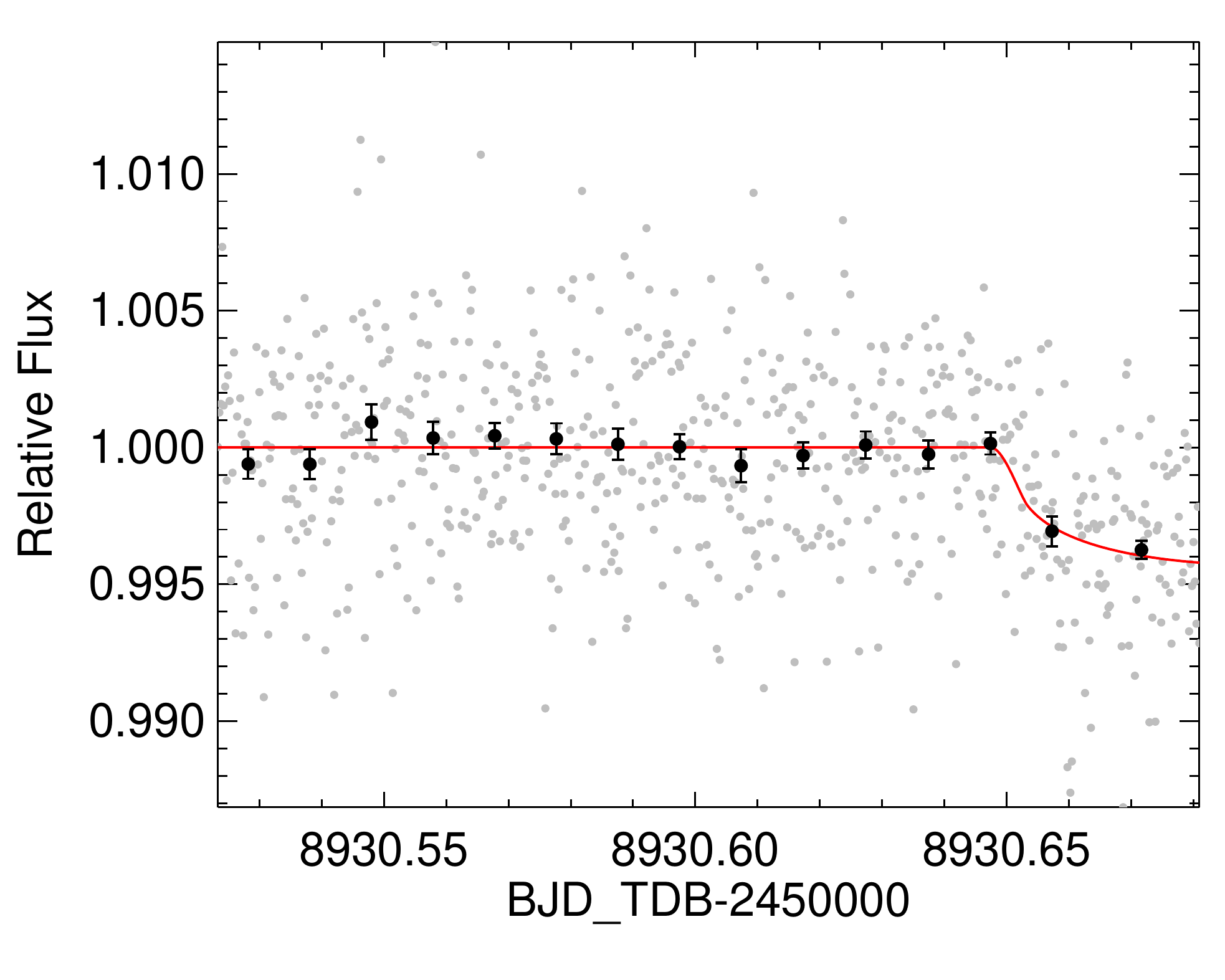}
    \caption{Single, detrended transits of TOI-1266\,b obtained with TRAPPIST-N in $z'$ (left), $V$ (centre), and with ARTEMIS in $r'$ (right) from the global analysis. Grey circles are un-binned data points while 15-min bins are shown as black circles. The best-fit model is shown in red.\label{fig:trapart_pf}}
\end{figure*}

\begin{figure*}
    \centering
    \includegraphics[width=0.32\textwidth]{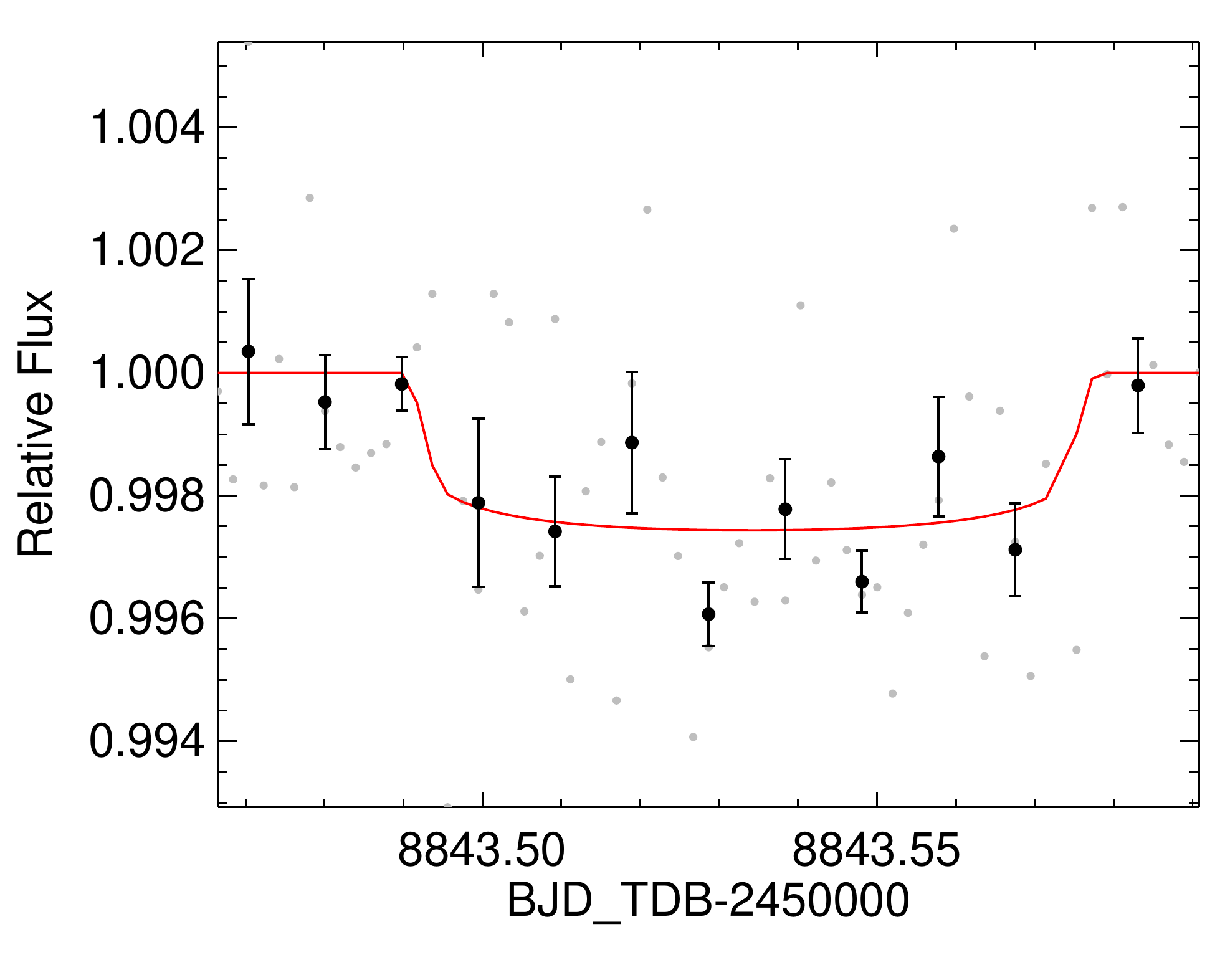}
    \includegraphics[width=0.32\textwidth]{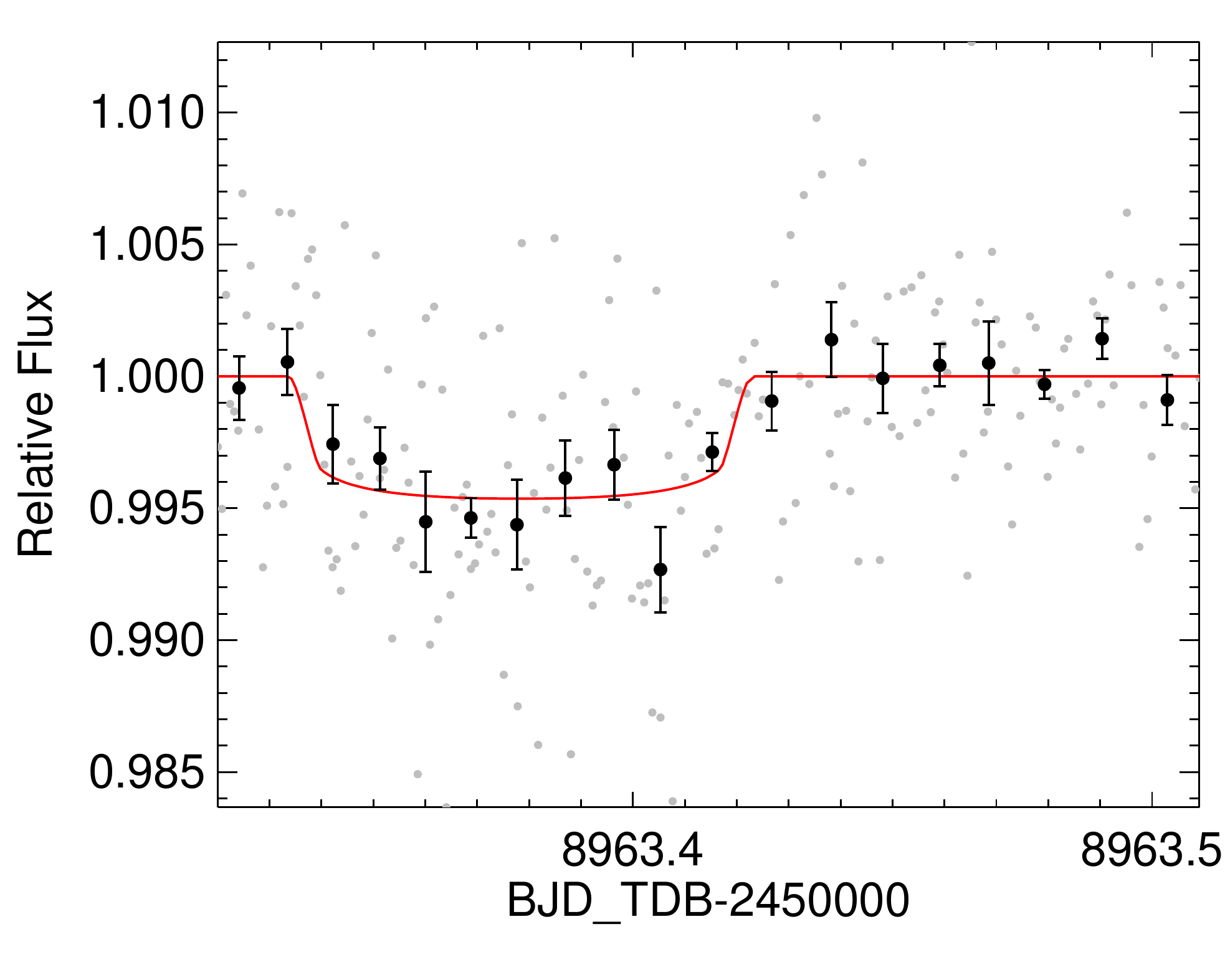}
    \includegraphics[width=0.32\textwidth]{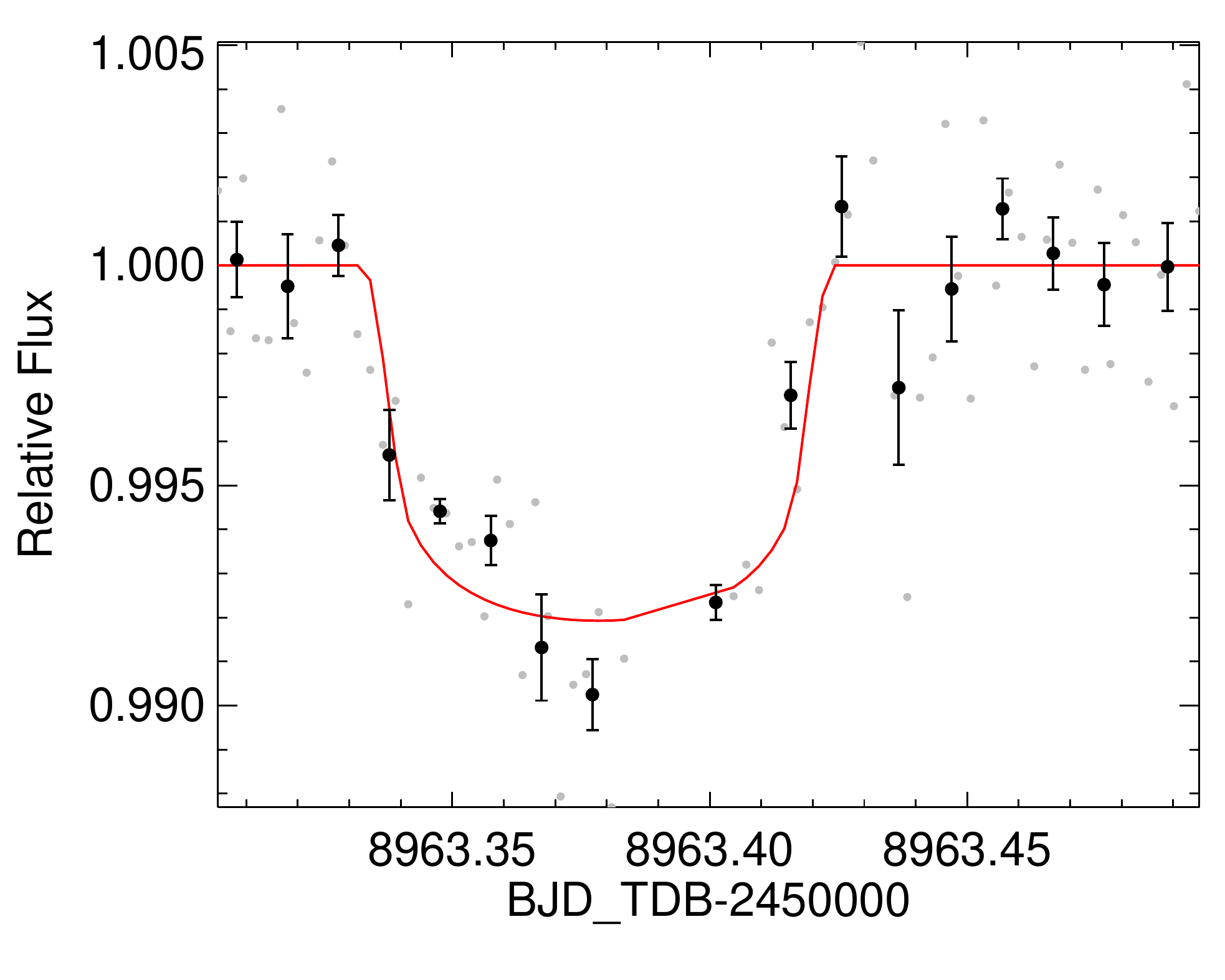}
    \caption{Single, detrended transits of TOI-1266\,b obtained with OAA (left), Kotizarovci (centre), and with ZRO (right) from the global analysis. Grey circles are un-binned data points while 15-min bins are shown as black circles. The best-fit model is shown in red.\label{fig:tfop_pf}}
\end{figure*}

We show in Table~\ref{tab:gares} the median and 1$\sigma$ credible intervals of the system parameter's posterior distribution functions. The corresponding light curves for both planets are shown in Fig.~\ref{fig:tess_pf}, \ref{fig:saintex_pf}, \ref{fig:trapart_pf}, and \ref{fig:tfop_pf}.

Our analysis shows a good agreement between the combined stellar density value of $\rho_\mathrm{\star}$ = $8.99\pm1.25$ g~cm$^{-3}$ derived from the photometry alone, and (1) the $\rho_\mathrm{\star}$  = $8.7\pm2.3$ g~cm$^{-3}$ derived from the SED+Mann analysis, as well as, (2) the $\rho_\mathrm{\star}$  = $8.04^{+0.66}_{-0.55}$ g~cm$^{-3}$ derived from the EXOFASTv2 analysis described in Sect.~\ref{sec:star}. The transit depths measured in four different bandpasses (TESS, $z'$, $r'$, and $V$) for TOI-1266\,b are all in agreement at the $\sim$2-$\sigma$ level. We find a good agreement as well for the transit depth of TOI-1266\,c albeit with only two bandpasses (\textit{TESS} and $z'$). We also repeated the same MCMC analysis, this time allowing $\sqrt{e}\cos \omega$ and $\sqrt{e} \sin \omega$ to vary, but did not find evidence for eccentric orbits for any of TOI-1266\,b and c,  using transit photometry alone.

\begin{table*}[!htbp] 
   \begin{center}
	\begin{tabular}{lcc}
    	\hline
    	\textbf{Parameter} & \textbf{TOI-1266\,b} & \textbf{TOI-1266\,c} \\ 
        \hline
        \hline
        \multicolumn{3}{c}{\textbf{Transit {\bf fitted} parameters}} \\ 
        \hline
        Transit depth, $(R_\mathrm{p} / R_\star)^2$ & $0.00276_{-0.00011}^{+0.00010}$ & $0.00120\pm0.00017$ \\ 
        \smallskip
        Transit duration, $T_{14}$ (days)&  $0.0887_{-0.0012}^{+0.0013}$ & $0.0911_{-0.0056}^{+0.0055}$  \\ 
        \smallskip
        Impact parameter, $b$ &  $0.38\pm0.12$ & $0.61_{-0.09}^{+0.08}$  \\ 
        \smallskip
        Mid-transit time, $T_{0}$ (BJD$_{\rm TDB}$) &  $2458821.74439_{-0.00055}^{+0.00054}$ & $2458821.5706_{-0.0029}^{+0.0034}$  \\ 
        \smallskip
        Orbital period, $P$ (days) &  $10.894843_{-0.000066}^{+0.000067}$ & $18.80151_{-0.00069}^{+0.00067}$   \\ 
        \smallskip
        Tr. depth diff., $\delta_\mathrm{TESS-SAINT-EX, z'}$ & $0.00106_{-0.00067}^{+0.00069}$ & $0.00089\pm0.00063$  \\
        \smallskip
        Tr. depth diff., $\delta_\mathrm{TESS-TRAPPIST-N, z'}$  & $0.00079_{-0.00075}^{+0.00071}$ &   \\
        \smallskip
        Tr. depth diff., $\delta_\mathrm{TESS-TRAPPIST-N, V}$  & $0.00067_{-0.00046}^{+0.00045}$ &   \\
        \smallskip
        Tr. depth diff., $\delta_\mathrm{TESS-ARTEMIS, r'}$  & $0.00049\pm0.00120$ &   \\
        \smallskip
        Tr. depth diff., $\delta_\mathrm{TESS-OAA, Ic}$  & $-0.00039_{-0.0013}^{+0.0016}$ &   \\
        \smallskip
        Tr. depth diff., $\delta_\mathrm{TESS-ZRO, clear}$  & $0.0035_{-0.0019}^{+0.0026}$ &   \\
        \smallskip
        Tr. depth diff., $\delta_\mathrm{TESS-Kotizarovci, TESS band}$  & $0.00192_{-0.00098}^{+0.00094}$ &   \\
        \hline
        \multicolumn{3}{c}{\textbf{Physical and orbital parameters}} \\ 
        \hline
        \smallskip
        Planet radius, $R_\mathrm{p}$ ($\mathrm{R_{\oplus}}$) & $2.37_{-0.12}^{+0.16}$ & $1.56_{-0.13}^{+0.15}$  \\
        \smallskip
        Semi-major axis, $a_\mathrm{p}$ (au)  & $0.0736_{-0.0017}^{+0.0016}$ & $0.1058_{-0.0024}^{+0.0023}$ \\ 
        \smallskip
        Orbital inclination, $i_\mathrm{p}$ (deg) & $89.5_{-0.2}^{+0.2}$ & $89.3_{-0.1}^{+0.1}$ \\ 
        \smallskip
        Irradiation, $S_{\mathrm{p}}$ ($S_{\oplus}$) & $4.9_{-0.8}^{+1.0}$ & $2.3_{-0.4}^{+0.5}$ \\ 
        \smallskip
        Equilibrium temperature, $T_\mathrm{eq}$ (K) & $413\pm20$ & $344\pm16$   \\
        \smallskip
        Planet mass (TTV), $M_\mathrm{p}$ ($\mathrm{M_{\oplus}}$) & $13.5_{-9.0}^{+11.0}$ ($<36.8$ at 2-$\sigma$) & $2.2_{-1.5}^{+2.0}$ ($<5.7$ at 2-$\sigma$)   \\
        \smallskip
        Orbital eccentricity (TTV), $e$ & $0.09_{-0.05}^{+0.06}$ ($<0.21$ at 2-$\sigma$) & $0.04\pm0.03$ ($<0.10$ at 2-$\sigma$)  \\
        \hline
        \end{tabular}
        \end{center}
   \caption{Global model fitted results along with mass and eccentricity estimates from the TTV analysis (see Sect.~\ref{subsec:ttv}). For each parameter, we indicate the median of the posterior distribution function, along with the 1-$\sigma$ credible intervals. The equilibrium temperature corresponds to a case with null Bond albedo and no heat recirculation from the day side to the nightside hemispheres of the planet.\label{tab:gares}}
\end{table*}

\subsection{Dynamical analysis}\label{subsec:dyn}

\subsubsection{Mass constraints from TTVs}
\label{subsec:ttv}
As the system is within 6\% of the second order 5:3 mean motion resonance (MMR) and within 14\% of the stronger first order 2:1 MMR, we attempt to constrain planet masses from the transit timing variations \citep{Agol:2005, Holman:2005, Agol:2017} measured in our combined photometric dataset. We have in total 13 transits of planet b and five transits of planet c (Table~\ref{tab:ttv}). We computed TTVs using a similar MCMC setup as in Sect.~\ref{sec:ga}, but allowing each transit timing to vary, using the same tests as the ones described in Sect.~\ref{sec:ga} to ensure that the chains have converged. 
We derived posterior distribution functions for each transit timing that we use for our TTV analysis described in this section. While our observations provide only a partial sampling of the TTV libration periods \citep{Steffen:2006, Lithwick:2012} of $\sim$ 106 and 68 days for the 5:3 and 2:1 MMR respectively, we can still place upper limits on the masses computed from N-body simulations.

\begin{figure}
    \centering
    \includegraphics[width=0.48\textwidth]{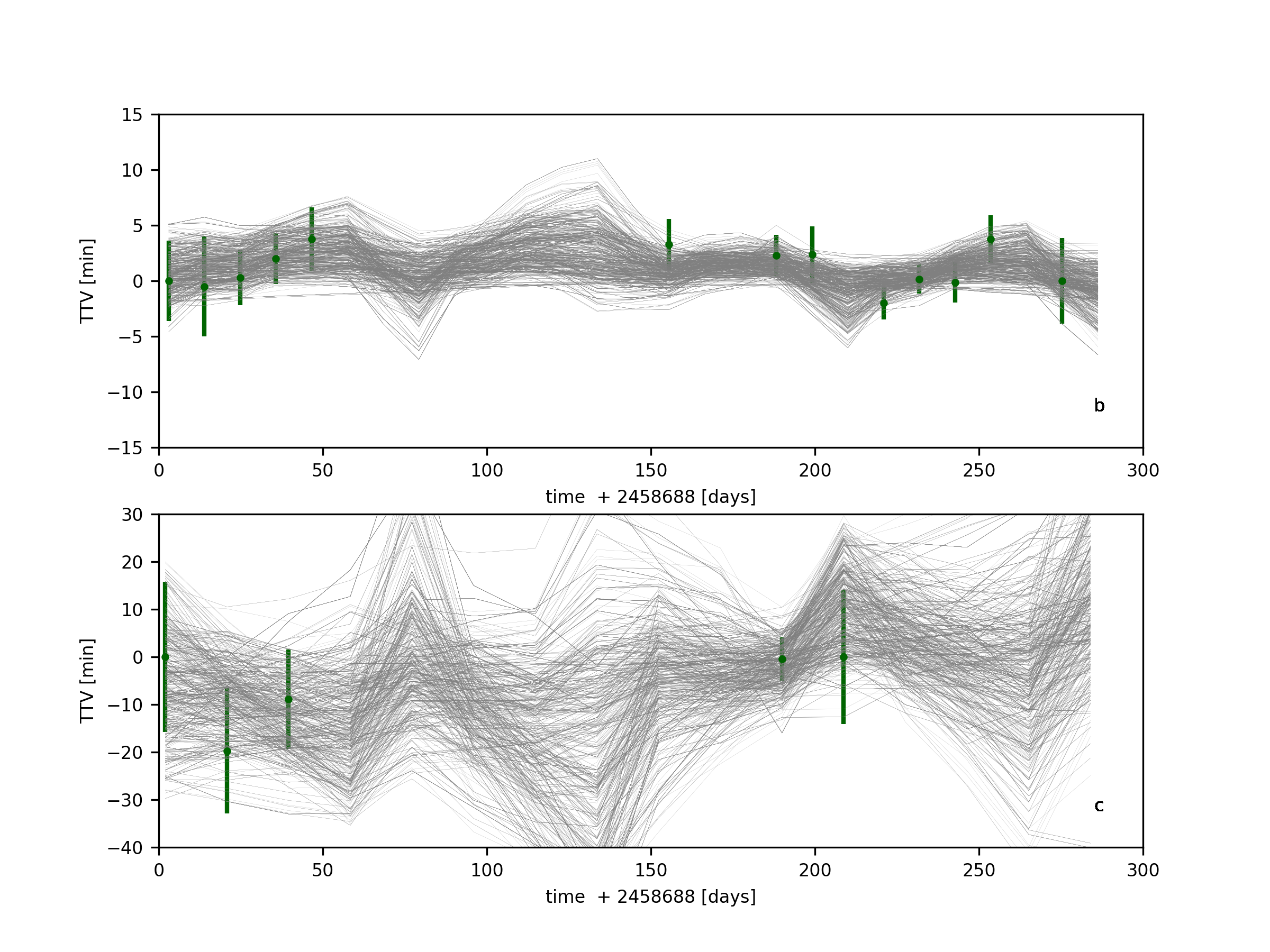}
    \caption{TTV signal of the two planets. In green are shown the observed transit times with the corresponding uncertainties, in black are shown the transit times from 1000 random MCMC samples.}
    \label{fig:TTV}
\end{figure}

\begin{figure*}
    \centering
    \includegraphics[width=0.48\textwidth]{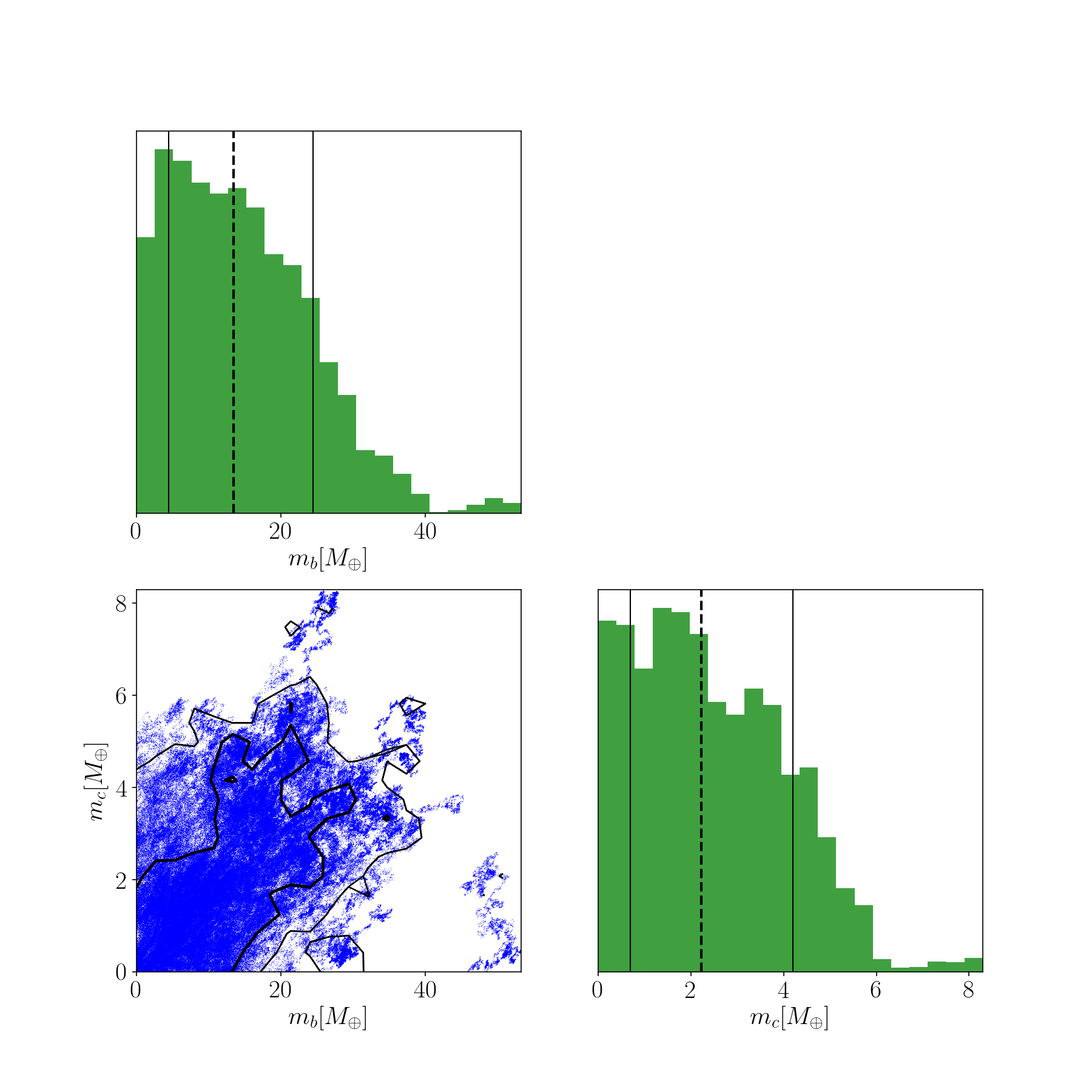}
    \includegraphics[width=0.48\textwidth]{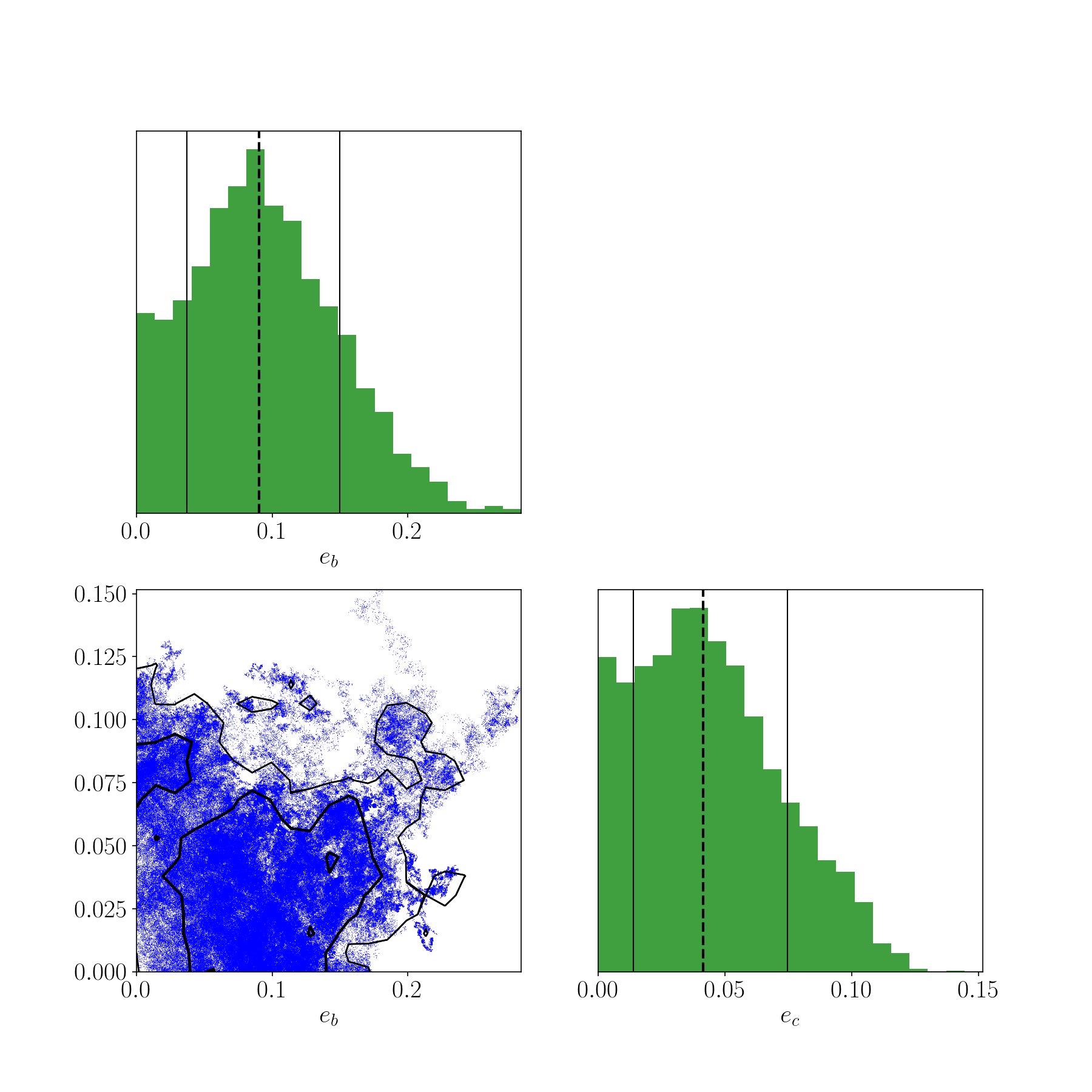}
    \caption{Posterior distribution for TOI-1266\,b and c masses (left), and eccentricities (right) from the TTV analysis. The histograms show the median (dashed lines) and the 1-$\sigma$ credible intervals (solid lines). We show in blue the posterior distributions with the 1- and 2-$\sigma$ contour lines. \label{fig:TTVM}}
    
\end{figure*}

We used the integrator GENGA (Gravitational Encounters in N-body simulations with Graphics processing unit Acceleration) \citep{Grimm:2014} together with a differential evolution Markov chain Monte Carlo (DEMCMC) method \citep{Braak:2006,Vrugt:2009}, as described in \cite{Grimm:2018}, to perform a TTV analysis of the transit timings reported in Table~\ref{tab:ttv}. The TTV signal is shown in Figure~\ref{fig:TTV} together with 1000 samples from the MCMC calculations. We find planetary masses of $M_\mathrm{p}$ = $13.5_{-9.0}^{+11.0}$ $\mathrm{M_{\oplus}}$ ($<36.8$ $\mathrm{M_{\oplus}}$ at 2-$\sigma$) for TOI-1266\,b, and $2.2_{-1.5}^{+2.0}$ $\mathrm{M_{\oplus}}$ ($<5.7$ $\mathrm{M_{\oplus}}$ at 2-$\sigma$) for TOI-1266\,c. We constrain the orbital eccentricities to $0.09_{-0.05}^{+0.06}$ ($<0.21$ at 2-$\sigma$) for TOI-1266\,b, and $0.04\pm0.03$ ($<0.10$ at 2-$\sigma$) for TOI-1266\,c. The posterior distributions of the derived masses and eccentricities are shown in Figure \ref{fig:TTVM}. We note that the arguments of periastron remain unconstrained. A visual inspection of Fig.~\ref{fig:TTV} shows that the libration period favoured by the fit is $\sim$70 days, which suggests that the system dynamics are more influenced by the 2:1 MMR than the 5:3 MMR. The small number of TOI-1266\,c transits combined with their shallow depth causes the derived mass for TOI-1266\,b to be less constrained by the data. During our exploration of the parameter space, we found another, lower statistical significance solution, yielding a similar mass for c but higher mass and smaller eccentricity for b. Further high-precision transits are, thus, required to improve these preliminary mass and eccentricity constraints. It is also worth mentioning that the expected radial-velocity semi-amplitude of approximately 4 m/s would enable mass measurements of TOI-1266\,b with Doppler spectroscopy, thus, complementing the TTV technique for the innermost planet.

\begingroup
\begin{table*}
\begin{center}
\begin{tabular}{l l l}
\toprule
Predicted timing & Observed difference & Source \\
BJD$_{\rm TDB}$-2450000 & days & \\
\midrule
        \hline
\multicolumn{3}{c}{\textit{TOI-1266\,b}} \\
        \hline
8691.00632	&  $-0.00138^{+0.00237}_{-0.00252}$  &  TESS \\
\smallskip
8701.90112	&  $-0.00164^{+0.00260}_{-0.00312}$  &    TESS \\
\smallskip
8712.79593	&  $-0.00101^{+0.00202}_{-0.00171}$  &    TESS \\
\smallskip
8723.69073	&  $0.00028^{+0.00159}_{-0.00158}$  &    TESS \\
\smallskip
8734.58553	&  $0.00160^{+0.00185}_{-0.00199}$  &  TESS \\
\smallskip
8843.53357	&  $0.00218^{+0.00158}_{-0.00142}$  &  OAA \\
\smallskip
8876.21798	&  $0.00178^{+0.00127}_{-0.00125}$  &  TESS \\  
\smallskip
8887.11278	&  $0.00192^{+0.00178}_{-0.00144}$  &  TESS \\  
\smallskip
8908.90239	&  $-0.00092^{+0.00086}_{-0.00103}$  &  TESS + SAINT-EX \\ 
\smallskip
8919.79719	&  $0.00066^{+0.00092}_{-0.00086}$  &  TESS \\
\smallskip
8930.69199	&  $0.00054^{+0.00106}_{-0.00127}$  &  ARTEMIS + TRAPPIST-N \\
\smallskip
8941.58680	&  $0.00334^{+0.00142}_{-0.00148}$  &  TRAPPIST-N \\
\smallskip
8963.37640	&  $0.00092^{+0.00210}_{-0.00267}$  &  Kotizarovci + ZRO \\
        \hline
        \hline
\multicolumn{3}{c}{\textit{TOI-1266\,c}} \\
        \hline
8689.96122	&  $0.01117^{+0.01095}_{-0.01084}$  &  TESS \\  
\smallskip
8708.76250	&  $-0.00347^{+0.00883}_{-0.00913}$  &  TESS \\ 
\smallskip
8727.56379	&  $0.00313^{+0.00726}_{-0.00656}$  &  TESS \\  
\smallskip
8877.97410	&  $0.00165^{+0.00293}_{-0.00321}$  &  TESS + SAINT-EX \\  
\smallskip
8896.77539	&  $0.00105^{+0.00979}_{-0.00702}$  &  TESS \\  
\hline 
\end{tabular}
\end{center}
\caption{
Transit timings used in the TTV analysis.\label{tab:ttv}}
 
\end{table*}
\endgroup

\subsubsection{Stability analysis}
\label{sec:stability}

With the current set of data, we found as a likely planetary architecture a system composed by two eccentric planets: a more massive inner planet, TOI-1266\,b, at 0.0736~au, and a lighter one, TOI-1266\,c, at 0.1058~au. Unfortunately, due to the large uncertainties in their masses, the nature of these planets remains unknown (see Table~\ref{tab:gares}). In this section, we seek to constrain the masses and eccentricities of the planets by exploring the global stability of the system. To achieve this, we made use of the  Mean Exponential Growth factor of Nearby Orbits, $Y(t)$ \citep[MEGNO;][]{Cincotta1999ConditionalEntropy,Cincotta2000SimpleI,Cincotta2003PhaseOrbits} parameter. This chaos index has been widely used to explore the stability of both the Solar System, and extrasolar planetary systems \cite[e.g.][]{Jenkins2009FirstDesert,Hinse2015PredictingSystem,Wood2017TheRings,Horner2019TheSolution}. In short, MEGNO evaluates the stability of a body's trajectory after a small perturbation of its initial conditions through its time-averaged mean value, $\langle Y(t) \rangle$, which amplifies any stochastic behaviour, thereby allowing one to distinguish between chaotic and quasi-periodic trajectories during the integration time: if $\langle Y(t) \rangle \rightarrow 2$ 
for $t\rightarrow \infty$, the motion is quasi-periodic; while if $\langle Y(t) \rangle \rightarrow \infty$ for $t\rightarrow \infty$, the system is chaotic. To investigate this, we used the MEGNO implementation with an N-body integrator {\scshape rebound} \citep{rein2012}, which employs the Wisdom-Holman WHfast code \citep{rein2015}.

\begin{figure}
\includegraphics[width=0.50\textwidth]{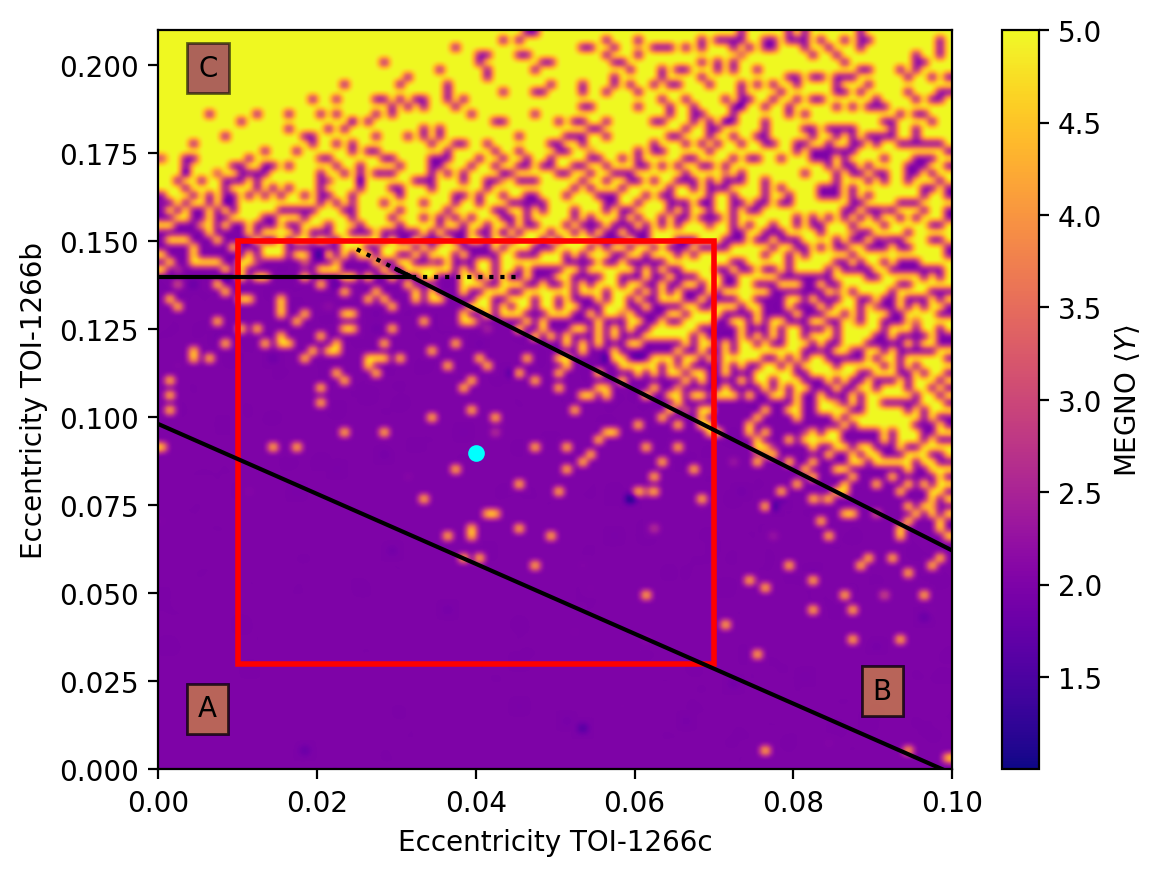}
\caption{Dynamical analysis of the TOI-1266 system based on a MEGNO-map.
The size of the map is 100$\times$100 pixels, which explores the $e_{b}$--$e_{c}$ parameter space at the 2-$\sigma$ uncertainty level.
When $\langle Y(t) \rangle \rightarrow 2$ (purple shaded regions), quasi-periodic orbits are found, while chaotic 
systems are found when $\langle Y(t) \rangle \rightarrow 5$ (yellow shaded regions). The configuration with the nominal values of $e_{b}$ and $e_{c}$ is shown by the
blue marker, while the 1-$\sigma$ uncertainty region is shown by the red box. The three areas, A (stable region), B (transition region), and C (unstable region) have been 
labelled, and are delimited by the solid black lines, which represent the conditions given by Eqs.~\ref{eq:A} and ~\ref{eq:B}.}
\label{fig:ecc}
\end{figure}

We performed two suites of simulations to explore if the system could actually be fully stable in the range of the 2-$\sigma$ uncertainties obtained for the masses and the eccentricities (see Section~\ref{subsec:ttv}). To address this question, we constructed two-dimensional MEGNO-maps in the $M_{b}$--$M_{c}$ and $e_{b}$--$e_{c}$ parameter spaces following \cite{james2019}.
Hence, in the first case, we explored planet masses ranging from 1 to 37~M$_{\oplus}$ for TOI-1266\,b, and from 1 to 6~M$_{\oplus}$ for TOI-1266\,c. In the second case, we explored planet eccentricities in the ranges of 0.0--0.21 and 0.0--0.1 for TOI-1266\,b and TOI-1266\,c, respectively. In both sets of simulations, we took 100 values from each range,  meaning that the size of the obtained MEGNO-maps were 100$\times$100 pixels.  Thus, we explored the $M_{b}$--$M_{c}$ and $e_{b}$--$e_{c}$ parameter spaces up to 10,000 times in total. In each case, we fixed the other parameters to the nominal values given in Table~\ref{tab:gares}. The integration time was set to 1 million orbits of the outermost planet, and the time-step was set to 5$\%$ of the orbital period of the innermost planet. We found that, concerning the masses, the system is fully stable
in the range of values studied here; hence we cannot set extra constraints on the planetary masses. On the other hand, we found that the full set of eccentricities explored is not permitted, and we identified regions with different behaviours where the system transitions from stable to unstable gradually towards the upper-right region of the $e_{b}$--$e_{c}$ parameter space (see Fig.~\ref{fig:ecc}). This allowed us to clearly identify three different regions (A, B, and C). First, where the system is fully stable (A), and where the mutual eccentricities follow the relationship given by:
\begin{equation}\label{eq:A}
    e_{b}+0.992e_{c}<0.098 . \\
\end{equation}

Then, the second is a transitional region where the system is still stable, but some instabilities appear (B). This region spans from the limit given by Eq.~\ref{eq:A}, and the upper limit given by: 
\begin{equation}
\begin{cases}\label{eq:B}
e_{b} < 0.140  & \text{if $e_{c} < 0.032$}, \\
e_{b} + 1.14e_{c} < 0.17 & \text{if $e_{c} > 0.032$}  .
\end{cases}
\end{equation}
 
Finally, region C, where the aforementioned relationships are violated, and instability is more likely. We found that the nominal values are stable in the transition region of the parameter space (i.e. region B), which may hint that these values could be more appropriately considered as upper limits to stability, where larger values would rapidly turn into unstable scenarios. This encouraged us to favour the hypothesis of low eccentricities for both planets in terms of long-term stability, with the most restrictive condition given by Eq.~\ref{eq:A}. Hereafter, for our dynamical purposes, we have adopted the nominal planetary masses and eccentricities given in Table~\ref{tab:gares}, which, as we demonstrated, are stable and, in the case of the eccentricities, may represent a realistic upper limits of their real values.

\begin{figure}
\includegraphics[width=\columnwidth]{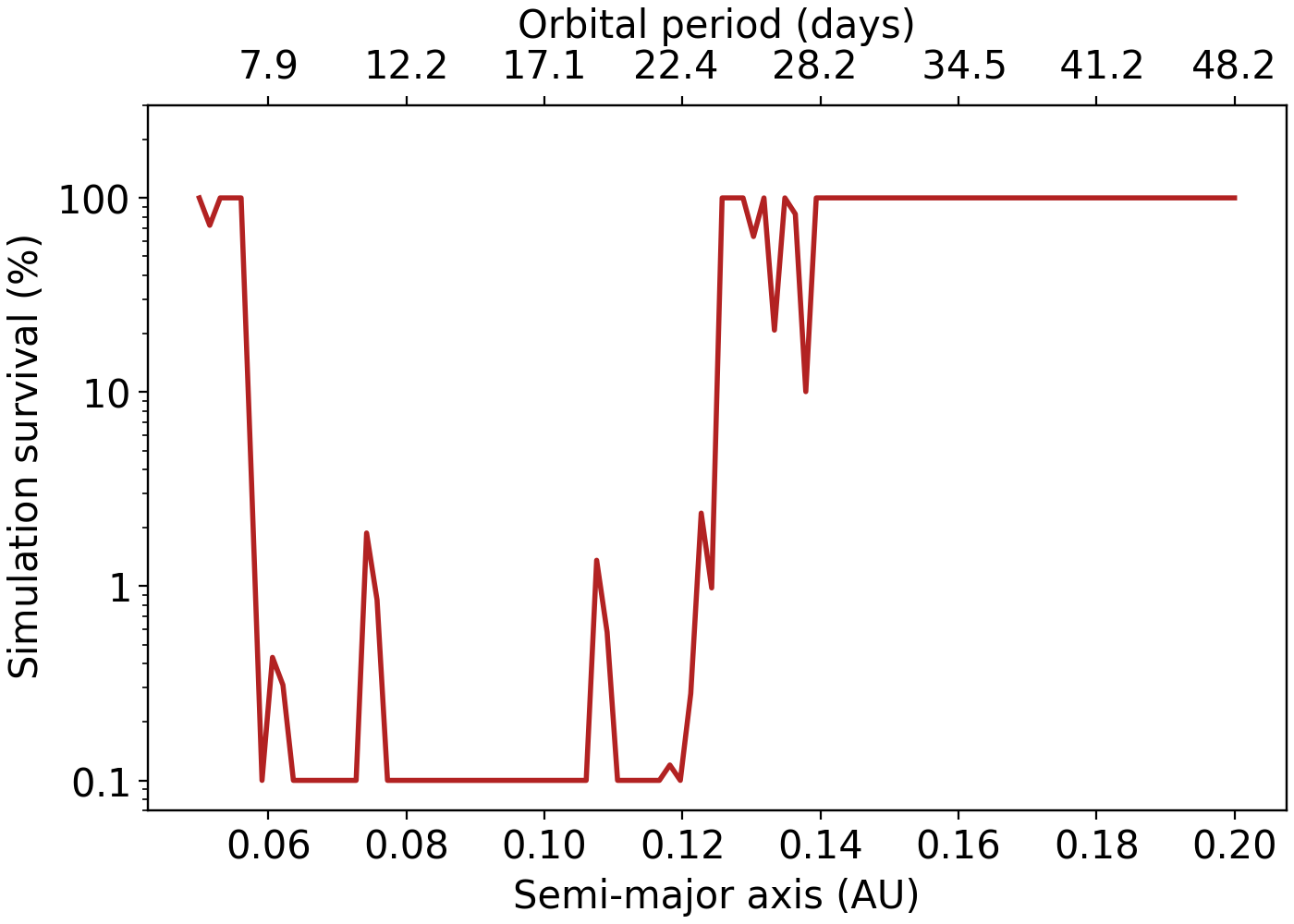}
\caption{Survival rate of an additional Earth-mass planet in the TOI-1266 system. The results of a dynamical stability analysis that placed an Earth-mass within the system in the semi-major axis range of 0.05--0.20~au and evaluated the overall system stability for a period of $10^6$ years (see description in Section~\ref{sec:stability}). These results demonstrate that there are few viable orbits allowed in the range 0.064--0.131~au, but that additional terrestrial planets may exist in the system outside of that range.}
\label{fig:stab}
\end{figure}

Using the stellar parameters given in Table~\ref{tab:starlit}, the planetary parameters provided in Table~\ref{tab:gares}, and the derived planetary masses described above, we conducted a dynamical analysis of the system with the goal of testing for the potential existence of additional planets, especially in the region where the TESS photometry is not accurate enough to detect them, as the case for planets smaller than 1.5~R$_{\oplus}$ with periods longer than 4~days (see Fig.~\ref{recovery}). For this purpose, we used the {\sc{mercury}} Integrator Package \citep{chambers1999} to perform the N-body integrations. The adopted methodology is similar to that used by \citet{kane2015b,kane2019c}, where a grid of initial conditions is used to explore the valid parameter space for possible additional planets within the system. The innermost planet has an orbital period of $\sim$10.8~d, and so we used a conservative time step of 0.2~days to ensure the perturbative reliability of the simulations. An initial simulation of the two known planets for $10^7$ years, equivalent to $2 \times 10^8$ orbits of the outer planet, demonstrated the intrinsic stability of the system described in Table~\ref{tab:gares}. We then inserted an Earth-mass planet in a circular orbit at several hundred locations within the semi-major axis range of 0.05--0.20~au to test for possible locations of additional planets within the system. These simulations were executed for $10^6$ years each, and the results were evaluated based on the survival of all planets. Non-survival means that one or more of the planets were ejected from the system or lost to the potential well of the host star. The results of this simulation are shown in Fig.~\ref{fig:stab}, where the semi-major axis (bottom) and orbital period (top) are shown on the $x$-axis, and the percentage of the simulation time during which all three planets survived is shown on the $y$-axis. These results show that the semi-major axis range of 0.058--0.138~au is a largely unstable region where, given an additional planet, it is highly unlikely that orbital integrity of the system may be retained. However, outside of this range, the viability of orbits is rapidly regained and so there is potential for numerous other locations where additional transiting or non-transiting planets may contribute to the overall system architecture.

\subsubsection{Tidal evolution}

TOI-1266\,b and TOI-1266\,c are close-in exoplanets, which, due to their proximity to their host star, may be affected by tides. While the size and maximum mass face values of TOI-1266\,c point to a terrestrial composition, the large uncertainty in the mass of TOI-1266\,b prevents us from making any definitive assertions regarding its nature. Therefore, in this section
we only focus on the tidal evolution of TOI-1266\,c. To quantify the influence of tides on this planet, we made use of the constant time-lag model (CTL), where a planet is considered as a weakly viscous fluid that is deformed due to gravitational effects \cite[see e.g.][]{Mignard1979TheI,Hut1981TidalSystems.,Eggleton1998TheFriction,Leconte2010IsEccentricity}. Tides affect each orbital parameter over a different time-scale: the obliquity and rotational period are the first to be altered, while the eccentricity and the semi-major axis are affected over longer periods of time \citep[see e.g.][]{Bolmont2014FormationSystem, barnes2017}. To study the effects of tides, we made use of {\sc{posidonius}} \citep{Blanco-Cuaresma2017StudyingRust}. The main free parameters that define the tidal dissipation of a given planet are the degree-2 potential Love number k$_{2}$ and its constant time lag $\Delta \tau$, where k$_{2}$ can have values between 0 and 1.5, and $\Delta \tau$ can span orders of magnitudes \citep{barnes2017}. Observations have revealed that the Earth's dissipation is completely dominated by the friction induced by its topography on tidal gravito-inertial waves that propagate in the oceans \citep{mathis2018}. Therefore, when exploring terrestrial exoplanets under the influence of tides, the Earth's reference value k$_{2,\oplus}\Delta\tau_{\oplus}$=213~s \citep{neron1997} is commonly adopted. Typically, 0.1$\times$k$_{2,\oplus}\Delta\tau_{\oplus}$ is used for planets without oceans, and 10$\times$k$_{2,\oplus}\Delta\tau_{\oplus}$ for volatile-rich planets \citep[see e.g.][]{Bolmont2014FormationSystem,Bolmont2015Mercury-T:Kepler-62}. This strategy allowed us to identify a range of possible tidal behaviours. 

For a close-in exoplanet, it is expected that tidal torques fix the rotation rate to a specific frequency, in a process called tidal locking.  A tidally-locked planet in a circular orbit will rotate synchronously, and the same side of the planet will always face the star. On the other hand, for a fluid planet, the rotation is pseudosynchronous, which means that its rotation tends to be as fast as the angular velocity at periapse (hence, the same side does not always face the star). Planets that are solid, like Mercury, can be locked because of a permanent deformation into a specific resonance\footnote{In Mercury's case, it's the 3:2, and because of Mercury's large e, that is near pseudosynchronous, but if Mercury had a smaller e, it could have gotten trapped into synchronous orbit.} \citep[see e.g.][]{Bolmont2015Mercury-T:Kepler-62,Hut1981TidalSystems.,barnes2017}.

To ascertain which of these scenarios applies to TOI-1266\,c, we followed the aforementioned strategy: we studied the evolution of the obliquity ($\epsilon$) and rotational period ($\mathrm{P}_{\mathrm{rot}}$) by considering a number of cases with: different initial planetary rotational periods of 10~hr, 100~hr, and 1000~hr, combined with obliquities of 15$\degr$, 45$\degr$, and 75$\degr$, 
for the three cases concerning tidal dissipation: (0.1, 1, and 10)$\times$k$_{2,\oplus}\Delta\tau_{\oplus}$. 

\begin{figure}
\includegraphics[width=\columnwidth]{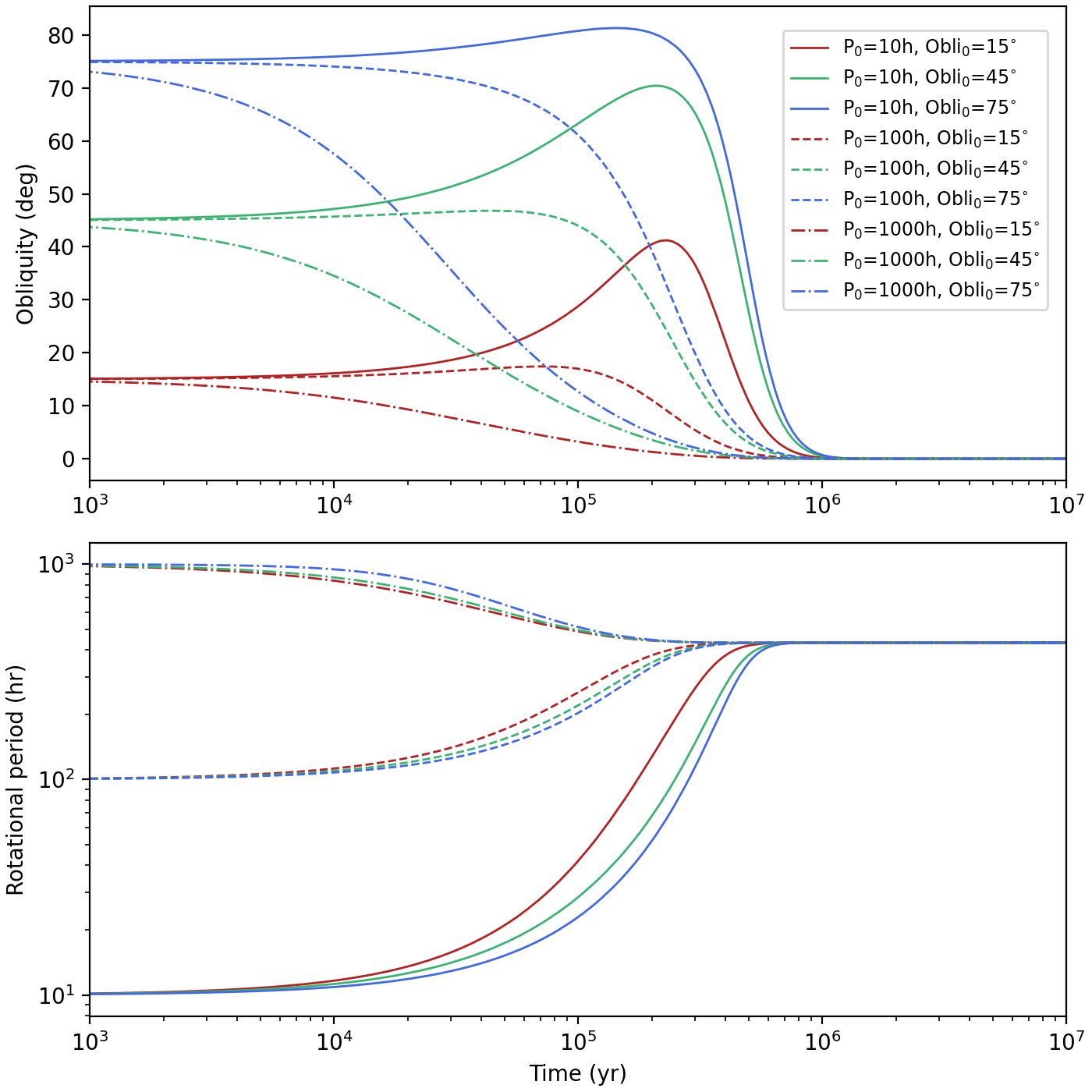}
\caption{Obliquity and rotational period evolution due to the effects of tides for planet TOI-1266\,c for the case of 1$\times$k$_{2,\oplus}\Delta\tau_{\oplus}$. Each set of line-style curves represents a different initial spin-rate, and each set of coloured curves represents a different initial obliquity. It is seems that for any possible combination of the initial conditions, within only a short time scale (compared with the age of the star), the planet is tidally locked.}\label{rot_obli}
\end{figure}

We found that in all cases, the planet is tidally locked rapidly: 10$^{7}$~yr for 0.1$\times$k$_{2,\oplus}\Delta\tau_{\oplus}$, 10$^{6}$~yr for 1$\times$k$_{2,\oplus}\Delta\tau_{\oplus}$, and 10$^{5}$~yr for 10$\times$k$_{2,\oplus}\Delta\tau_{\oplus}$. The results for the particular case of 1$\times$k$_{2,\oplus}\Delta\tau_{\oplus}$ are displayed in Fig.~\ref{rot_obli}. We note that the presence of a relatively large moon orbiting the planet may provoke chaotic fluctuations of its obliquity, or even impart it with another value \citep[see e.g. ][]{Laskar1993ThePlanets,Lissauer2012ObliquityEarth}. However, it has been found that compact planetary systems are unlikely to have moons \citep{Lissauer:1985,kane2017}, which encourages us to tentatively consider that TOI-1266\,c is tidally locked and fairly well 
aligned with the host star. 

Circularisation of the orbits due to the effects of tides is a slow process, which may last from hundreds to billion years \cite[see e.g.][]{Bolmont2015Mercury-T:Kepler-62,barnes2017,pozuelos2020}.
From our results of the global model, we found that the planet TOI-1266\,c may have a certain level of eccentricity with a median value of 0.04. This may be provoked by the lack of time to  circularise the orbit by tides (which is not 
explored in this study) or due to the architecture of the system, where the planets are close to the 5:3 and 2:1 MMRs, which can excite the orbits and consequently induce a marginal level of eccentricity. Hence, both planets might be experiencing some level of tidal heating. In this context, we computed for TOI-1266\,c the tidal heating for the nominal eccentricity of 0.04 for (0.1, 1, and 10)$\times$k$_{2,\oplus}\Delta\tau_{\oplus}$ configurations. We found that the planet is heated by 0.5~W~m$^{-2}$ for 10$\times$k$_{2,\oplus}\Delta\tau_{\oplus}$, 0.05~W~m$^{-2}$ for 1$\times$k$_{2,\oplus}\Delta\tau_{\oplus}$, and 0.005~W~m$^{-2}$ for 0.1$\times$k$_{2,\oplus}\Delta\tau_{\oplus}$. 

Taken together, our results suggest that it is likely that the outermost planet TOI-1266\,c is tidally locked, but due to its small but non-zero eccentricity, which may persist through MMR excitations, it may not have the same side always facing the star. In a recent paper by \cite{pozuelos2020}, the authors adapted the general description of the flux received by a planet given by \cite{kane2017b} for the configuration found here, for a tidally-locked planet on a non-circular orbit. It was found that in such a configuration, the orbital phase is no longer relevant, and the distribution of flux along the latitude ($\beta$) is the same for the whole orbit, which depends only on the planet's eccentricity. In our case, assuming the nominal value of eccentricity found in our models (i.e. 0.04) we found 3805--3242~W~m$^{-2}$ (2.79--2.37~F$_{\oplus}$)  at the equator,  2691--2293~W~m$^{-2}$ (1.97--1.68~F$_{\oplus}$) at $\beta=\pm45\degr$, and $\sim$0~W~m$^{-2}$ (0.0~F$_{\oplus}$) at the poles. These values suggest that the stellar flux received by the planet may vary by $\sim$15$\%$ along its orbit. Based on dynamical arguments, this variation of $\sim$15$\%$ may be considered as an upper limit. Indeed, we demonstrated that the nominal values of the eccentricities might be considered as upper limits, where the system tends to more stable scenarios for low eccentricities (see Fig.~\ref{fig:ecc}). At the 1-$\sigma$ uncertainty level, the minimum eccentricity of TOI-1266\,c is 0.01. In such a case, the variation of the flux along the orbit is only $\sim$4$\%$. Hence, a plausible set of values for the variation of the flux along the orbit ranges from 4 to 15$\%$.

\section{Discussion}\label{sec:disc}

\subsection{TOI-1266\,b and c in the mass-radius diagram}

\begin{figure*}
\includegraphics[width=\textwidth]{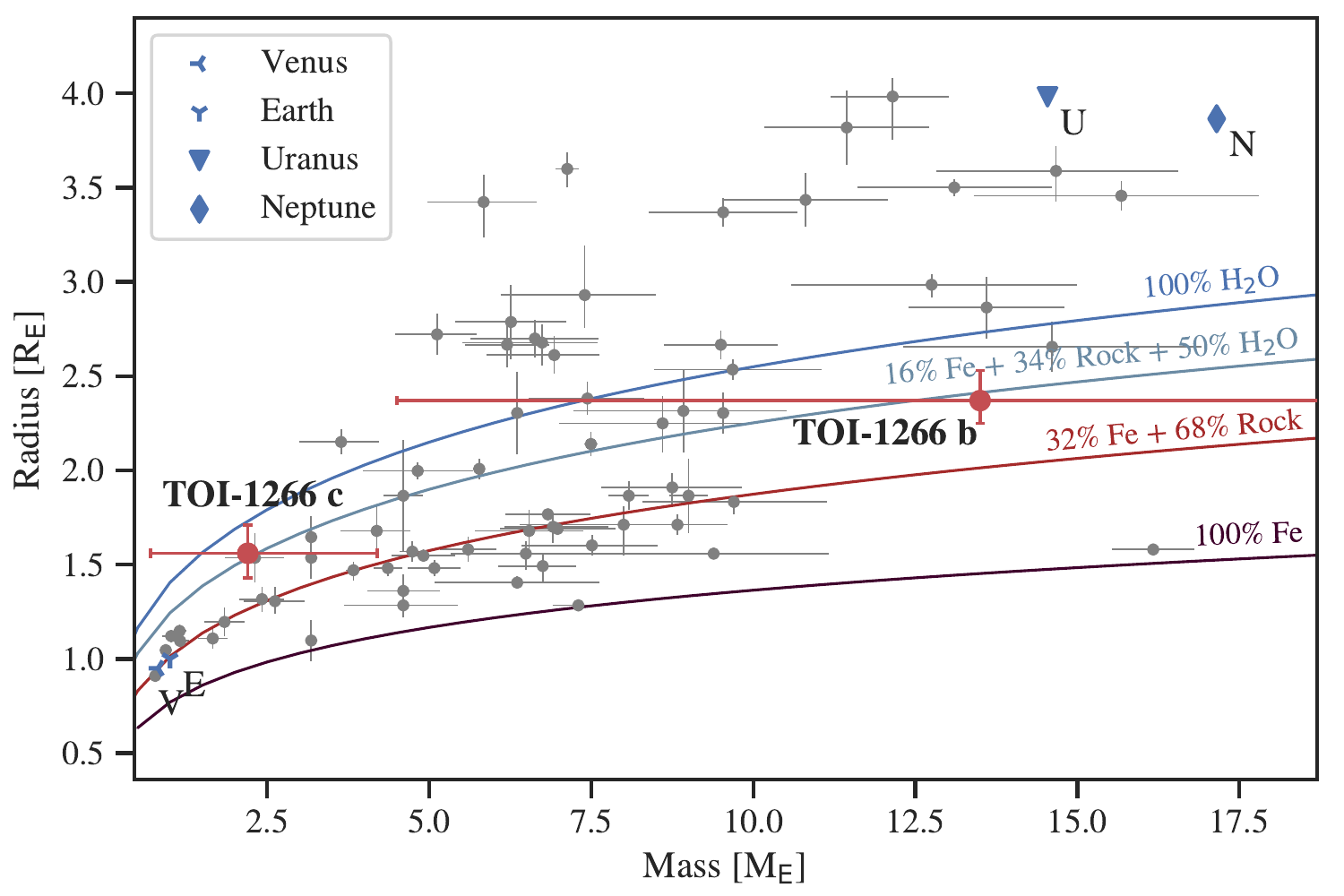}
\caption{TOI-1266\,b and c on the mass-radius diagram based on the radii and masses derived from the transit photometry and TTV analyses. Alongside, we show mass radius curves for pure iron (brown), an earth like composition of 32\% iron, and 68\% rock (red), a mixture of 50\% water with an earth like core (grey blue), and for pure water (blue). The data of the background planets were taken from the NASA Exoplanet Archive (retrieved on 2020 July 15). We show only those planets that have a relative mass and radius uncertainty of less than 20\%.}\label{fig:mrd}
\end{figure*}

In this section, we use the mass and radii constraints on TOI-1266\,b and c derived from the transit photometry to investigate the location of these planets in the mass-radius diagram. We emphasise that the mass constraints will be improved once additional transit timings, or precise radial-velocity timeseries become available. However, Figure~\ref{fig:mrd} provides a first assessment on whether the derived planet properties are consistent with usual mass-radius relationships. In this figure, the composition curves were calculated with model B of \citet{michel_2020}. We used for iron the equation of state of \citet{hakim_2018}, for rock the model of \citet{sotin_2007} and for water the equation of state of \citet{mazevet_2019}.

\subsection{TOI-1266 and the radius valley}
\label{sec:ev}
With radii of about 2.4 and 1.6 $R_{\oplus}$, TOI-1266\,b and c span the so-called radius valley \citep{Fulton:2017}. 
Since atmospheric evaporation is a possible explanation for the origin of the valley \citep{Owen2013,Lopez2013,Jin2014},
this feature is also known as the `evaporation valley'. We note, however, that the origin for the radius valley feature is currently unconstrained and that other studies advocate, for instance, for core-powered mass loss as the driving mechanism for this pattern \citep{Ginzburg:2018,Gupta:2020}. The recent discovery of a change in the fraction of planets above and below the valley over $\sim$\,Gyr timescales has indeed been interpreted in favour of the core-powered mass loss scenario for some solar-type stars \citep{Berger:2020}, although there is evidence that the formation pathway may be different in low-mass stars \citep{Cloutier2020a}.

Planets in multiple systems with dissimilar radii, like TOI-1266\,b \& c, Kepler-36\,b \& c \citep{Carter2012}  
or TOI-402\,b \& c \citep{Dumusque2019},
are of particular interest to understand the origin of the radius valley, as they allow to study the differential evolutionary history of the planets, and to check whether evaporation can self-consistently explain all the planets in a system \citep{Lopez:2012a,Lopez:2013a,Owen:2019}. 

To understand the implication of the hypothesis that evaporation has shaped the distinct radii of TOI-1266\,b \& c, we simulated their long-term thermodynamical evolution (cooling, contraction, atmospheric escape) with the Bern evolution model \texttt{completo21} 
\citep{Mordasini2012,Jin2014,Jin2018}.

The model simulates the long-term evolution of the planets after the dissipation of the protoplanetary  disc by solving the internal structure equations of the planets. It includes XUV-driven atmospheric photoevaporation in the radiation-recombination and energy-limited regimes 
\citep{MurrayClay2009}. 

The planets consist of a solid core and a H/He envelope, and we assumed that their cores have an Earth-like 2:1 silicate:iron composition described by the polytropic equation of state (EOS) of 
\citet{Seager2007}.

The gaseous envelope consists of H/He, described by the EOS of 
\citet{Saumon1995}, and the opacity corresponds to a condensate-free solar-composition gas 
\citep{Freedman2014}.

As in \citet{Mordasini2020}, we simulated the evolution of 6\,000 planets on a grid of semi-major axis and mass. The initial conditions (i.e. the post-formation envelope --- core mass ratio and luminosity) are the same as in the nominal case considered in \citet{Mordasini2020}, but the stellar mass is now 0.5 $M_{\rm \odot}$. The stellar XUV luminosity as a function of time was taken from \citet{McDonald2019}.  

\begin{figure}
\centering
\resizebox{\hsize}{!}{\includegraphics[]{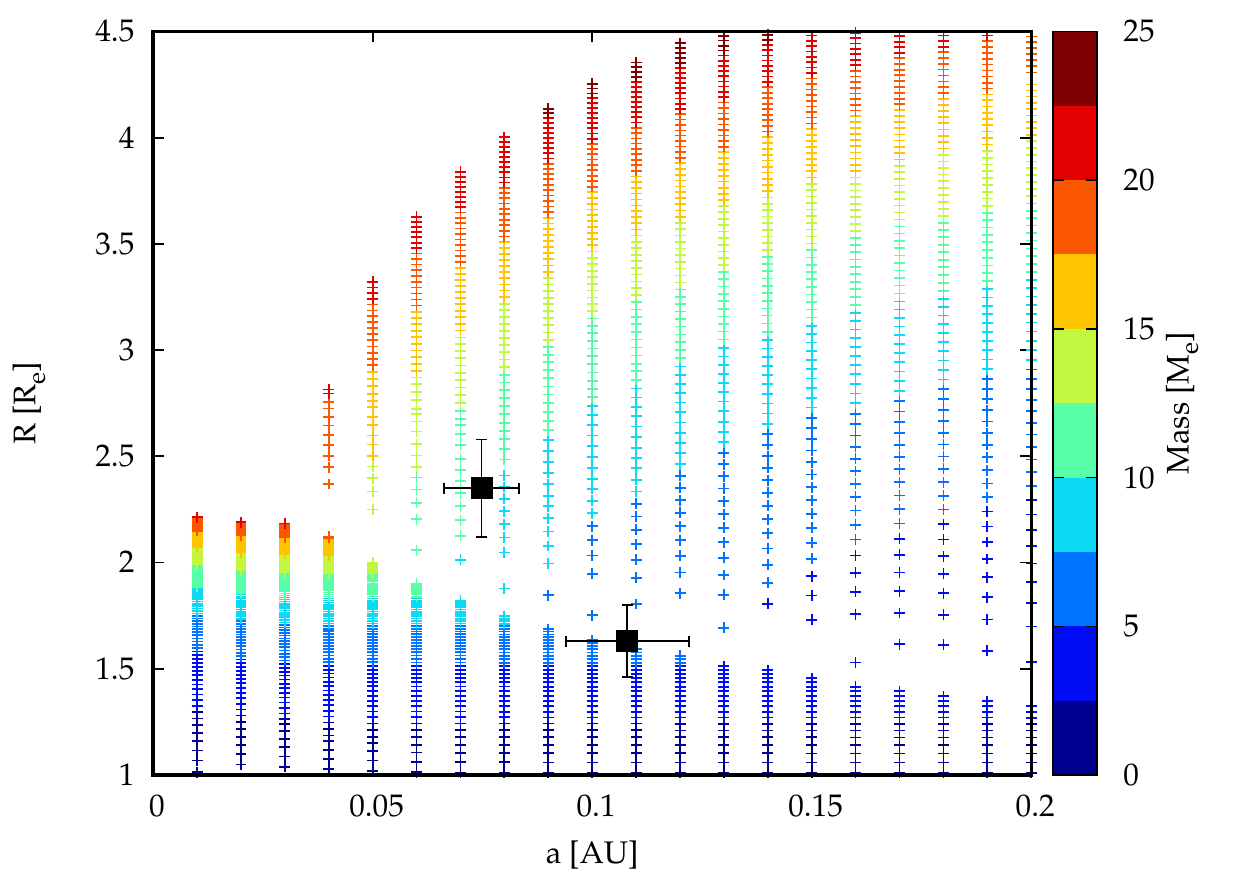}}
\caption{Comparison of TOI-1266\,b \& c with evaporation models. The coloured dots show the position in the plane of semi-major axis versus radius of simulated planets evolving under the effect of cooling, contraction, and atmospheric escape around a 0.5 $M_{\rm \odot}$ star at an age of 10 Gyr. One sees the radius valley and the sub-Neptunian desert. The colours show the planet mass. Black symbols show TOI-1266\,b \& c.}
\label{fig:ar}
\end{figure}

Figure~\ref{fig:ar} shows the result in the plane of orbital distance versus planet radius at an age of 10 Gyr, where the evaporation valley is apparent. Planets above this threshold retain some H/He, while those below become bare rocky cores. In the top left corner, the region devoid of planets corresponds to the sub-Neptunian desert 
\citep{Lundkvist2016,Mazeh2016,Bourrier2018}. 

TOI-1266\,b \& c are also shown as black squares. Planet b is located clearly above the valley, while planet c's median radius sits just underneath. This position corresponds to the most massive (and largest) cores at a given semi-major axis that have lost their H/He.

The interesting property of TOI-1266 is that the inner planet is larger than the outer one. In the context of evaporation, this can be explained if the inner planet is also more massive than the outer one. The higher mass allows  for the planet to keep its H/He envelope even at a higher XUV flux. We thus studied the masses of model planets that are compatible with TOI-1266\,b \& c in terms of orbital distance and radius in Figure~\ref{fig:ar}, and find that these model planets have masses of about $10\pm1.5$ $M_{\oplus}$ for planet b, and $6^{+1}_{-2}$ $M_{\oplus}$ for planet c, which is consistent with the masses derived using our TTV measurements (Sect.
~\ref{subsec:ttv}). These specific numbers depend on model assumptions, such as the core and envelope composition, the initial conditions, or the evaporation model. But the general result, namely that the inner planet should be more massive, is robust. 

We note that Figure~\ref{fig:ar} has been obtained with an evaporation model that assumes a constant evaporation efficiency in the energy limited domain. It is  known \citep{Owen2017} that this yields a steeper slope (R vs a) for the evaporation valley than the one found in models which calculate the evaporation efficiency self-consistently \citep{Owen2012}.

This shallower slope is also in better agreement with observations \citep{Mordasini2020}. This affects the prediction for the mass ratios of the two planets; in particular, a shallow slope means that the two masses can be more similar than estimated above. If we assume that the slope of the valley is instead the same as the slope observed for FGK stars \citep{VanEylen2018,Cloutier2020a}, and not the one predicted in the model, then we can estimate that the inner planet must have a mass that is at least 25\% higher than the mass of the outer planet \citep{Mordasini2020}. Then, the inner planet can keep some H/He.   

\subsection{Potential for atmospheric characterisation}

\begin{figure*}
\includegraphics[width=\textwidth]{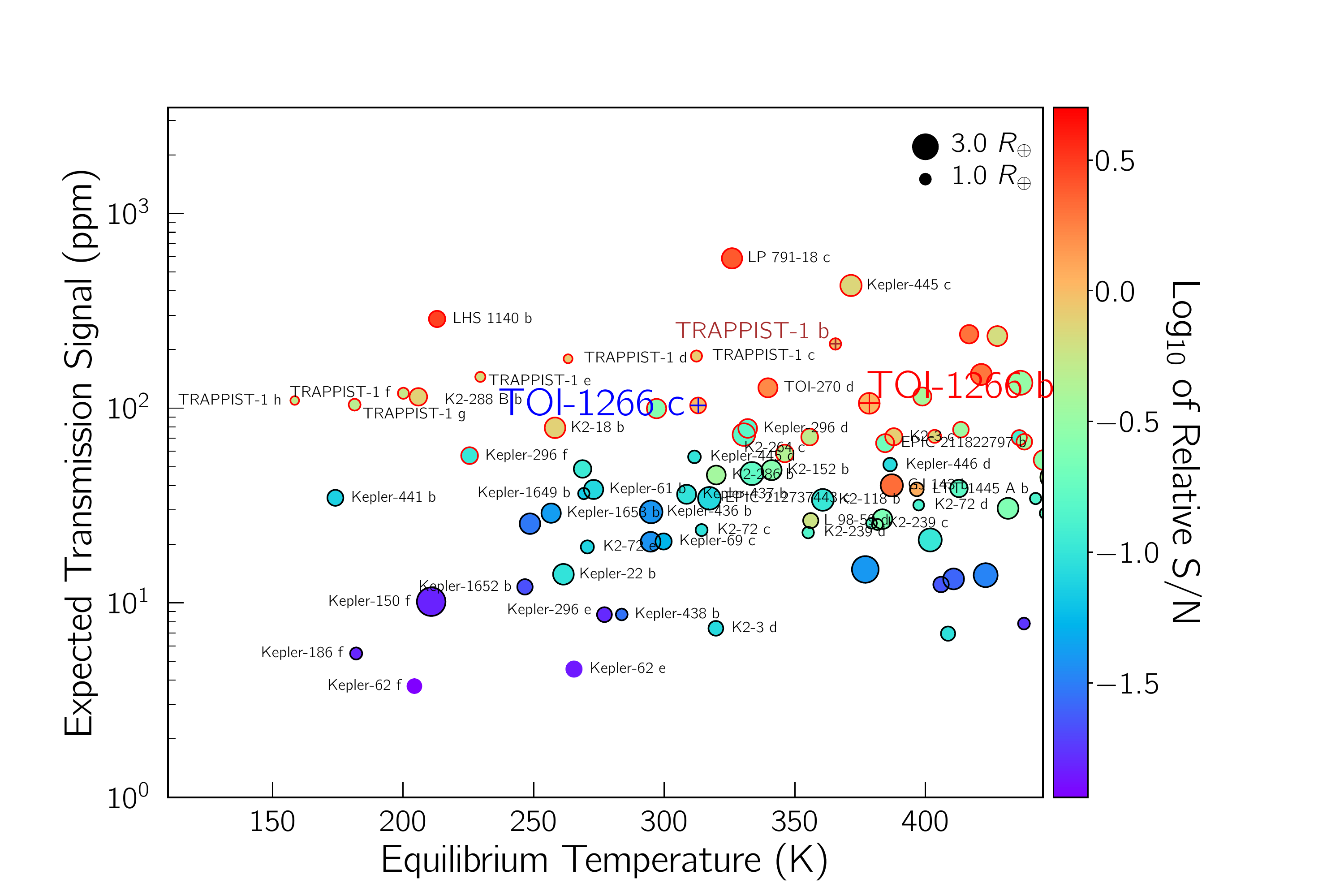}
\caption{Most promising sub-Neptune-sized planets for atmospheric characterisation. Point  colours illustrate the  S/N of a \textit{JWST/NIRSPEC} observation relative to TRAPPIST-1~b.   S/N below 1/100th of TRAPPIST-1~b have been removed to enhance readability of the figure. The planets for which the presence of an atmosphere could be assessed by \textit{JWST} within $\sim$ 100hrs are encircled in red, if their atmospheric signals are above \textit{JWST}'s threshold of $\sim$ 50 ppm. The size of the circle is proportional to the size of the planet.}\label{fig:atmsnr}
\end{figure*}

The potential for atmospheric characterisation of an exoplanetary system relates directly to the size and brightness of the host: the smaller and brighter the host, the larger the signal in transmission and photon count (i.e.  S/N), all other parameters being equal. TOI-1266’s relatively small size and proximity makes it a remarkably good host for the atmospheric study of super-Earth-sized and larger planets.  In order to quantify and  contextualise its prospect for atmospheric study, we followed the same approach as in \citet{Gillon:2016}, focusing here on temperate sub-Neptune-sized planets following the NASA Exoplanet Archive\footnote{\href{https://exoplanetarchive.ipac.caltech.edu}{https://exoplanetarchive.ipac.caltech.edu}}. Figure~\ref{fig:atmsnr} reports the planets’ signal in transmission, which we derive as follows:
\begin{equation}
\begin{split}
S &=\frac{2 R_p h_{\rm eff}}{R_*^2}, {\rm with}\\
h_{\rm eff} &= \frac{7 k T}{\mu g},
\label{eqn:AtmS}
\end{split}
\end{equation}
where $R_p$ is the planetary radius, $R_*$ is the stellar radius, $h_{\rm eff}$ is the effective atmospheric height, $\mu$ is the atmospheric mean molecular mass, $T$ is the atmospheric temperature, and $g$ is the local gravity. We assume $h_{\rm eff}$ to cover seven atmospheric scale heights, assuming atmospheres down to $\sim$0.1bar. We assume the atmospheric mean molecular mass to be 2.3 amu for planets larger than 1.6 $R_{\oplus}$, which we model as sub-Neptunes, and 20 amu for the smaller planets, which we model as terrestrial.  We assume the atmospheric temperature to be the equilibrium temperature for a Bond albedo of 0. For the planets with missing masses, we estimated $g$ using the statistical model from \citet{Chen:2017}. 

TOI-1266\,b and c’s projected signals in transmission are $\sim$220 and $\sim$100 ppm, which are significantly above \textit{JWST}’s plausible noise floor level \citep[20--50 ppm;][]{Greene:2016}. In order to quantify this further, we also report in Figure~\ref{fig:atmsnr} a relative  S/N scaling the signal amplitude with the hosts' brightness in the J band and using TRAPPIST-1~b’s  S/N as a reference. We find that TOI-1266’s planets compare favourably to TRAPPIST-1b in terms of potential for atmospheric exploration. In fact, assessing the presence of a clear hydrogen-dominated atmosphere as assumed for the purpose of this discussion would require 1 and 4 transits for planet b and c, respectively, with JWST NIRSpec Prism mode.

However, this preliminary assessment could be complicated by at least two factors.
A first obstacle of transmission spectroscopy is refraction, which bends starlight away from the line of the sight to the observer. In a transmission spectrum, this has a similar effect to an opaque cloud with a `cloud-top pressure' \citep{Sidis2010}, which introduces a spectral continuum that mutes the strength of spectral features. To determine if this effect is significant, we use equation (14) and Table 1 of \citet{Robinson2017} to estimate the pressure corresponding to this refraction continuum.  For nitrogen-dominated atmospheres, we estimate this pressure to be 1.3 and 0.8 bar for TOI-1266\,b and TOI-1266\,c, respectively, if we assume $g \sim 10^3$ cm s$^{-2}$ and $T \sim T_{\rm eq}$.  For carbon dioxide-dominated atmospheres, we estimate this pressure to be 0.7 and 0.4 bar for TOI-1266\,b and TOI-1266\,c, respectively.  Since transmission spectroscopy probes pressures $\sim 1$--10 mbar or lower, we therefore expect refraction to have a negligible effect on the shape of the transmission spectra for TOI-1266\,b and TOI-1266\,c.  Given the low equilibrium temperatures, a second obstacle may be the presence of clouds or hazes that are likely to strongly shape the transmission spectra of these weakly-irradiated exoplanets \citep[see e.g.][and references therein]{Crossfield:2017}.

We finally note that the planets of TOI-1266 are attractive targets for transmission spectroscopy due to the small geometric size of and lack of activity of the host star. More active M dwarfs like TRAPPIST-1 pose significant challenges to transmission spectroscopy, since the stellar surface features emit different spectra from the mean photosphere, introducing a degenerate signal that one must disentangle to correctly identify the signal from the exoplanet atmosphere \citep{robin, Morris2018c, Morris2018d, Ducrot2018, Wakeford2019}. TOI-1266 has no photometric evidence of these confounding starspot signatures, and therefore, makes a clean case for transmission spectroscopy.

\subsection{Prospects for radial velocity follow-up and additional TESS data}

Radial velocities (RVs) are likely to become available for this target in the future. It is critical to obtain precise masses and thus bulk densities for planets with sub-Neptune radii to identify if they are evaporated giants or giant rocks, so that we may identify model atmospheres to apply to the transmission spectroscopy. This requirement for a reliable mass constraint will be one of the fundamental limitations in choosing sub-Neptune targets for observations with JWST  \citep{Batalha:2017}. For this system, we might have some certainty about whether or not these are evaporated giants, which will strengthen our interpretations of the transmission spectra.

In addition, we might we be able to measure the Rossiter-McLaughlin effect for this system, due to its brightness and lack of confounding stellar activity. The obliquity of the system would be a valuable addition to the recent observation by \citet{Hirano:2020} that the TRAPPIST-1 planets are well-aligned with the host star's spin. 

We finally note that as of writing, TOI-1266 will not be observed by TESS during Sectors 23-26 nor during Extended Mission 1.

\section{Conclusions}\label{sec:concl}

This study reports on the discovery and preliminary characterisation of the TOI-1266 system that hosts a super-Earth and a sub-Neptune around a M3 dwarf. Our analysis combines photometry obtained from space- and ground-based facilities with a careful treatment of instrumental systematics and correlated noise for each dataset. The resulting data enable us to compute preliminary mass measurements for both planets, investigate their tidal evolution and search for additional companions in the system.
Along with other recently discovered TESS exoplanets \citep[e.g.][]{Cloutier2020,Cloutier2020a}, TOI-1266 will likely become a key system to better understand the nature of the radius valley around early to mid-M dwarfs. First, its orbital architecture, influenced by the 2:1 mean-motion resonance, will facilitate the measurement of precise planetary masses with high-precision photometry (TTV) and Doppler spectroscopy (radial velocities). Second, the host brightness is such that the system will be observable in most JWST modes, hence providing a large wavelength coverage. Third, the outer planet TOI-1266\,c has an irradiation level that is similar to that of Venus and is also a favourable target for atmospheric characterisation. We also note that TOI-1266\,c may be tidally-locked but with a non-circular orbit, resulting in incident stellar flux varying at a few percent level only along its orbit. Such a configuration would also lead to more homogeneous longitudinal temperature differences, in stark contrast with the bulk of small transiting exoplanets discovered so far.

\begin{acknowledgements}
We warmly thank the entire technical staff of the Observatorio Astron\'omico Nacional at San Pedro M\'artir in M\'exico for their unfailing support to SAINT-EX operations, namely: 
U.~Cese\~na, A.~C\'ordova, B.~Garc\'ia, C.A.~Guerrero, F.~Guill\'en, J.A.~Hern\'andez, B.~Hern\'andez, E.~L\'opez, B.~Mart\'inez, G.~Melgoza, F.~Montalvo, S.~Monrroy, J.C.~Narvaez, J.M.~Nu\~nez, J.L.~Ochoa, I.~Plauch\'u, F.~Quiroz, H.~Serrano, T.~Verdugo. We gratefully acknowledge the support from the Embassy of Mexico in Bern to the SAINT-EX project. We also thank the past members of the SAINT-EX team B.~Courcol, E.~Rose and K.~Housen for their help in the course of the project. We are grateful to the anonymous referee for a thorough and helpful review of our paper. We thank Jonathan Irwin for his help on the TRES data analysis.
B.-O.D. acknowledges support from the Swiss National Science Foundation (PP00P2-163967). This work has been carried out within the frame of the National Centre for Competence in Research PlanetS supported by the Swiss National Science Foundation. 
Y.G.M.C acknowledges support from UNAM-PAPIIT IN-107518. R.P. and E.J. acknowledge DGAPA for their postdoctoral fellowships.
The research leading to these results has received funding from the ARC grant for Concerted Research Actions, financed by the Wallonia-Brussels Federation. TRAPPIST is funded by the Belgian Fund for Scientific Research (Fonds National de la Recherche Scientifique, FNRS) under the grant FRFC 2.5.594.09.F, with the participation of the Swiss National Science Fundation (SNF). TRAPPIST-North is a project funded by the University of Liege, and performed in collaboration with Cadi Ayyad University of Marrakesh. MG and EJ are F.R.S.-FNRS Senior Research Associates. J.C.S. acknowledges funding support from Spanish public funds for research under projects ESP2017-87676-2-2 and RYC-2012-09913 (Ram\'on y Cajal programme) of the Spanish Ministry of Science and Education.
This paper includes data collected by the TESS mission. We acknowledge the use of public TOI Release data from pipelines at the TESS Science Office and at the TESS Science Processing Operations Center. Funding for the TESS mission is provided by the NASA Explorer Program. Resources supporting this work were provided by the NASA High-End Computing (HEC) Program through the NASA Advanced Supercomputing (NAS) Division at Ames Research Center for the production of the SPOC data products.
This work is based upon observations carried out at the Observatorio Astron\'omico Nacional on the Sierra de San Pedro M\'artir (OAN-SPM), Baja California, M\'exico. 
This research has been partly funded by the Spanish State Research Agency (AEI) Projects No.ESP2017-87676-C5-1-R and No. MDM-2017-0737 Unidad de Excelencia "Mar\'ia de Maeztu"- Centro de Astrobiolog\'ia (INTA-CSIC). 
This paper is based on observations collected at Centro Astron\'omico Hispano en Andaluc\'ia (CAHA) at Calar Alto, operated jointly by Instituto de Astrof\'isica de Andaluc\'ia (CSIC) and Junta de Andaluc\'ia.
The research leading to these results has received funding from the European Research Council under the European Union's Seventh Framework Programme (FP/2007-2013) ERC Grant Agreement n$^o$ 336480 (SPECULOOS). D.H. acknowledges support from the Alfred P. Sloan Foundation, the National Aeronautics and Space Administration (80NSSC18K1585, 80NSSC19K0379), and the National Science Foundation (AST-1717000).
We finally acknowledge support from the \textit{Antoine de Saint-Exup\'ery Youth Foundation} through their involvement in the educational programme of the SAINT-EX Observatory.

\end{acknowledgements}

\bibliographystyle{aa}
\bibliography{toi1266.bib}

\begin{thebibliography}{185}
\expandafter\ifx\csname natexlab\endcsname\relax\def\natexlab#1{#1}\fi

\bibitem[{{Agol} \& {Fabrycky}(2017)}]{Agol:2017}
{Agol}, E. \& {Fabrycky}, D.~C. 2017, {Transit-Timing and Duration Variations
  for the Discovery and Characterization of Exoplanets}, 7

\bibitem[{Agol {et~al.}(2005)Agol, Steffen, Sari, \& Clarkson}]{Agol:2005}
Agol, E., Steffen, J., Sari, R., \& Clarkson, W. 2005, Monthly Notices of the
  Royal Astronomical Society, 359, 567

\bibitem[{{Aigrain} {et~al.}(2016){Aigrain}, {Parviainen}, \&
  {Pope}}]{aigrain2016}
{Aigrain}, S., {Parviainen}, H., \& {Pope}, B.~J.~S. 2016, \mnras, 459, 2408

\bibitem[{{Alam} {et~al.}(2015){Alam}, {Albareti}, {Allende Prieto}, {Anders},
  {Anderson}, {Anderton}, {Andrews}, {Armengaud}, {Aubourg}, {Bailey}, \&
  et~al.}]{Alam:2015}
{Alam}, S., {Albareti}, F.~D., {Allende Prieto}, C., {et~al.} 2015, \apjs, 219,
  12

\bibitem[{{Aller} {et~al.}(2020){Aller}, {Lillo-Box}, {Jones}, {Miranda}, \&
  {Barcel{\'o} Forteza}}]{aller20}
{Aller}, A., {Lillo-Box}, J., {Jones}, D., {Miranda}, L.~F., \& {Barcel{\'o}
  Forteza}, S. 2020, \aap, 635, A128

\bibitem[{{Astropy Collaboration} {et~al.}(2013){Astropy Collaboration},
  {Robitaille}, {Tollerud}, {Greenfield}, {Droettboom}, {Bray}, {Aldcroft},
  {Davis}, {Ginsburg}, {Price-Whelan}, {Kerzendorf}, {Conley}, {Crighton},
  {Barbary}, {Muna}, {Ferguson}, {Grollier}, {Parikh}, {Nair}, {Unther},
  {Deil}, {Woillez}, {Conseil}, {Kramer}, {Turner}, {Singer}, {Fox}, {Weaver},
  {Zabalza}, {Edwards}, {Azalee Bostroem}, {Burke}, {Casey}, {Crawford},
  {Dencheva}, {Ely}, {Jenness}, {Labrie}, {Lim}, {Pierfederici}, {Pontzen},
  {Ptak}, {Refsdal}, {Servillat}, \& {Streicher}}]{astropy:2013}
{Astropy Collaboration}, {Robitaille}, T.~P., {Tollerud}, E.~J., {et~al.} 2013,
  \aap, 558, A33

\bibitem[{{Baraffe} {et~al.}(2015){Baraffe}, {Homeier}, {Allard}, \&
  {Chabrier}}]{baraffe2015}
{Baraffe}, I., {Homeier}, D., {Allard}, F., \& {Chabrier}, G. 2015, \aap, 577,
  A42

\bibitem[{{Barkaoui} {et~al.}(2019){Barkaoui}, {Burdanov}, {Hellier}, {Gillon},
  {Smalley}, {Maxted}, {Lendl}, {Triaud}, {Anderson}, {McCormac}, {Jehin},
  {Almleaky}, {Armstrong}, {Benkhaldoun}, {Bouchy}, {Brown}, {Cameron},
  {Daassou}, {Delrez}, {Ducrot}, {Foxell}, {Murray}, {Nielsen}, {Pepe},
  {Pollacco}, {Pozuelos}, {Queloz}, {Segransan}, {Udry}, {Thompson}, \&
  {West}}]{barkaoui2019}
{Barkaoui}, K., {Burdanov}, A., {Hellier}, C., {et~al.} 2019, \aj, 157, 43

\bibitem[{Barnes(2017)}]{barnes2017}
Barnes, R. 2017, Celestial Mechanics and Dynamical Astronomy, 129, 509

\bibitem[{Batalha {et~al.}(2017)Batalha, Kempton, \& Mbarek}]{Batalha:2017}
Batalha, N.~E., Kempton, E. M.-R., \& Mbarek, R. 2017, The Astrophysical
  Journal, 836, L5

\bibitem[{{Berger} {et~al.}(2020){Berger}, {Huber}, {Gaidos}, {van Saders}, \&
  {Weiss}}]{Berger:2020}
{Berger}, T.~A., {Huber}, D., {Gaidos}, E., {van Saders}, J.~L., \& {Weiss},
  L.~M. 2020, AJ, in press (arXiv:2005.14671), arXiv:2005.14671

\bibitem[{Blanco-Cuaresma \& Bolmont(2017)}]{Blanco-Cuaresma2017StudyingRust}
Blanco-Cuaresma, S. \& Bolmont, E. 2017, EWASS Special Session 4 (2017):
  Star-planet interactions (EWASS-SS4-2017

\bibitem[{Bolmont {et~al.}(2015)Bolmont, Raymond, Leconte, Hersant, \&
  Correia}]{Bolmont2015Mercury-T:Kepler-62}
Bolmont, E., Raymond, S.~N., Leconte, J., Hersant, F., \& Correia, A. C.~M.
  2015, 116

\bibitem[{Bolmont {et~al.}(2014)Bolmont, Raymond, von Paris, Selsis, Hersant,
  Quintana, \& Barclay}]{Bolmont2014FormationSystem}
Bolmont, E., Raymond, S.~N., von Paris, P., {et~al.} 2014, The Astrophysical
  Journal, 793, 3

\bibitem[{{Bourrier} {et~al.}(2018){Bourrier}, {Lecavelier des Etangs},
  {Ehrenreich}, {Sanz-Forcada}, {Allart}, {Ballester}, {Buchhave}, {Cohen},
  {Deming}, {Evans}, {Garc{\'\i}a Mu{\~n}oz}, {Henry}, {Kataria}, {Lavvas},
  {Lewis}, {L{\'o}pez-Morales}, {Marley}, {Sing}, \& {Wakeford}}]{Bourrier2018}
{Bourrier}, V., {Lecavelier des Etangs}, A., {Ehrenreich}, D., {et~al.} 2018,
  \aap, 620, A147

\bibitem[{Bradley {et~al.}(2019)Bradley, Sip{\H o}cz, Robitaille, Tollerud,
  Vin{\'{\i}}cius, Deil, Barbary, G{\"u}nther, Cara, Busko, Conseil,
  Droettboom, Bostroem, Bray, Bratholm, Wilson, Craig, Barentsen, Pascual,
  Donath, Greco, Perren, Lim, \& Kerzendorf}]{Bradley_2019_2533376}
Bradley, L., Sip{\H o}cz, B., Robitaille, T., {et~al.} 2019, astropy/photutils:
  v0.6

\bibitem[{{Brown} {et~al.}(2018){Brown}, {Vallenari}, {Prusti}, {de Bruijne},
  {Babusiaux}, {Bailer-Jones}, {Biermann}, {Evans}, {Eyer}, \&
  et~al.}]{Brown:2018}
{Brown}, A.~G.~A., {Vallenari}, A., {Prusti}, T., {et~al.} 2018, \aap, 616, A1

\bibitem[{{Buchhave} {et~al.}(2010){Buchhave}, {Bakos}, {Hartman}, {Torres},
  {Kov{\'a}cs}, {Latham}, {Noyes}, {Esquerdo}, {Everett}, {Howard}, {Marcy},
  {Fischer}, {Johnson}, {Andersen}, {F{\H{u}}r{\'e}sz}, {Perumpilly},
  {Sasselov}, {Stefanik}, {B{\'e}ky}, {L{\'a}z{\'a}r}, {Papp}, \&
  {S{\'a}ri}}]{Buchhave:2010}
{Buchhave}, L.~A., {Bakos}, G.~{\'A}., {Hartman}, J.~D., {et~al.} 2010, \apj,
  720, 1118

\bibitem[{{Carter} {et~al.}(2012){Carter}, {Agol}, {Chaplin}, {Basu},
  {Bedding}, {Buchhave}, {Christensen-Dalsgaard}, {Deck}, {Elsworth},
  {Fabrycky}, {Ford}, {Fortney}, {Hale}, {Handberg}, {Hekker}, {Holman},
  {Huber}, {Karoff}, {Kawaler}, {Kjeldsen}, {Lissauer}, {Lopez}, {Lund},
  {Lundkvist}, {Metcalfe}, {Miglio}, {Rogers}, {Stello}, {Borucki}, {Bryson},
  {Christiansen}, {Cochran}, {Geary}, {Gilliland}, {Haas}, {Hall}, {Howard},
  {Jenkins}, {Klaus}, {Koch}, {Latham}, {MacQueen}, {Sasselov}, {Steffen},
  {Twicken}, \& {Winn}}]{Carter2012}
{Carter}, J.~A., {Agol}, E., {Chaplin}, W.~J., {et~al.} 2012, Science, 337, 556

\bibitem[{{Chambers}(1999)}]{chambers1999}
{Chambers}, J.~E. 1999, \mnras, 304, 793

\bibitem[{{Chen} \& {Kipping}(2017)}]{Chen:2017}
{Chen}, J. \& {Kipping}, D. 2017, \apj, 834, 17

\bibitem[{{Choi} {et~al.}(2016){Choi}, {Dotter}, {Conroy}, {Cantiello},
  {Paxton}, \& {Johnson}}]{Choi:2016}
{Choi}, J., {Dotter}, A., {Conroy}, C., {et~al.} 2016, \apj, 823, 102

\bibitem[{Cincotta \& Sim{\'{o}}(1999)}]{Cincotta1999ConditionalEntropy}
Cincotta, P. \& Sim{\'{o}}, C. 1999, Celestial Mechanics and Dynamical
  Astronomy, 73, 195

\bibitem[{Cincotta {et~al.}(2003)Cincotta, Giordano, \&
  Sim{\'{o}}}]{Cincotta2003PhaseOrbits}
Cincotta, P.~M., Giordano, C.~M., \& Sim{\'{o}}, C. 2003, Physica D: Nonlinear
  Phenomena, 182, 151

\bibitem[{Cincotta \& Sim{\'{o}}(2000)}]{Cincotta2000SimpleI}
Cincotta, P.~M. \& Sim{\'{o}}, C. 2000, Astronomy and Astrophysics Supplement
  Series, 147, 205

\bibitem[{{Cloutier} {et~al.}(2020){Cloutier}, {Eastman}, {Rodriguez},
  {Astudillo-Defru}, {Bonfils}, {Mortier}, {Watson}, {Stalport}, {Pinamonti},
  {Lienhard}, {Harutyunyan}, {Damasso}, {Latham}, {Collins}, {Massey}, {Irwin},
  {Winters}, {Charbonneau}, {Ziegler}, {Matthews}, {Crossfield}, {Kreidberg},
  {Quinn}, {Ricker}, {Vanderspek}, {Seager}, {Winn}, {Jenkins}, {Vezie},
  {Udry}, {Twicken}, {Tenenbaum}, {Sozzetti}, {S{\'e}gransan}, {Schlieder},
  {Sasselov}, {Santos}, {Rice}, {Rackham}, {Poretti}, {Piotto}, {Phillips},
  {Pepe}, {Molinari}, {Mignon}, {Micela}, {Melo}, {de Medeiros}, {Mayor},
  {Matson}, {Martinez Fiorenzano}, {Mann}, {Magazz{\'u}}, {Lovis},
  {L{\'o}pez-Morales}, {Lopez}, {Lissauer}, {L{\'e}pine}, {Law}, {Kielkopf},
  {Johnson}, {Jensen}, {Howell}, {Gonzales}, {Ghedina}, {Forveille},
  {Figueira}, {Dumusque}, {Dressing}, {Doyon}, {D{\'\i}az}, {Di Fabrizio},
  {Delfosse}, {Cosentino}, {Conti}, {Collins}, {Collier Cameron}, {Ciardi},
  {Caldwell}, {Burke}, {Buchhave}, {Brice{\~n}o}, {Boyd}, {Bouchy}, {Beichman},
  {Artigau}, \& {Almenara}}]{Cloutier2020}
{Cloutier}, R., {Eastman}, J.~D., {Rodriguez}, J.~E., {et~al.} 2020, arXiv
  e-prints, arXiv:2003.01136

\bibitem[{{Cloutier} \& {Menou}(2020)}]{Cloutier2020a}
{Cloutier}, R. \& {Menou}, K. 2020, \aj, 159, 211

\bibitem[{{Collins} {et~al.}(2017){Collins}, {Kielkopf}, {Stassun}, \&
  {Hessman}}]{Collins:2017}
{Collins}, K.~A., {Kielkopf}, J.~F., {Stassun}, K.~G., \& {Hessman}, F.~V.
  2017, \aj, 153, 77

\bibitem[{{Crossfield} \& {Kreidberg}(2017)}]{Crossfield:2017}
{Crossfield}, I. J.~M. \& {Kreidberg}, L. 2017, \aj, 154, 261

\bibitem[{{Cutri} \& {et al.}(2013)}]{Cutri:2013}
{Cutri}, R.~M. \& {et al.} 2013, VizieR Online Data Catalog, II/328

\bibitem[{{Cutri} {et~al.}(2003){Cutri}, {Skrutskie}, {van Dyk}, {Beichman},
  {Carpenter}, {Chester}, {Cambresy}, {Evans}, {Fowler}, {Gizis}, {Howard},
  {Huchra}, {Jarrett}, {Kopan}, {Kirkpatrick}, {Light}, {Marsh}, {McCallon},
  {Schneider}, {Stiening}, {Sykes}, {Weinberg}, {Wheaton}, {Wheelock}, \&
  {Zacarias}}]{Cutri:2003}
{Cutri}, R.~M., {Skrutskie}, M.~F., {van Dyk}, S., {et~al.} 2003, {2MASS All
  Sky Catalog of point sources.}

\bibitem[{{Delrez} {et~al.}(2018){Delrez}, {Gillon}, {Queloz}, {Demory},
  {Almleaky}, {de Wit}, {Jehin}, {Triaud}, {Barkaoui}, {Burdanov}, {Burgasser},
  {Ducrot}, {McCormac}, {Murray}, {Silva Fernandes}, {Sohy}, {Thompson}, {Van
  Grootel}, {Alonso}, {Benkhaldoun}, \& {Rebolo}}]{Delrez:2018b}
{Delrez}, L., {Gillon}, M., {Queloz}, D., {et~al.} 2018, in Society of
  Photo-Optical Instrumentation Engineers (SPIE) Conference Series, Vol. 10700,
  Society of Photo-Optical Instrumentation Engineers (SPIE) Conference Series,
  107001I

\bibitem[{{Demory} {et~al.}(2012){Demory}, {Gillon}, {Seager}, {Benneke},
  {Deming}, \& {Jackson}}]{Demory:2012}
{Demory}, B.-O., {Gillon}, M., {Seager}, S., {et~al.} 2012, \apjl, 751, L28

\bibitem[{{Dorn} {et~al.}(2017){Dorn}, {Venturini}, {Khan}, {Heng}, {Alibert},
  {Helled}, {Rivoldini}, \& {Benz}}]{Dorn2017}
{Dorn}, C., {Venturini}, J., {Khan}, A., {et~al.} 2017, \aap, 597, A37

\bibitem[{{Dotter}(2016)}]{Dotter:2016}
{Dotter}, A. 2016, \apjs, 222, 8

\bibitem[{{Ducrot} {et~al.}(2018){Ducrot}, {Sestovic}, {Morris}, {Gillon},
  {Triaud}, {De Wit}, {Thimmarayappa}, {Agol}, {Almleaky}, {Burdanov},
  {Burgasser}, {Delrez}, {Demory}, {Jehin}, {Leconte}, {McCormac}, {Murray},
  {Queloz}, {Selsis}, {Thompson}, \& {Van Grootel}}]{Ducrot2018}
{Ducrot}, E., {Sestovic}, M., {Morris}, B.~M., {et~al.} 2018, \aj, 156, 218

\bibitem[{{Dumusque} {et~al.}(2019){Dumusque}, {Turner}, {Dorn}, {Eastman},
  {Allart}, {Adibekyan}, {Sousa}, {Santos}, {Mordasini}, {Bourrier}, {Bouchy},
  {Coffinet}, {Davies}, {D{\'\i}az}, {Fausnaugh}, {Glidden}, {Guerrero},
  {Henze}, {Jenkins}, {Latham}, {Lovis}, {Mayor}, {Pepe}, {Quintana}, {Ricker},
  {Rowden}, {Segransan}, {Mascare{\~n}o}, {Seager}, {Twicken}, {Udry}, {Vand
  erspek}, \& {Winn}}]{Dumusque2019}
{Dumusque}, X., {Turner}, O., {Dorn}, C., {et~al.} 2019, \aap, 627, A43

\bibitem[{{Eastman} {et~al.}(2019){Eastman}, {Rodriguez}, {Agol}, {Stassun},
  {Beatty}, {Vanderburg}, {Gaudi}, {Collins}, \& {Luger}}]{Eastman:2019}
{Eastman}, J.~D., {Rodriguez}, J.~E., {Agol}, E., {et~al.} 2019, arXiv
  e-prints, arXiv:1907.09480

\bibitem[{Eggleton {et~al.}(1998)Eggleton, Kiseleva, \&
  Hut}]{Eggleton1998TheFriction}
Eggleton, P.~P., Kiseleva, L.~G., \& Hut, P. 1998, The Astrophysical Journal,
  499, 853

\bibitem[{Eisner {et~al.}(2020)Eisner, Barrag{\'{a}}n, Aigrain, Lintott,
  Miller, Zicher, Boyajian, Brice{\~{n}}o, Bryant, Christiansen, Feinstein,
  Flor-Torres, Fridlund, Gandolfi, Gilbert, Guerrero, Jenkins, Jones,
  Kristiansen, Vanderburg, Law, L{\'{o}}pez-S{\'{a}}nchez, Mann, Safron,
  Schwamb, Stassun, Osborn, Wang, Zic, Ziegler, Barnet, Bean, Bundy, Chetnik,
  Dawson, Garstone, Stenner, Huten, Larish, Melanson, Mitchell, Moore, Peltsch,
  Rogers, Schuster, Smith, Simister, Tanner, Terentev, \&
  Tsymbal}]{Eisner2019PlanetOrbit}
Eisner, N.~L., Barrag{\'{a}}n, O., Aigrain, S., {et~al.} 2020, Monthly Notices
  of the Royal Astronomical Society

\bibitem[{{Freedman} {et~al.}(2014){Freedman}, {Lustig-Yaeger}, {Fortney},
  {Lupu}, {Marley}, \& {Lodders}}]{Freedman2014}
{Freedman}, R.~S., {Lustig-Yaeger}, J., {Fortney}, J.~J., {et~al.} 2014, \apjs,
  214, 25

\bibitem[{{Fried}(1978)}]{fried1978}
{Fried}, D.~L. 1978, Journal of the Optical Society of America (1917-1983), 68,
  1651

\bibitem[{{Fulton} {et~al.}(2017){Fulton}, {Petigura}, {Howard}, {Isaacson},
  {Marcy}, {Cargile}, {Hebb}, {Weiss}, {Johnson}, {Morton}, {Sinukoff},
  {Crossfield}, \& {Hirsch}}]{Fulton:2017}
{Fulton}, B.~J., {Petigura}, E.~A., {Howard}, A.~W., {et~al.} 2017, \aj, 154,
  109

\bibitem[{Furesz(2008)}]{Furesz:2008}
Furesz, G. 2008, PhD thesis, University of Szeged, Hungary

\bibitem[{{Gaia Collaboration} {et~al.}(2018){Gaia Collaboration}, {Brown},
  {Vallenari}, {Prusti}, {de Bruijne}, {Babusiaux}, {Bailer-Jones}, {Biermann},
  {Evans}, {Eyer}, {Jansen}, {Jordi}, {Klioner}, {Lammers}, {Lindegren},
  {Luri}, {Mignard}, {Panem}, {Pourbaix}, {Randich}, {Sartoretti}, {Siddiqui},
  {Soubiran}, {van Leeuwen}, {Walton}, {Arenou}, {Bastian}, {Cropper},
  {Drimmel}, {Katz}, {Lattanzi}, {Bakker}, {Cacciari}, {Casta{\~n}eda},
  {Chaoul}, {Cheek}, {De Angeli}, {Fabricius}, {Guerra}, {Holl}, {Masana},
  {Messineo}, {Mowlavi}, {Nienartowicz}, {Panuzzo}, {Portell}, {Riello},
  {Seabroke}, {Tanga}, {Th{\'e}venin}, {Gracia-Abril}, {Comoretto},
  {Garcia-Reinaldos}, {Teyssier}, {Altmann}, {Andrae}, {Audard},
  {Bellas-Velidis}, {Benson}, {Berthier}, {Blomme}, {Burgess}, {Busso},
  {Carry}, {Cellino}, {Clementini}, {Clotet}, {Creevey}, {Davidson}, {De
  Ridder}, {Delchambre}, {Dell'Oro}, {Ducourant},
  {Fern{\'a}ndez-Hern{\'a}ndez}, {Fouesneau}, {Fr{\'e}mat}, {Galluccio},
  {Garc{\'\i}a-Torres}, {Gonz{\'a}lez-N{\'u}{\~n}ez}, {Gonz{\'a}lez-Vidal},
  {Gosset}, {Guy}, {Halbwachs}, {Hambly}, {Harrison}, {Hern{\'a}ndez},
  {Hestroffer}, {Hodgkin}, {Hutton}, {Jasniewicz}, {Jean-Antoine-Piccolo},
  {Jordan}, {Korn}, {Krone-Martins}, {Lanzafame}, {Lebzelter}, {L{\"o}ffler},
  {Manteiga}, {Marrese}, {Mart{\'\i}n-Fleitas}, {Moitinho}, {Mora}, {Muinonen},
  {Osinde}, {Pancino}, {Pauwels}, {Petit}, {Recio-Blanco}, {Richards},
  {Rimoldini}, {Robin}, {Sarro}, {Siopis}, {Smith}, {Sozzetti}, {S{\"u}veges},
  {Torra}, {van Reeven}, {Abbas}, {Abreu Aramburu}, {Accart}, {Aerts},
  {Altavilla}, {{\'A}lvarez}, {Alvarez}, {Alves}, {Anderson}, {Andrei},
  {Anglada Varela}, {Antiche}, {Antoja}, {Arcay}, {Astraatmadja}, {Bach},
  {Baker}, {Balaguer-N{\'u}{\~n}ez}, {Balm}, {Barache}, {Barata}, {Barbato},
  {Barblan}, {Barklem}, {Barrado}, {Barros}, {Barstow}, {Bartholom{\'e}
  Mu{\~n}oz}, {Bassilana}, {Becciani}, {Bellazzini}, {Berihuete}, {Bertone},
  {Bianchi}, {Bienaym{\'e}}, {Blanco-Cuaresma}, {Boch}, {Boeche}, {Bombrun},
  {Borrachero}, {Bossini}, {Bouquillon}, {Bourda}, {Bragaglia}, {Bramante},
  {Breddels}, {Bressan}, {Brouillet}, {Br{\"u}semeister}, {Brugaletta},
  {Bucciarelli}, {Burlacu}, {Busonero}, {Butkevich}, {Buzzi}, {Caffau},
  {Cancelliere}, {Cannizzaro}, {Cantat-Gaudin}, {Carballo}, {Carlucci},
  {Carrasco}, {Casamiquela}, {Castellani}, {Castro-Ginard}, {Charlot},
  {Chemin}, {Chiavassa}, {Cocozza}, {Costigan}, {Cowell}, {Crifo}, {Crosta},
  {Crowley}, {Cuypers}, {Dafonte}, {Damerdji}, {Dapergolas}, {David}, {David},
  {de Laverny}, {De Luise}, {De March}, {de Martino}, {de Souza}, {de Torres},
  {Debosscher}, {del Pozo}, {Delbo}, {Delgado}, {Delgado}, {Di Matteo},
  {Diakite}, {Diener}, {Distefano}, {Dolding}, {Drazinos}, {Dur{\'a}n},
  {Edvardsson}, {Enke}, {Eriksson}, {Esquej}, {Eynard Bontemps}, {Fabre},
  {Fabrizio}, {Faigler}, {Falc{\~a}o}, {Farr{\`a}s Casas}, {Federici},
  {Fedorets}, {Fernique}, {Figueras}, {Filippi}, {Findeisen}, {Fonti},
  {Fraile}, {Fraser}, {Fr{\'e}zouls}, {Gai}, {Galleti}, {Garabato},
  {Garc{\'\i}a-Sedano}, {Garofalo}, {Garralda}, {Gavel}, {Gavras}, {Gerssen},
  {Geyer}, {Giacobbe}, {Gilmore}, {Girona}, {Giuffrida}, {Glass}, {Gomes},
  {Granvik}, {Gueguen}, {Guerrier}, {Guiraud}, {Guti{\'e}rrez-S{\'a}nchez},
  {Haigron}, {Hatzidimitriou}, {Hauser}, {Haywood}, {Heiter}, {Helmi}, {Heu},
  {Hilger}, {Hobbs}, {Hofmann}, {Holland}, {Huckle}, {Hypki}, {Icardi},
  {Jan{\ss}en}, {Jevardat de Fombelle}, {Jonker}, {Juh{\'a}sz}, {Julbe},
  {Karampelas}, {Kewley}, {Klar}, {Kochoska}, {Kohley}, {Kolenberg},
  {Kontizas}, {Kontizas}, {Koposov}, {Kordopatis}, {Kostrzewa-Rutkowska},
  {Koubsky}, {Lambert}, {Lanza}, {Lasne}, {Lavigne}, {Le Fustec}, {Le
  Poncin-Lafitte}, {Lebreton}, {Leccia}, {Leclerc}, {Lecoeur-Taibi},
  {Lenhardt}, {Leroux}, {Liao}, {Licata}, {Lindstr{\o}m}, {Lister}, {Livanou},
  {Lobel}, {L{\'o}pez}, {Managau}, {Mann}, {Mantelet}, {Marchal}, {Marchant},
  {Marconi}, {Marinoni}, {Marschalk{\'o}}, {Marshall}, {Martino}, {Marton},
  {Mary}, {Massari}, {Matijevi{\v{c}}}, {Mazeh}, {McMillan}, {Messina},
  {Michalik}, {Millar}, {Molina}, {Molinaro}, {Moln{\'a}r}, {Montegriffo},
  {Mor}, {Morbidelli}, {Morel}, {Morris}, {Mulone}, {Muraveva}, {Musella},
  {Nelemans}, {Nicastro}, {Noval}, {O'Mullane}, {Ord{\'e}novic},
  {Ord{\'o}{\~n}ez-Blanco}, {Osborne}, {Pagani}, {Pagano}, {Pailler},
  {Palacin}, {Palaversa}, {Panahi}, {Pawlak}, {Piersimoni}, {Pineau}, {Plachy},
  {Plum}, {Poggio}, {Poujoulet}, {Pr{\v{s}}a}, {Pulone}, {Racero}, {Ragaini},
  {Rambaux}, {Ramos-Lerate}, {Regibo}, {Reyl{\'e}}, {Riclet}, {Ripepi}, {Riva},
  {Rivard}, {Rixon}, {Roegiers}, {Roelens}, {Romero-G{\'o}mez}, {Rowell},
  {Royer}, {Ruiz-Dern}, {Sadowski}, {Sagrist{\`a} Sell{\'e}s}, {Sahlmann},
  {Salgado}, {Salguero}, {Sanna}, {Santana-Ros}, {Sarasso}, {Savietto},
  {Schultheis}, {Sciacca}, {Segol}, {Segovia}, {S{\'e}gransan}, {Shih},
  {Siltala}, {Silva}, {Smart}, {Smith}, {Solano}, {Solitro}, {Sordo}, {Soria
  Nieto}, {Souchay}, {Spagna}, {Spoto}, {Stampa}, {Steele},
  {Steidelm{\"u}ller}, {Stephenson}, {Stoev}, {Suess}, {Surdej}, {Szabados},
  {Szegedi-Elek}, {Tapiador}, {Taris}, {Tauran}, {Taylor}, {Teixeira},
  {Terrett}, {Teyssand ier}, {Thuillot}, {Titarenko}, {Torra Clotet}, {Turon},
  {Ulla}, {Utrilla}, {Uzzi}, {Vaillant}, {Valentini}, {Valette}, {van Elteren},
  {Van Hemelryck}, {van Leeuwen}, {Vaschetto}, {Vecchiato}, {Veljanoski},
  {Viala}, {Vicente}, {Vogt}, {von Essen}, {Voss}, {Votruba}, {Voutsinas},
  {Walmsley}, {Weiler}, {Wertz}, {Wevers}, {Wyrzykowski}, {Yoldas},
  {{\v{Z}}erjal}, {Ziaeepour}, {Zorec}, {Zschocke}, {Zucker}, {Zurbach}, \&
  {Zwitter}}]{gaia18}
{Gaia Collaboration}, {Brown}, A.~G.~A., {Vallenari}, A., {et~al.} 2018, \aap,
  616, A1

\bibitem[{Gavel {et~al.}(2014)Gavel, Kupke, Dillon, Norton, Ratliff, Cabak,
  Phillips, Rockosi, McGurk, Srinath, {et~al.}}]{gavel2014shaneao}
Gavel, D., Kupke, R., Dillon, D., {et~al.} 2014, in Adaptive Optics Systems IV,
  Vol. 9148, International Society for Optics and Photonics, 914805

\bibitem[{Gelman \& Rubin(1992)}]{Gelman:1992}
Gelman \& Rubin. 1992, Statistical Science, 7, 457

\bibitem[{Giacalone \& Dressing(2020{\natexlab{a}})}]{giacalone2020}
Giacalone, S. \& Dressing, C.~D. 2020{\natexlab{a}}, arXiv preprint
  arXiv:2002.00691

\bibitem[{Giacalone \& Dressing(2020{\natexlab{b}})}]{2020ascl.soft02004G}
Giacalone, S. \& Dressing, C.~D. 2020{\natexlab{b}}, {triceratops: Candidate
  exoplanet rating tool}

\bibitem[{{Gillon} {et~al.}(2013){Gillon}, {Anderson}, {Collier-Cameron},
  {Doyle}, {Fumel}, {Hellier}, {Jehin}, {Lendl}, {Maxted}, {Montalb{\'a}n},
  {Pepe}, {Pollacco}, {Queloz}, {S{\'e}gransan}, {Smith}, {Smalley},
  {Southworth}, {Triaud}, {Udry}, \& {West}}]{gillon2013}
{Gillon}, M., {Anderson}, D.~R., {Collier-Cameron}, A., {et~al.} 2013, \aap,
  552, A82

\bibitem[{{Gillon} {et~al.}(2014){Gillon}, {Demory}, {Madhusudhan}, {Deming},
  {Seager}, {Zsom}, {Knutson}, {Lanotte}, {Bonfils}, {D{\'e}sert}, {Delrez},
  {Jehin}, {Fraine}, {Magain}, \& {Triaud}}]{Gillon:2014a}
{Gillon}, M., {Demory}, B.-O., {Madhusudhan}, N., {et~al.} 2014, \aap, 563, A21

\bibitem[{{Gillon} {et~al.}(2016){Gillon}, {Jehin}, {Lederer}, {Delrez}, {de
  Wit}, {Burdanov}, {Van Grootel}, {Burgasser}, {Triaud}, {Opitom}, {Demory},
  {Sahu}, {Bardalez Gagliuffi}, {Magain}, \& {Queloz}}]{Gillon:2016}
{Gillon}, M., {Jehin}, E., {Lederer}, S.~M., {et~al.} 2016, \nat, 533, 221

\bibitem[{{Gillon} {et~al.}(2012){Gillon}, {Triaud}, {Fortney}, {Demory},
  {Jehin}, {Lendl}, {Magain}, {Kabath}, {Queloz}, {Alonso}, {Anderson},
  {Collier Cameron}, {Fumel}, {Hebb}, {Hellier}, {Lanotte}, {Maxted},
  {Mowlavi}, \& {Smalley}}]{Gillon:2012a}
{Gillon}, M., {Triaud}, A.~H.~M.~J., {Fortney}, J.~J., {et~al.} 2012, \aap,
  542, A4

\bibitem[{{Ginzburg} {et~al.}(2018){Ginzburg}, {Schlichting}, \&
  {Sari}}]{Ginzburg:2018}
{Ginzburg}, S., {Schlichting}, H.~E., \& {Sari}, R. 2018, \mnras, 476, 759

\bibitem[{{Greene} {et~al.}(2016){Greene}, {Line}, {Montero}, {Fortney},
  {Lustig-Yaeger}, \& {Luther}}]{Greene:2016}
{Greene}, T.~P., {Line}, M.~R., {Montero}, C., {et~al.} 2016, \apj, 817, 17

\bibitem[{{Grimm} {et~al.}(2018){Grimm}, {Demory}, {Gillon}, {Dorn}, {Agol},
  {Burdanov}, {Delrez}, {Sestovic}, {Triaud}, {Turbet}, {Bolmont}, {Caldas},
  {de Wit}, {Jehin}, {Leconte}, {Raymond}, {Van Grootel}, {Burgasser}, {Carey},
  {Fabrycky}, {Heng}, {Hernandez}, {Ingalls}, {Lederer}, {Selsis}, \&
  {Queloz}}]{Grimm:2018}
{Grimm}, S.~L., {Demory}, B.-O., {Gillon}, M., {et~al.} 2018, \aap, 613, A68

\bibitem[{{Grimm} \& {Stadel}(2014)}]{Grimm:2014}
{Grimm}, S.~L. \& {Stadel}, J.~G. 2014, \apj, 796, 23

\bibitem[{{G{\"u}nther} {et~al.}(2019){G{\"u}nther}, {Pozuelos}, {Dittmann},
  {Dragomir}, {Kane}, {Daylan}, {Feinstein}, {Huang}, {Morton}, {Bonfanti},
  {Bouma}, {Burt}, {Collins}, {Lissauer}, {Matthews}, {Montet}, {Vand erburg},
  {Wang}, {Winters}, {Ricker}, {Vanderspek}, {Latham}, {Seager}, {Winn},
  {Jenkins}, {Armstrong}, {Barkaoui}, {Batalha}, {Bean}, {Caldwell}, {Ciardi},
  {Collins}, {Crossfield}, {Fausnaugh}, {Furesz}, {Gan}, {Gillon}, {Guerrero},
  {Horne}, {Howell}, {Ireland }, {Isopi}, {Jehin}, {Kielkopf}, {Lepine},
  {Mallia}, {Matson}, {Myers}, {Palle}, {Quinn}, {Relles}, {Rojas-Ayala},
  {Schlieder}, {Sefako}, {Shporer}, {Su{\'a}rez}, {Tan}, {Ting}, {Twicken}, \&
  {Waite}}]{Gunther2019}
{G{\"u}nther}, M.~N., {Pozuelos}, F.~J., {Dittmann}, J.~A., {et~al.} 2019,
  Nature Astronomy, 3, 1099

\bibitem[{{Gupta} \& {Schlichting}(2020)}]{Gupta:2020}
{Gupta}, A. \& {Schlichting}, H.~E. 2020, \mnras, 493, 792

\bibitem[{Hakim {et~al.}(2018)Hakim, Rivoldini, Van~Hoolst, Cottenier, Jaeken,
  Chust, \& Steinle-Neumann}]{hakim_2018}
Hakim, K., Rivoldini, A., Van~Hoolst, T., {et~al.} 2018, Icarus, 313, 61

\bibitem[{Hauschildt {et~al.}(1999)Hauschildt, Allard, \&
  Baron}]{Hauschildt:1999}
Hauschildt, P.~H., Allard, F., \& Baron, E. 1999, \apj, 512, 377

\bibitem[{{Hawley} {et~al.}(2014){Hawley}, {Davenport}, {Kowalski},
  {Wisniewski}, {Hebb}, {Deitrick}, \& {Hilton}}]{Hawley:2014}
{Hawley}, S.~L., {Davenport}, J. R.~A., {Kowalski}, A.~F., {et~al.} 2014, \apj,
  797, 121

\bibitem[{Heller {et~al.}(2019)Heller, Hippke, \&
  Rodenbeck}]{Heller2019TransitK2}
Heller, R., Hippke, M., \& Rodenbeck, K. 2019, Astronomy {\&} Astrophysics,
  627, A66

\bibitem[{Hinse {et~al.}(2015)Hinse, Haghighipour, Kostov, \&
  G{\'{o}}zdziewski}]{Hinse2015PredictingSystem}
Hinse, T.~C., Haghighipour, N., Kostov, V.~B., \& G{\'{o}}zdziewski, K. 2015,
  Astrophysical Journal, 799

\bibitem[{Hippke {et~al.}(2019)Hippke, David, Mulders, \&
  Heller}]{Hippke2019Python}
Hippke, M., David, T.~J., Mulders, G.~D., \& Heller, R. 2019, The Astronomical
  Journal, 158, 143

\bibitem[{Hippke \& Heller(2019)}]{Hippke2019TransitPlanets}
Hippke, M. \& Heller, R. 2019, Astronomy and Astrophysics, 623, A39

\bibitem[{{Hirano} {et~al.}(2020){Hirano}, {Gaidos}, {Winn}, {Dai}, {Fukui},
  {Kuzuhara}, {Kotani}, {Tamura}, {Hjorth}, {Albrecht}, {Huber}, {Bolmont},
  {Harakawa}, {Hodapp}, {Ishizuka}, {Jacobson}, {Konishi}, {Kudo}, {Kurokawa},
  {Nishikawa}, {Omiya}, {Serizawa}, {Ueda}, \& {Weiss}}]{Hirano:2020}
{Hirano}, T., {Gaidos}, E., {Winn}, J.~N., {et~al.} 2020, \apjl, 890, L27

\bibitem[{Holman(2005)}]{Holman:2005}
Holman, M.~J. 2005, Science, 307, 1288

\bibitem[{{Holman} \& {Wiegert}(1999)}]{Holman1999}
{Holman}, M.~J. \& {Wiegert}, P.~A. 1999, \aj, 117, 621

\bibitem[{{Hormuth} {et~al.}(2008){Hormuth}, {Brandner}, {Hippler}, \&
  {Henning}}]{hormuth08}
{Hormuth}, F., {Brandner}, W., {Hippler}, S., \& {Henning}, T. 2008, Journal of
  Physics Conference Series, 131, 012051

\bibitem[{Horner {et~al.}(2019)Horner, Wittenmyer, Wright, Hinse, Marshall,
  Kane, Clark, Mengel, Agnew, \& Johns}]{Horner2019TheSolution}
Horner, J., Wittenmyer, R.~A., Wright, D.~J., {et~al.} 2019, The Astronomical
  Journal, 158, 100

\bibitem[{{Husser} {et~al.}(2013){Husser}, {Wende-von Berg}, {Dreizler},
  {Homeier}, {Reiners}, {Barman}, \& {Hauschildt}}]{Husser:2013}
{Husser}, T.~O., {Wende-von Berg}, S., {Dreizler}, S., {et~al.} 2013, \aap,
  553, A6

\bibitem[{Hut(1981)}]{Hut1981TidalSystems.}
Hut, P. 1981, Astronomy and Astrophysics, 99, 126

\bibitem[{Janson {et~al.}(2011)Janson, Bonavita, Klahr, Lafreniere,
  Jayawardhana, \& Zinnecker}]{janson2011high}
Janson, M., Bonavita, M., Klahr, H., {et~al.} 2011, The Astrophysical Journal,
  736, 89

\bibitem[{{Jehin} {et~al.}(2018){Jehin}, {Gillon}, {Queloz}, {Delrez},
  {Burdanov}, {Murray}, {Sohy}, {Ducrot}, {Sebastian}, {Thompson}, {McCormac},
  {Almleaky}, {Burgasser}, {Demory}, {de Wit}, {Barkaoui}, {Pozuelos},
  {Triaud}, \& {Grootel}}]{Jehin:2018}
{Jehin}, E., {Gillon}, M., {Queloz}, D., {et~al.} 2018, The Messenger, 174, 2

\bibitem[{{Jehin} {et~al.}(2011){Jehin}, {Gillon}, {Queloz}, {Magain},
  {Manfroid}, {Chantry}, {Lendl}, {Hutsem{\'e}kers}, \& {Udry}}]{jehin2011}
{Jehin}, E., {Gillon}, M., {Queloz}, D., {et~al.} 2011, The Messenger, 145, 2

\bibitem[{{Jenkins}(2002)}]{Jenkins2002}
{Jenkins}, J.~M. 2002, \apj, 575, 493

\bibitem[{{Jenkins} {et~al.}(2017){Jenkins}, {Tenenbaum}, {Seader}, {Burke},
  {McCauliff}, {Smith}, {Twicken}, \& {Chandrasekaran}}]{Jenkins:2017}
{Jenkins}, J.~M., {Tenenbaum}, P., {Seader}, S., {et~al.} 2017, {Kepler Data
  Processing Handbook: Transiting Planet Search}, Kepler Science Document

\bibitem[{{Jenkins} {et~al.}(2016){Jenkins}, {Twicken}, {McCauliff},
  {Campbell}, {Sanderfer}, {Lung}, {Mansouri-Samani}, {Girouard}, {Tenenbaum},
  {Klaus}, {Smith}, {Caldwell}, {Chacon}, {Henze}, {Heiges}, {Latham},
  {Morgan}, {Swade}, {Rinehart}, \& {Vanderspek}}]{Jenkins2016}
{Jenkins}, J.~M., {Twicken}, J.~D., {McCauliff}, S., {et~al.} 2016, Society of
  Photo-Optical Instrumentation Engineers (SPIE) Conference Series, Vol. 9913,
  {The TESS science processing operations center}, 99133E

\bibitem[{Jenkins {et~al.}(2009)Jenkins, Jones, Go{\'{z}}dziewski, Migaszewski,
  Barnes, Jones, Rojo, Pinfield, Day-Jones, \& Hoyer}]{Jenkins2009FirstDesert}
Jenkins, J.~S., Jones, H.~R., Go{\'{z}}dziewski, K., {et~al.} 2009, Monthly
  Notices of the Royal Astronomical Society, 398, 911

\bibitem[{{Jenkins} {et~al.}(2019){Jenkins}, {Pozuelos}, {Tuomi},
  {Berdi{\~n}as}, {D{\'\i}az}, {Vines}, {Su{\'a}rez}, \& {Pe{\~n}a
  Rojas}}]{Jenkins2019}
{Jenkins}, J.~S., {Pozuelos}, F.~J., {Tuomi}, M., {et~al.} 2019, \mnras, 490,
  5585

\bibitem[{Jenkins {et~al.}(2019)Jenkins, Pozuelos, Tuomi, Berdi{\~{n}}as,
  D{\'{i}}az, Vines, Su{\'{a}}rez, \& Pe{\~{n}}a~Rojas}]{james2019}
Jenkins, J.~S., Pozuelos, F.~J., Tuomi, M., {et~al.} 2019, Monthly Notices of
  the Royal Astronomical Society, 490, 5585

\bibitem[{{Jensen}(2013)}]{jensen2013}
{Jensen}, E. 2013, {Tapir: A web interface for transit/eclipse observability},
  Astrophysics Source Code Library

\bibitem[{{Jin} \& {Mordasini}(2018)}]{Jin2018}
{Jin}, S. \& {Mordasini}, C. 2018, \apj, 853, 163

\bibitem[{{Jin} {et~al.}(2014){Jin}, {Mordasini}, {Parmentier}, {van Boekel},
  {Henning}, \& {Ji}}]{Jin2014}
{Jin}, S., {Mordasini}, C., {Parmentier}, V., {et~al.} 2014, \apj, 795, 65

\bibitem[{{Jofr{\'e}} {et~al.}(2015){Jofr{\'e}}, {Petrucci}, {Saffe}, {Saker},
  {Artur de la Villarmois}, {Chavero}, {G{\'o}mez}, \& {Mauas}}]{Jofre2015}
{Jofr{\'e}}, E., {Petrucci}, R., {Saffe}, C., {et~al.} 2015, \aap, 574, A50

\bibitem[{{Kane}(2015)}]{kane2015b}
{Kane}, S.~R. 2015, \apjl, 814, L9

\bibitem[{{Kane}(2017)}]{kane2017}
{Kane}, S.~R. 2017, \apjl, 839, L19

\bibitem[{{Kane}(2019)}]{kane2019c}
{Kane}, S.~R. 2019, \aj, 158, 72

\bibitem[{{Kane} \& {Torres}(2017)}]{kane2017b}
{Kane}, S.~R. \& {Torres}, S.~M. 2017, \aj, 154, 204

\bibitem[{{Kiseleva} {et~al.}(1998){Kiseleva}, {Eggleton}, \&
  {Mikkola}}]{Kiseleva1998}
{Kiseleva}, L.~G., {Eggleton}, P.~P., \& {Mikkola}, S. 1998, \mnras, 300, 292

\bibitem[{Kolbl {et~al.}(2014)Kolbl, Marcy, Isaacson, \&
  Howard}]{kolbl2014detection}
Kolbl, R., Marcy, G.~W., Isaacson, H., \& Howard, A.~W. 2014, The Astronomical
  Journal, 149, 18

\bibitem[{{Kostov} {et~al.}(2019){Kostov}, {Schlieder}, {Barclay}, {Quintana},
  {Col{\'o}n}, {Brand e}, {Collins}, {Feinstein}, {Hadden}, {Kane},
  {Kreidberg}, {Kruse}, {Lam}, {Matthews}, {Montet}, {Pozuelos}, {Stassun},
  {Winters}, {Ricker}, {Vanderspek}, {Latham}, {Seager}, {Winn}, {Jenkins},
  {Afanasev}, {Armstrong}, {Arney}, {Boyd}, {Barentsen}, {Barkaoui}, {Batalha},
  {Beichman}, {Bayliss}, {Burke}, {Burdanov}, {Cacciapuoti}, {Carson},
  {Charbonneau}, {Christiansen}, {Ciardi}, {Clampin}, {Collins}, {Conti},
  {Coughlin}, {Covone}, {Crossfield}, {Delrez}, {Domagal-Goldman}, {Dressing},
  {Ducrot}, {Essack}, {Everett}, {Fauchez}, {Foreman-Mackey}, {Gan}, {Gilbert},
  {Gillon}, {Gonzales}, {Hamann}, {Hedges}, {Hocutt}, {Hoffman}, {Horch},
  {Horne}, {Howell}, {Hynes}, {Ireland }, {Irwin}, {Isopi}, {Jensen}, {Jehin},
  {Kaltenegger}, {Kielkopf}, {Kopparapu}, {Lewis}, {Lopez}, {Lissauer}, {Mann},
  {Mallia}, {Mandell}, {Matson}, {Mazeh}, {Monsue}, {Moran}, {Moran}, {Morley},
  {Morris}, {Muirhead}, {Mukai}, {Mullally}, {Mullally}, {Murray}, {Narita},
  {Palle}, {Pidhorodetska}, {Quinn}, {Relles}, {Rinehart}, {Ritsko},
  {Rodriguez}, {Rowden}, {Rowe}, {Sebastian}, {Sefako}, {Shahaf}, {Shporer},
  {Ta{\~n}{\'o}n Reyes}, {Tenenbaum}, {Ting}, {Twicken}, {van Belle}, {Vega},
  {Volosin}, {Walkowicz}, \& {Youngblood}}]{Kostov2019}
{Kostov}, V.~B., {Schlieder}, J.~E., {Barclay}, T., {et~al.} 2019, \aj, 158, 32

\bibitem[{{Kozai}(1962)}]{Kozai1962}
{Kozai}, Y. 1962, \aj, 67, 579

\bibitem[{{Kubyshkina} {et~al.}(2019){Kubyshkina}, {Cubillos}, {Fossati},
  {Erkaev}, {Johnstone}, {Kislyakova}, {Lammer}, {Lendl}, {Odert}, \&
  {G{\"u}del}}]{Kubyshkina2019}
{Kubyshkina}, D., {Cubillos}, P.~E., {Fossati}, L., {et~al.} 2019, \apj, 879,
  26

\bibitem[{Lang {et~al.}(2010)Lang, Hogg, Mierle, Blanton, \&
  Roweis}]{Lang_2010}
Lang, D., Hogg, D.~W., Mierle, K., Blanton, M., \& Roweis, S. 2010, The
  Astronomical Journal, 139, 1782–1800

\bibitem[{Laskar \& Robutel(1993)}]{Laskar1993ThePlanets}
Laskar, J. \& Robutel, P. 1993, Nature, 361, 608

\bibitem[{{Latham} {et~al.}(2011){Latham}, {Rowe}, {Quinn}, {Batalha},
  {Borucki}, {Brown}, {Bryson}, {Buchhave}, {Caldwell}, {Carter},
  {Christiansen}, {Ciardi}, {Cochran}, {Dunham}, {Fabrycky}, {Ford}, {Gautier},
  {Gilliland}, {Holman}, {Howell}, {Ibrahim}, {Isaacson}, {Jenkins}, {Koch},
  {Lissauer}, {Marcy}, {Quintana}, {Ragozzine}, {Sasselov}, {Shporer},
  {Steffen}, {Welsh}, \& {Wohler}}]{Latham:2011a}
{Latham}, D.~W., {Rowe}, J.~F., {Quinn}, S.~N., {et~al.} 2011, \apjl, 732, L24

\bibitem[{Leconte {et~al.}(2010)Leconte, Chabrier, Baraffe, \&
  Levrard}]{Leconte2010IsEccentricity}
Leconte, J., Chabrier, G., Baraffe, I., \& Levrard, B. 2010, Astronomy and
  Astrophysics, 516, A64

\bibitem[{{Li} {et~al.}(2019){Li}, {Tenenbaum}, {Twicken}, {Burke}, {Jenkins},
  {Quintana}, {Rowe}, \& {Seader}}]{Li:2019}
{Li}, J., {Tenenbaum}, P., {Twicken}, J.~D., {et~al.} 2019, \pasp, 131, 024506

\bibitem[{{Lidov}(1962)}]{Lidov1962}
{Lidov}, M.~L. 1962, \planss, 9, 719

\bibitem[{{Lightkurve Collaboration} {et~al.}(2018){Lightkurve Collaboration},
  {Cardoso}, {Hedges}, {Gully-Santiago}, {Saunders}, {Cody}, {Barclay}, {Hall},
  {Sagear}, {Turtelboom}, {Zhang}, {Tzanidakis}, {Mighell}, {Coughlin}, {Bell},
  {Berta-Thompson}, {Williams}, {Dotson}, \& {Barentsen}}]{lightkurve:2018}
{Lightkurve Collaboration}, {Cardoso}, J.~V.~d.~M., {Hedges}, C., {et~al.}
  2018, {Lightkurve: Kepler and TESS time series analysis in Python},
  Astrophysics Source Code Library

\bibitem[{Lightkurve~Collaboration {et~al.}(2018)Lightkurve~Collaboration,
  Cardoso, Hedges, Gully-Santiago, Saunders, Cody, Barclay, Hall, Sagear,
  Turtelboom, Zhang, Tzanidakis, Mighell, Coughlin, Bell, Berta-Thompson,
  Williams, Dotson, Barentsen, Collaboration, Cardoso, Hedges, Gully-Santiago,
  Saunders, Cody, Barclay, Hall, Sagear, Turtelboom, Zhang, Tzanidakis,
  Mighell, Coughlin, Bell, Berta-Thompson, Williams, Dotson, \&
  Barentsen}]{LightkurveCollaboration2018Lightkurve:Python}
Lightkurve~Collaboration, L., Cardoso, J. V. d.~M., Hedges, C., {et~al.} 2018,
  ascl, ascl:1812.013

\bibitem[{{Lillo-Box} {et~al.}(2012){Lillo-Box}, {Barrado}, \&
  {Bouy}}]{lillo-box12}
{Lillo-Box}, J., {Barrado}, D., \& {Bouy}, H. 2012, \aap, 546, A10

\bibitem[{{Lillo-Box} {et~al.}(2014){Lillo-Box}, {Barrado}, \&
  {Bouy}}]{lillo-box14b}
{Lillo-Box}, J., {Barrado}, D., \& {Bouy}, H. 2014, \aap, 566, A103

\bibitem[{Lissauer {et~al.}(2012)Lissauer, Barnes, \&
  Chambers}]{Lissauer2012ObliquityEarth}
Lissauer, J.~J., Barnes, J.~W., \& Chambers, J.~E. 2012, Icarus, 217, 77

\bibitem[{{Lissauer} \& {Cuzzi}(1985)}]{Lissauer:1985}
{Lissauer}, J.~J. \& {Cuzzi}, J.~N. 1985, in Protostars and Planets II, ed.
  D.~C. {Black} \& M.~S. {Matthews}, 920--956

\bibitem[{{Lissauer} {et~al.}(2012){Lissauer}, {Marcy}, {Rowe}, {Bryson},
  {Adams}, {Buchhave}, {Ciardi}, {Cochran}, {Fabrycky}, {Ford}, {Fressin},
  {Geary}, {Gilliland }, {Holman}, {Howell}, {Jenkins}, {Kinemuchi}, {Koch},
  {Morehead}, {Ragozzine}, {Seader}, {Tanenbaum}, {Torres}, \&
  {Twicken}}]{Lissauer2012}
{Lissauer}, J.~J., {Marcy}, G.~W., {Rowe}, J.~F., {et~al.} 2012, \apj, 750, 112

\bibitem[{Lissauer {et~al.}(2012)Lissauer, Marcy, Rowe, Bryson, Adams,
  Buchhave, Ciardi, Cochran, Fabrycky, Ford, {et~al.}}]{lissauer2012almost}
Lissauer, J.~J., Marcy, G.~W., Rowe, J.~F., {et~al.} 2012, The Astrophysical
  Journal, 750, 112

\bibitem[{{Lithwick} {et~al.}(2012){Lithwick}, {Xie}, \& {Wu}}]{Lithwick:2012}
{Lithwick}, Y., {Xie}, J., \& {Wu}, Y. 2012, \apj, 761, 122

\bibitem[{{Lopez} \& {Fortney}(2013{\natexlab{a}})}]{Lopez2013}
{Lopez}, E.~D. \& {Fortney}, J.~J. 2013{\natexlab{a}}, \apj, 776, 2

\bibitem[{{Lopez} \& {Fortney}(2013{\natexlab{b}})}]{Lopez:2013a}
{Lopez}, E.~D. \& {Fortney}, J.~J. 2013{\natexlab{b}}, \apj, 776, 2

\bibitem[{{Lopez} {et~al.}(2012){Lopez}, {Fortney}, \& {Miller}}]{Lopez:2012a}
{Lopez}, E.~D., {Fortney}, J.~J., \& {Miller}, N. 2012, \apj, 761, 59

\bibitem[{{Lundkvist} {et~al.}(2016){Lundkvist}, {Kjeldsen}, {Albrecht},
  {Davies}, {Basu}, {Huber}, {Justesen}, {Karoff}, {Silva Aguirre}, {van
  Eylen}, {Vang}, {Arentoft}, {Barclay}, {Bedding}, {Campante}, {Chaplin},
  {Christensen-Dalsgaard}, {Elsworth}, {Gilliland}, {Handberg}, {Hekker},
  {Kawaler}, {Lund}, {Metcalfe}, {Miglio}, {Rowe}, {Stello}, {Tingley}, \&
  {White}}]{Lundkvist2016}
{Lundkvist}, M.~S., {Kjeldsen}, H., {Albrecht}, S., {et~al.} 2016, Nature
  Communications, 7, 11201

\bibitem[{{Maldonado} {et~al.}(2015){Maldonado}, {Affer}, {Micela},
  {Scandariato}, {Damasso}, {Stelzer}, {Barbieri}, {Bedin}, {Biazzo},
  {Bignamini}, {Borsa}, {Claudi}, {Covino}, {Desidera}, {Esposito}, {Gratton},
  {Gonz{\'a}lez Hern{\'a}ndez}, {Lanza}, {Maggio}, {Molinari}, {Pagano},
  {Perger}, {Pillitteri}, {Piotto}, {Poretti}, {Prisinzano}, {Rebolo}, {Ribas},
  {Shkolnik}, {Southworth}, {Sozzetti}, \& {Su{\'a}rez
  Mascare{\~n}o}}]{Maldonado2015}
{Maldonado}, J., {Affer}, L., {Micela}, G., {et~al.} 2015, \aap, 577, A132

\bibitem[{{Mann} {et~al.}(2019){Mann}, {Dupuy}, {Kraus}, {Gaidos}, {Ansdell},
  {Ireland}, {Rizzuto}, {Hung}, {Dittmann}, {Factor}, {Feiden}, {Martinez},
  {Ru{\'\i}z-Rodr{\'\i}guez}, \& {Thao}}]{Mann:2019}
{Mann}, A.~W., {Dupuy}, T., {Kraus}, A.~L., {et~al.} 2019, \apj, 871, 63

\bibitem[{Marois {et~al.}(2006)Marois, Lafreniere, Doyon, Macintosh, \&
  Nadeau}]{marois2006angular}
Marois, C., Lafreniere, D., Doyon, R., Macintosh, B., \& Nadeau, D. 2006, The
  Astrophysical Journal, 641, 556

\bibitem[{{Mathis}(2018)}]{mathis2018}
{Mathis}, S. 2018, {Tidal Star-Planet Interactions: A Stellar and Planetary
  Perspective}, 24

\bibitem[{{Mazeh} {et~al.}(2016){Mazeh}, {Holczer}, \& {Faigler}}]{Mazeh2016}
{Mazeh}, T., {Holczer}, T., \& {Faigler}, S. 2016, \aap, 589, A75

\bibitem[{Mazevet {et~al.}(2019)Mazevet, Licari, Chabrier, \&
  Potekhin}]{mazevet_2019}
Mazevet, S., Licari, A., Chabrier, G., \& Potekhin, A.~Y. 2019, Astronomy \&
  Astrophysics, 621, A128

\bibitem[{{McCormac} {et~al.}(2013){McCormac}, {Pollacco}, {Skillen}, {Faedi},
  {Todd}, \& {Watson}}]{McCormac:2013}
{McCormac}, J., {Pollacco}, D., {Skillen}, I., {et~al.} 2013, \pasp, 125, 548

\bibitem[{{McDonald} {et~al.}(2019){McDonald}, {Kreidberg}, \&
  {Lopez}}]{McDonald2019}
{McDonald}, G.~D., {Kreidberg}, L., \& {Lopez}, E. 2019, \apj, 876, 22

\bibitem[{McGurk {et~al.}(2014)McGurk, Rockosi, Gavel, Kupke, Peck, Pfister,
  Ward, Deich, Gates, Gates, {et~al.}}]{mcgurk2014commissioning}
McGurk, R., Rockosi, C., Gavel, D., {et~al.} 2014, in Adaptive Optics Systems
  IV, Vol. 9148, International Society for Optics and Photonics, 91483A

\bibitem[{Michel {et~al.}(2020)Michel, Haldemann, Mordasini, \&
  Alibert}]{michel_2020}
Michel, A., Haldemann, J., Mordasini, C., \& Alibert, Y. 2020, Astronomy \&
  Astrophysics, 639, A66

\bibitem[{Mignard(1979)}]{Mignard1979TheI}
Mignard, F. 1979, The Moon and the Planets, 20, 301

\bibitem[{{Modirrousta-Galian} {et~al.}(2020){Modirrousta-Galian}, {Locci}, \&
  {Micela}}]{Modirrousta2020}
{Modirrousta-Galian}, D., {Locci}, D., \& {Micela}, G. 2020, \apj, 891, 158

\bibitem[{{Mordasini}(2020)}]{Mordasini2020}
{Mordasini}, C. 2020, arXiv e-prints, arXiv:2002.02455

\bibitem[{{Mordasini} {et~al.}(2012){Mordasini}, {Alibert}, {Klahr}, \&
  {Henning}}]{Mordasini2012}
{Mordasini}, C., {Alibert}, Y., {Klahr}, H., \& {Henning}, T. 2012, \aap, 547,
  A111

\bibitem[{{Morris} {et~al.}(2018{\natexlab{a}}){Morris}, {Agol}, {Davenport},
  \& {Hawley}}]{Morris2018d}
{Morris}, B.~M., {Agol}, E., {Davenport}, J. R.~A., \& {Hawley}, S.~L.
  2018{\natexlab{a}}, \apj, 857, 39

\bibitem[{{Morris} {et~al.}(2018{\natexlab{b}}){Morris}, {Agol}, {Hebb}, \&
  {Hawley}}]{robin}
{Morris}, B.~M., {Agol}, E., {Hebb}, L., \& {Hawley}, S.~L. 2018{\natexlab{b}},
  \aj, 156, 91

\bibitem[{{Morris} {et~al.}(2018{\natexlab{c}}){Morris}, {Agol}, {Hebb},
  {Hawley}, {Gillon}, {Ducrot}, {Delrez}, {Ingalls}, \& {Demory}}]{Morris2018c}
{Morris}, B.~M., {Agol}, E., {Hebb}, L., {et~al.} 2018{\natexlab{c}}, \apjl,
  863, L32

\bibitem[{{Morris} {et~al.}(2017){Morris}, {Twicken}, {Smith}, {Clarke},
  {Jenkins}, {Bryson}, {Girouard}, \& {Klaus}}]{Morris2017}
{Morris}, R.~L., {Twicken}, J.~D., {Smith}, J.~C., {et~al.} 2017, {Kepler Data
  Processing Handbook: Photometric Analysis}, Kepler Science Document

\bibitem[{Morton(2012)}]{morton2012}
Morton, T.~D. 2012, The Astrophysical Journal, 761, 6

\bibitem[{{Morton}(2015)}]{2015ascl.soft03011M}
{Morton}, T.~D. 2015, {VESPA: False positive probabilities calculator},
  Astrophysics Source Code Library

\bibitem[{{Morton} {et~al.}(2016){Morton}, {Bryson}, {Coughlin}, {Rowe},
  {Ravichandran}, {Petigura}, {Haas}, \& {Batalha}}]{Morton:2016}
{Morton}, T.~D., {Bryson}, S.~T., {Coughlin}, J.~L., {et~al.} 2016, \apj, 822,
  86

\bibitem[{Morzinski {et~al.}(2015)Morzinski, Males, Skemer, Close, Hinz,
  Rodigas, Puglisi, Esposito, Riccardi, Pinna,
  {et~al.}}]{morzinski2015magellan}
Morzinski, K.~M., Males, J.~R., Skemer, A.~J., {et~al.} 2015, The Astrophysical
  Journal, 815, 108

\bibitem[{{Murray-Clay} {et~al.}(2009){Murray-Clay}, {Chiang}, \&
  {Murray}}]{MurrayClay2009}
{Murray-Clay}, R.~A., {Chiang}, E.~I., \& {Murray}, N. 2009, \apj, 693, 23

\bibitem[{{Neron de Surgy} \& {Laskar}(1997)}]{neron1997}
{Neron de Surgy}, O. \& {Laskar}, J. 1997, \aap, 318, 975

\bibitem[{{Newton} {et~al.}(2016){Newton}, {Irwin}, {Charbonneau},
  {Berta-Thompson}, {Dittmann}, \& {West}}]{Newton:2016}
{Newton}, E.~R., {Irwin}, J., {Charbonneau}, D., {et~al.} 2016, \apj, 821, 93

\bibitem[{{Newton} {et~al.}(2018){Newton}, {Mondrik}, {Irwin}, {Winters}, \&
  {Charbonneau}}]{Newton:2018}
{Newton}, E.~R., {Mondrik}, N., {Irwin}, J., {Winters}, J.~G., \&
  {Charbonneau}, D. 2018, \aj, 156, 217

\bibitem[{Newville {et~al.}(2014)Newville, Stensitzki, Allen, \&
  Ingargiola}]{newville_matthew_2014_11813}
Newville, M., Stensitzki, T., Allen, D.~B., \& Ingargiola, A. 2014, {LMFIT:
  Non-Linear Least-Square Minimization and Curve-Fitting for Python}

\bibitem[{Nielsen {et~al.}(2008)Nielsen, Close, Biller, Masciadri, \&
  Lenzen}]{nielsen2008constraints}
Nielsen, E.~L., Close, L.~M., Biller, B.~A., Masciadri, E., \& Lenzen, R. 2008,
  The Astrophysical Journal, 674, 466

\bibitem[{{Owen} \& {Campos Estrada}(2020)}]{Owen:2019}
{Owen}, J.~E. \& {Campos Estrada}, B. 2020, \mnras, 491, 5287

\bibitem[{{Owen} \& {Jackson}(2012)}]{Owen2012}
{Owen}, J.~E. \& {Jackson}, A.~P. 2012, \mnras, 425, 2931

\bibitem[{{Owen} \& {Wu}(2013)}]{Owen2013}
{Owen}, J.~E. \& {Wu}, Y. 2013, \apj, 775, 105

\bibitem[{{Owen} \& {Wu}(2017)}]{Owen2017}
{Owen}, J.~E. \& {Wu}, Y. 2017, \apj, 847, 29

\bibitem[{{Parviainen} \& {Aigrain}(2015)}]{Parviainen:2015}
{Parviainen}, H. \& {Aigrain}, S. 2015, \mnras, 453, 3821

\bibitem[{Pedregosa {et~al.}(2011)Pedregosa, Varoquaux, Gramfort, Michel,
  Thirion, Grisel, Blondel, Prettenhofer, Weiss, Dubourg, Vanderplas, Passos,
  Cournapeau, Brucher, Perrot, \& Duchesnay}]{scikit-learn}
Pedregosa, F., Varoquaux, G., Gramfort, A., {et~al.} 2011, Journal of Machine
  Learning Research, 12, 2825

\bibitem[{Pont {et~al.}(2006)Pont, Zucker, \& Queloz}]{Pont:2006}
Pont, F., Zucker, S., \& Queloz, D. 2006, Monthly Notices of the Royal
  Astronomical Society, 373, 231

\bibitem[{{Pozuelos} {et~al.}(2020){Pozuelos}, {Su{\'a}rez}, {de El{\'\i}a},
  {Berdi{\~n}as}, {Bonfanti}, {Dugaro}, {Gillon}, {Jehin}, {G{\"u}nther}, {Van
  Grootel}, {Garcia}, {Thuillier}, {Delrez}, \& {Rod{\'o}n}}]{pozuelos2020}
{Pozuelos}, F.~J., {Su{\'a}rez}, J.~C., {de El{\'\i}a}, G.~C., {et~al.} 2020,
  arXiv e-prints, arXiv:2006.09403

\bibitem[{{Press} {et~al.}(1992){Press}, {Teukolsky}, {Vetterling}, \&
  {Flannery}}]{Press:1992}
{Press}, W.~H., {Teukolsky}, S.~A., {Vetterling}, W.~T., \& {Flannery}, B.~P.
  1992, {Numerical recipes in FORTRAN. The art of scientific computing}

\bibitem[{{Price-Whelan} {et~al.}(2018){Price-Whelan}, {Sip{\H{o}}cz},
  {G{\"u}nther}, {Lim}, {Crawford}, {Conseil}, {Shupe}, {Craig}, {Dencheva},
  {Ginsburg}, {VanderPlas}, {Bradley}, {P{\'e}rez-Su{\'a}rez}, {de Val-Borro},
  {Paper Contributors}, {Aldcroft}, {Cruz}, {Robitaille}, {Tollerud},
  {Coordination Committee}, {Ardelean}, {Babej}, {Bach}, {Bachetti}, {Bakanov},
  {Bamford}, {Barentsen}, {Barmby}, {Baumbach}, {Berry}, {Biscani}, {Boquien},
  {Bostroem}, {Bouma}, {Brammer}, {Bray}, {Breytenbach}, {Buddelmeijer},
  {Burke}, {Calderone}, {Cano Rodr{\'\i}guez}, {Cara}, {Cardoso}, {Cheedella},
  {Copin}, {Corrales}, {Crichton}, {D{\textquoteright}Avella}, {Deil},
  {Depagne}, {Dietrich}, {Donath}, {Droettboom}, {Earl}, {Erben}, {Fabbro},
  {Ferreira}, {Finethy}, {Fox}, {Garrison}, {Gibbons}, {Goldstein}, {Gommers},
  {Greco}, {Greenfield}, {Groener}, {Grollier}, {Hagen}, {Hirst}, {Homeier},
  {Horton}, {Hosseinzadeh}, {Hu}, {Hunkeler}, {Ivezi{\'c}}, {Jain}, {Jenness},
  {Kanarek}, {Kendrew}, {Kern}, {Kerzendorf}, {Khvalko}, {King}, {Kirkby},
  {Kulkarni}, {Kumar}, {Lee}, {Lenz}, {Littlefair}, {Ma}, {Macleod},
  {Mastropietro}, {McCully}, {Montagnac}, {Morris}, {Mueller}, {Mumford},
  {Muna}, {Murphy}, {Nelson}, {Nguyen}, {Ninan}, {N{\"o}the}, {Ogaz}, {Oh},
  {Parejko}, {Parley}, {Pascual}, {Patil}, {Patil}, {Plunkett}, {Prochaska},
  {Rastogi}, {Reddy Janga}, {Sabater}, {Sakurikar}, {Seifert}, {Sherbert},
  {Sherwood-Taylor}, {Shih}, {Sick}, {Silbiger}, {Singanamalla}, {Singer},
  {Sladen}, {Sooley}, {Sornarajah}, {Streicher}, {Teuben}, {Thomas},
  {Tremblay}, {Turner}, {Terr{\'o}n}, {van Kerkwijk}, {de la Vega}, {Watkins},
  {Weaver}, {Whitmore}, {Woillez}, {Zabalza}, \& {Contributors}}]{astropy:2018}
{Price-Whelan}, A.~M., {Sip{\H{o}}cz}, B.~M., {G{\"u}nther}, H.~M., {et~al.}
  2018, \aj, 156, 123

\bibitem[{{Reddy} {et~al.}(2006){Reddy}, {Lambert}, \& {Allende
  Prieto}}]{Reddy2006}
{Reddy}, B.~E., {Lambert}, D.~L., \& {Allende Prieto}, C. 2006, \mnras, 367,
  1329

\bibitem[{Rein \& Liu(2012)}]{rein2012}
Rein, H. \& Liu, S.~F. 2012, Astronomy and Astrophysics, 537

\bibitem[{Rein \& Tamayo(2015)}]{rein2015}
Rein, H. \& Tamayo, D. 2015, Monthly Notices of the Royal Astronomical Society,
  452, 376

\bibitem[{{Ricker} {et~al.}(2015){Ricker}, {Winn}, {Vanderspek}, {Latham},
  {Bakos}, {Bean}, {Berta-Thompson}, {Brown}, {Buchhave}, {Butler}, {Butler},
  {Chaplin}, {Charbonneau}, {Christensen-Dalsgaard}, {Clampin}, {Deming},
  {Doty}, {De Lee}, {Dressing}, {Dunham}, {Endl}, {Fressin}, {Ge}, {Henning},
  {Holman}, {Howard}, {Ida}, {Jenkins}, {Jernigan}, {Johnson}, {Kaltenegger},
  {Kawai}, {Kjeldsen}, {Laughlin}, {Levine}, {Lin}, {Lissauer}, {MacQueen},
  {Marcy}, {McCullough}, {Morton}, {Narita}, {Paegert}, {Palle}, {Pepe},
  {Pepper}, {Quirrenbach}, {Rinehart}, {Sasselov}, {Sato}, {Seager},
  {Sozzetti}, {Stassun}, {Sullivan}, {Szentgyorgyi}, {Torres}, {Udry}, \&
  {Villasenor}}]{Ricker:2015}
{Ricker}, G.~R., {Winn}, J.~N., {Vanderspek}, R., {et~al.} 2015, Journal of
  Astronomical Telescopes, Instruments, and Systems, 1, 014003

\bibitem[{{Robinson} {et~al.}(2017){Robinson}, {Fortney}, \&
  {Hubbard}}]{Robinson2017}
{Robinson}, T.~D., {Fortney}, J.~J., \& {Hubbard}, W.~B. 2017, \apj, 850, 128

\bibitem[{{Saumon} {et~al.}(1995){Saumon}, {Chabrier}, \& {van
  Horn}}]{Saumon1995}
{Saumon}, D., {Chabrier}, G., \& {van Horn}, H.~M. 1995, \apjs, 99, 713

\bibitem[{{Scargle}(1982)}]{scargle:1982}
{Scargle}, J.~D. 1982, \apj, 263, 835

\bibitem[{{Schlafly} \& {Finkbeiner}(2011)}]{Schlafly:2011}
{Schlafly}, E.~F. \& {Finkbeiner}, D.~P. 2011, \apj, 737, 103

\bibitem[{{Seager} {et~al.}(2007){Seager}, {Kuchner}, {Hier-Majumder}, \&
  {Militzer}}]{Seager2007}
{Seager}, S., {Kuchner}, M., {Hier-Majumder}, C.~A., \& {Militzer}, B. 2007,
  \apj, 669, 1279

\bibitem[{{Sidis} \& {Sari}(2010)}]{Sidis2010}
{Sidis}, O. \& {Sari}, R. 2010, \apj, 720, 904

\bibitem[{{Smith} {et~al.}(2012){Smith}, {Stumpe}, {Van Cleve}, {Jenkins},
  {Barclay}, {Fanelli}, {Girouard}, {Kolodziejczak}, {McCauliff}, {Morris}, \&
  {Twicken}}]{Smith2012}
{Smith}, J.~C., {Stumpe}, M.~C., {Van Cleve}, J.~E., {et~al.} 2012, \pasp, 124,
  1000

\bibitem[{Sotin {et~al.}(2007)Sotin, Grasset, \& Mocquet}]{sotin_2007}
Sotin, C., Grasset, O., \& Mocquet, A. 2007, Icarus, 191, 337

\bibitem[{{Stassun} {et~al.}(2017){Stassun}, {Collins}, \&
  {Gaudi}}]{Stassun:2017}
{Stassun}, K.~G., {Collins}, K.~A., \& {Gaudi}, B.~S. 2017, \aj, 153, 136

\bibitem[{{Stassun} {et~al.}(2018){Stassun}, {Oelkers}, {Pepper}, {Paegert},
  {De Lee}, {Torres}, {Latham}, {Charpinet}, {Dressing}, {Huber}, {Kane},
  {L{\'e}pine}, {Mann}, {Muirhead}, {Rojas-Ayala}, {Silvotti}, {Fleming},
  {Levine}, \& {Plavchan}}]{Stassun:2018}
{Stassun}, K.~G., {Oelkers}, R.~J., {Pepper}, J., {et~al.} 2018, \aj, 156, 102

\bibitem[{{Stassun} \& {Torres}(2016)}]{Stassun:2016}
{Stassun}, K.~G. \& {Torres}, G. 2016, \aj, 152, 180

\bibitem[{{Stassun} \& {Torres}(2018)}]{Stassun:2018a}
{Stassun}, K.~G. \& {Torres}, G. 2018, \apj, 862, 61

\bibitem[{Steffen(2006)}]{Steffen:2006}
Steffen, J. 2006, 107

\bibitem[{{Strehl}(1902)}]{strehl1902}
{Strehl}, K. 1902, Astronomische Nachrichten, 158, 89

\bibitem[{{Stumpe} {et~al.}(2014){Stumpe}, {Smith}, {Catanzarite}, {Van Cleve},
  {Jenkins}, {Twicken}, \& {Girouard}}]{Stumpe2014}
{Stumpe}, M.~C., {Smith}, J.~C., {Catanzarite}, J.~H., {et~al.} 2014, \pasp,
  126, 100

\bibitem[{{Stumpe} {et~al.}(2012){Stumpe}, {Smith}, {Van Cleve}, {Twicken},
  {Barclay}, {Fanelli}, {Girouard}, {Jenkins}, {Kolodziejczak}, {McCauliff}, \&
  {Morris}}]{Stumpe2012}
{Stumpe}, M.~C., {Smith}, J.~C., {Van Cleve}, J.~E., {et~al.} 2012, \pasp, 124,
  985

\bibitem[{{Ter Braak}(2006)}]{Braak:2006}
{Ter Braak}, C. J.~F. 2006, Statistics and Computing, 16, 239

\bibitem[{{Twicken} {et~al.}(2018){Twicken}, {Catanzarite}, {Clarke},
  {Girouard}, {Jenkins}, {Klaus}, {Li}, {McCauliff}, {Seader}, {Tenenbaum},
  {Wohler}, {Bryson}, {Burke}, {Caldwell}, {Haas}, {Henze}, \&
  {Sanderfer}}]{Twicken:2018}
{Twicken}, J.~D., {Catanzarite}, J.~H., {Clarke}, B.~D., {et~al.} 2018, \pasp,
  130, 064502

\bibitem[{{Twicken} {et~al.}(2010){Twicken}, {Clarke}, {Bryson}, {Tenenbaum},
  {Wu}, {Jenkins}, {Girouard}, \& {Klaus}}]{Twicken2010}
{Twicken}, J.~D., {Clarke}, B.~D., {Bryson}, S.~T., {et~al.} 2010, Society of
  Photo-Optical Instrumentation Engineers (SPIE) Conference Series, Vol. 7740,
  {Photometric analysis in the Kepler Science Operations Center pipeline},
  774023

\bibitem[{{Van Eylen} {et~al.}(2018){Van Eylen}, {Agentoft}, {Lundkvist},
  {Kjeldsen}, {Owen}, {Fulton}, {Petigura}, \& {Snellen}}]{VanEylen2018}
{Van Eylen}, V., {Agentoft}, C., {Lundkvist}, M.~S., {et~al.} 2018, \mnras,
  479, 4786

\bibitem[{{Vogt} {et~al.}(1994){Vogt}, {Allen}, {Bigelow}, {Bresee}, {Brown},
  {Cantrall}, {Conrad}, {Couture}, {Delaney}, {Epps}, {Hilyard}, {Hilyard},
  {Horn}, {Jern}, {Kanto}, {Keane}, {Kibrick}, {Lewis}, {Osborne},
  {Pardeilhan}, {Pfister}, {Ricketts}, {Robinson}, {Stover}, {Tucker}, {Ward},
  \& {Wei}}]{1994SPIE.2198..362V}
{Vogt}, S.~S., {Allen}, S.~L., {Bigelow}, B.~C., {et~al.} 1994, Society of
  Photo-Optical Instrumentation Engineers (SPIE) Conference Series, Vol. 2198,
  {HIRES: the high-resolution echelle spectrometer on the Keck 10-m Telescope},
  ed. D.~L. {Crawford} \& E.~R. {Craine}, 362

\bibitem[{Vrugt {et~al.}(2009)Vrugt, Braak, Diks, Robinson, Hyman, \&
  Higdon}]{Vrugt:2009}
Vrugt, J.~A., Braak, C. J. F.~T., Diks, C. G.~H., {et~al.} 2009, in VRUGT ET
  AL.: TREATMENT OF FORCING DATA ERROR USING MCMC SAMPLING

\bibitem[{{Wakeford} {et~al.}(2019){Wakeford}, {Lewis}, {Fowler}, {Bruno},
  {Wilson}, {Moran}, {Valenti}, {Batalha}, {Filippazzo}, {Bourrier},
  {H{\"o}rst}, {Lederer}, \& {de Wit}}]{Wakeford2019}
{Wakeford}, H.~R., {Lewis}, N.~K., {Fowler}, J., {et~al.} 2019, \aj, 157, 11

\bibitem[{Wang {et~al.}(2015)Wang, Fischer, Xie, \& Ciardi}]{wang2015influence}
Wang, J., Fischer, D.~A., Xie, J.-W., \& Ciardi, D.~R. 2015, The Astrophysical
  Journal, 813, 130

\bibitem[{Winn {et~al.}(2008)Winn, Holman, Torres, McCullough, Johns-Krull,
  Latham, Shporer, Mazeh, Garcia-Melendo, Foote, Esquerdo, \&
  Everett}]{Winn:2008b}
Winn, J.~N., Holman, M.~J., Torres, G., {et~al.} 2008, \apj, 683, 1076

\bibitem[{Wood {et~al.}(2017)Wood, Horner, Hinse, \&
  Marsden}]{Wood2017TheRings}
Wood, J., Horner, J., Hinse, T.~C., \& Marsden, S.~C. 2017, The Astronomical
  Journal, 153, 245

\bibitem[{{Wu} {et~al.}(2007){Wu}, {Murray}, \& {Ramsahai}}]{Wu2007}
{Wu}, Y., {Murray}, N.~W., \& {Ramsahai}, J.~M. 2007, \apj, 670, 820

\bibitem[{Yee {et~al.}(2017)Yee, Petigura, \& Von~Braun}]{yee2017precision}
Yee, S.~W., Petigura, E.~A., \& Von~Braun, K. 2017, The Astrophysical Journal,
  836, 77

\bibitem[{{Zacharias} {et~al.}(2013){Zacharias}, {Finch}, {Girard}, {Henden},
  {Bartlett}, {Monet}, \& {Zacharias}}]{Zacharias:2013}
{Zacharias}, N., {Finch}, C.~T., {Girard}, T.~M., {et~al.} 2013, \aj, 145, 44

\end{thebibliography}

\end{document}